\documentclass[12pt]{article}

\pdfoutput=1
\usepackage{color}
\usepackage{epsfig, palatino}
\usepackage{pstricks,pst-node,pst-tree}
\usepackage{epic}
\usepackage{mathrsfs}
\usepackage{physics}
\usepackage{ae} 
\usepackage[T1]{fontenc}
\usepackage[ansinew]{inputenc}
\usepackage{amsmath}

\usepackage{ytableau}
\usepackage{mdframed}

\usepackage{amssymb}
\usepackage{graphicx}
\usepackage{ulem}
\usepackage{color}
\definecolor{darkblue}{cmyk}{0.9,0.9,0,0}
\usepackage[colorlinks=true,linkcolor=darkblue,citecolor=darkblue,urlcolor=darkblue]{hyperref}
\usepackage{cite}
\usepackage{hyperref}
\usepackage{wasysym}
\usepackage{varioref}
\usepackage{makeidx}
\usepackage[english]{babel}
\usepackage{simplewick}
\usepackage{array}
\usepackage{multirow}

\newcounter{mycounter} 

\usepackage{lipsum}
\usepackage{slashed}

\usepackage{animate}
\usepackage{mdframed}
\usepackage[font={small}]{caption}
\usepackage{nicematrix}

\newcommand{\comment}[1]{}

\newcommand{\beq}{\begin{equation}}
\newcommand{\eeq}{\end{equation}}
\newcommand{\beqq}{\begin{equation*}}
\newcommand{\eeqq}{\end{equation*}}
\newcommand\beqa{\begin{eqnarray}}
\newcommand\eeqa{\end{eqnarray}}
\newcommand\beqaa{\begin{eqnarray*}}
\newcommand\eeqaa{\end{eqnarray*}}
\newcommand\bea{\begin{array}}
\newcommand\eea{\end{array}}

\newcommand{\nn}{\nonumber}

\newcommand{\diag}[1]{{\rm diag}(#1)} 
\newcommand{\neqa}{\nonumber\end{eqnarray}} 
\newcommand{\la}[1]{\label{#1}}

\renewcommand{\d}{\partial}

\newcommand{\<}{{\langle}}
\renewcommand{\>}{{\rangle}}

\newcommand{\re}{\relax{\rm I\kern-.18em R}}

\renewcommand{\sp}{p\hspace{-.40em}/}

\definecolor{darkgreen}{rgb}{0.0, 0.45, 0.0}

\def\Xint#1{\mathchoice
{\XXint\displaystyle\textstyle{#1}}
{\XXint\textstyle\scriptstyle{#1}}
{\XXint\scriptstyle\scriptscriptstyle{#1}}
{\XXint\scriptscriptstyle\scriptscriptstyle{#1}}
\!\int}
\def\XXint#1#2#3{{\setbox0=\hbox{$#1{#2#3}{\int}$}
\vcenter{\hbox{$#2#3$}}\kern-.5\wd0}}

\def\dashint{\Xint-}

\def\tr{{\rm tr~}}

\def\su2{{SU(2)}}

\def\eps{{\epsilon}}

\def\a{{\alpha}}

\def\[{\left[}
\def\]{\right]}

\def\a{\alpha}

\def\<{\langle}
\def\>{\rangle}

\def\sG{\,\slash\!\!\!\! G}

\def\i2{\frac{i}{2}}

\def\spi{\relax{\rm \pi\kern-0.5em /}}
\def\sA{\relax{\rm A\kern-0.5em /}}
\def\sp{\relax{\rm p\kern-0.5em /}}
\def\sd{\relax{\rm \d\kern-0.5em /}}
\def\sk{\relax{\rm k\kern-0.5em /}}
\def\sn{\relax{\rm n\kern-0.5em /}}
\def\sl{\relax{\rm l\kern-0.5em /}}
\def\sP{\relax{\rm P\kern-0.7em /}}
\def\sBethe{\relax{\rm \Bethe\kern-0.5em /}}

\def\2F1{\,_2{\rm F}_1}

\makeindex

\def\Xint#1{\mathchoice
   {\XXint\displaystyle\textstyle{#1}}%
   {\XXint\textstyle\scriptstyle{#1}}%
   {\XXint\scriptstyle\scriptscriptstyle{#1}}%
   {\XXint\scriptscriptstyle\scriptscriptstyle{#1}}%
   \!\int}
\def\XXint#1#2#3{{\setbox0=\hbox{$#1{#2#3}{\int}$}
     \vcenter{\hbox{$#2#3$}}\kern-.5\wd0}}

\def\dashint{\Xint-}

\begin{document}

\thispagestyle{empty}

\renewcommand{\thefootnote}{\fnsymbol{footnote}}
\setcounter{page}{1}
\setcounter{footnote}{0}
\setcounter{figure}{0}

\begin{center}
$$$$
{\Large\textbf{\mathversion{bold}
Huge BPS Operators and Fluid Dynamics in $\mathcal{N}=4$ SYM
}\par}

\vspace{1.0cm}

\textrm{Vladimir Kazakov$^\text{\tiny 1}$\footnote{\tt  Vladimir.Kazakov@ens.fr}, Harish Murali$^\text{\tiny 2,\tiny3}$\footnote{\tt  harish02murali@gmail.com}, Pedro Vieira$^\text{\tiny 2,\tiny 4}$\footnote{\tt  pedrogvieira@gmail.com}}

\vspace{1.2cm}
\footnotesize{\textit{
$^\text{\tiny 1}$Laboratoire de Physique de l'\'Ecole Normale Sup\'erieure, CNRS,
Universit\'e PSL, Sorbonne Universit\'es,\\
24 rue Lhomond, 75005 Paris, France
\\
$^\text{\tiny 2}$Perimeter Institute for Theoretical Physics,
Waterloo, Ontario N2L 2Y5, Canada\\
$^\text{\tiny 3}$Department of Physics and Astronomy, University of Waterloo, Waterloo, Ontario, N2L 3G1, Canada
$^\text{\tiny 4}$ICTP South American Institute for Fundamental Research, IFT-UNESP, S\~ao Paulo, SP Brazil 01440-070 \\
}  
\vspace{4mm}
}

\par\vspace{1.5cm}

\textbf{Abstract}\vspace{2mm}
\end{center}
In the bulk dual of holography, huge operators correspond to sources so heavy that they fully backreact on the space-time geometry. Here we study the correlation function of three such huge operators when they are given by $1/2$ BPS operators in $\mathcal{N}=4$ SYM theory, dual to IIB Strings in $AdS_5 \times S^5$. We unveil simple matrix model representations for these correlators which we can sometimes solve analytically. For general huge operators, we translate these matrix model expressions into a $1+1$ dimensional hydrodynamical fluid problem. This fluid is integrable thus unveiling a novel integrable sector of the $AdS/CFT$ duality in a full fledged gravitational regime, very far from the usual free string planar regime where integrability reigns supreme. We explain how an adiabatic deformation method can be developed to yield the solution to an integrable discrete formulation of these fluids -- the rational Calogero-Moser Model -- so we can access the general three point correlation functions of generic huge $1/2$-BPS operators. Everything will be done on the gauge theory side of the duality. It would be fascinating to find the holographic dual of these matrix models and fluids.  
\noindent \\

\setcounter{page}{1}
\renewcommand{\thefootnote}{\arabic{footnote}}
\setcounter{footnote}{0}

\setcounter{tocdepth}{2}

 \def\nref#1{{(\ref{#1})}}
\newpage
\tableofcontents
\parskip 5pt plus 1pt   \jot = 1.5ex

\newpage

\section{Introduction} 

\textit{\small What is the microscopic description of a space-time geometry? How can we derive its dynamics from this underlying description? Can we think of spacetime as a single micro-state? Which one? Does it matter? How much does it matter? For which quantities does it matter? What about more esoteric objects like universe decays, the probability that a given geometry tunnels into two disconnected geometries, a baby universe creation of sorts. What induces it and how universal are such decays? How unlikely are such processes and what do they teach us about the quantum nature of gravity? 
}

Such questions lead themselves to great passionate discussions with a plethora of diverse (albeit definite) conclusions.

One of the fascinating prospects on AdS/CFT is that we might be able to translate such interesting speculative questions into bland, well defined problems in conformal field theory, with mathematically sharp answers.  

In a CFT we have operators and their correlation functions. Operators are called \textit{huge}/\textit{small} when their conformal dimension is \textit{huge}/\textit{small}. In the dual AdS picture, a small operator behaves as a probe moving on a fixed geometry while a huge operator can backreact on the geometry itself.\footnote{We could further subdivide probes into heavy and light. Light probes, like gravitons or other very light states, are quantum. Heavy probes like highly exciting strings or D-branes are big classical objects (but not as big as the \textit{Huge} operators which can backreact on the geometry).   Correlation functions of three light operators were considered already in the most early days of $AdS/CFT$, see e.g. \cite{Witten:1998qj,Freedman:1998tz}. Correlation functions of three heavy operators were only considered much more recently, see e.g. \cite{Janik:2010gc,Janik:2011bd,Kazama:2011cp,Minahan:2012fh,Aprile:2020luw,Jiang:2019xdz,Coronado:2018cxj}. Inspiring as these might be, most of the tools developed there do \textit{not} translate to the case at hand of \textit{huge} geometry deforming operators. For those we need to develop new tools; that is the main subject of this paper.}
A transition involving three geometries should be related to a three point function of three huge operators \cite{Abajian:2023jye,Abajian:2023bqv}. A transition between a geometry into another geometry could be related to a three point function involving two huge operators and a small operator. The expectation value of a probe in a fixed geometry should be related to the same setup with the two huge operators being very similar. And so on. In the end, all these fascinating questions should become sharp questions about correlators. 

\begin{figure}[t]
    \centering
    \includegraphics[width=\textwidth]{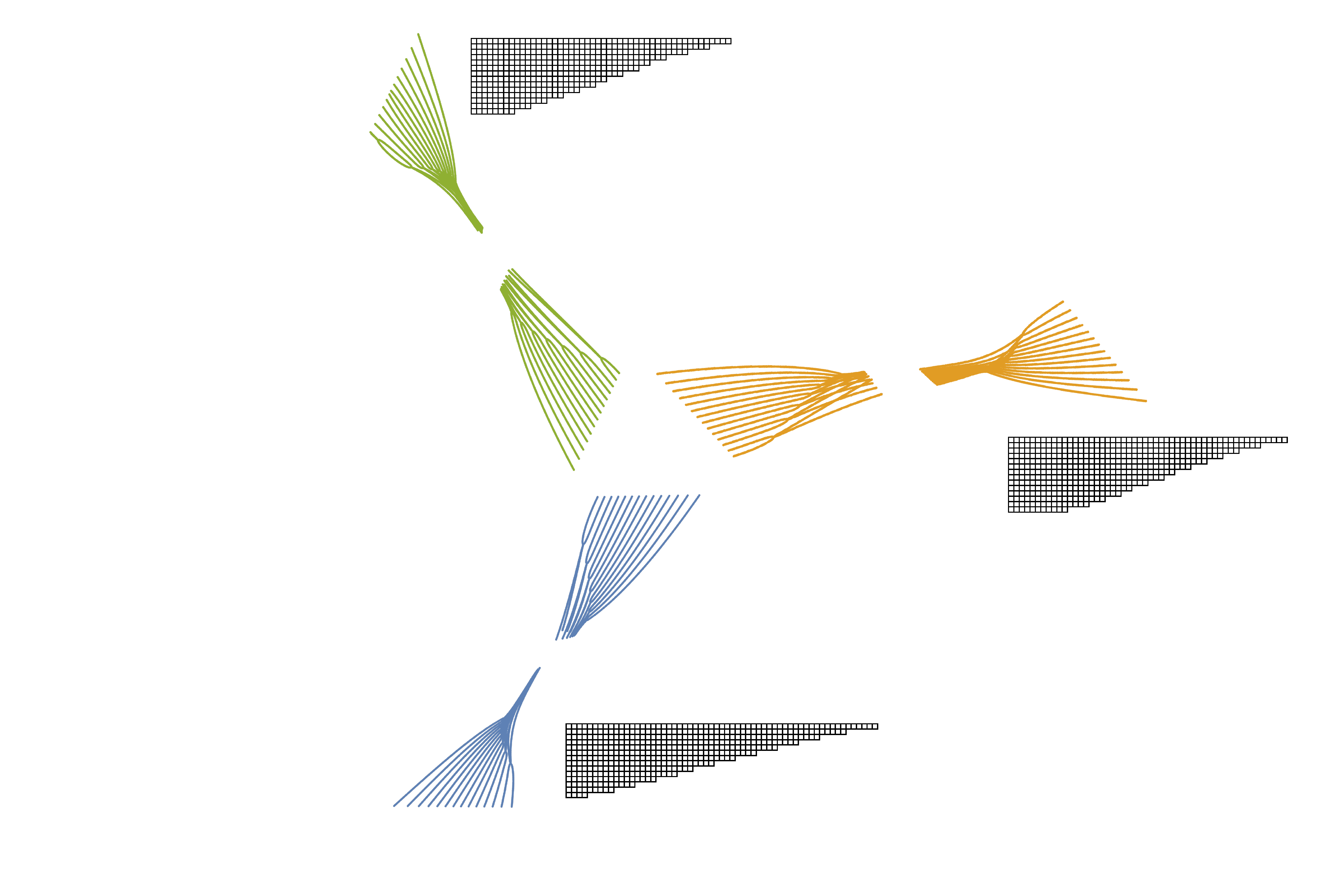}
    \caption{An example of the type of (complex) one dimensional fluid flows studied in this paper. They start at the outer edges with densities given by the densities of the three Young Tableaux shown, and are glued together at the middle by a regularity condition. The lines drawn here are positions $x_i(t)$ of the fluid bits whose locations are governed by an integrable discrete model known as the Calogero-Moser rational model, see (\ref{discrete}).}
    \label{fig:asymmetricFlow}
\end{figure}

Here we will study a particular set of operators, called half-BPS operators $O_i$, in a particular gauge theory, called $\mathcal{N}=4$ super Yang-Mills theory. These operators are a tiny subset of all possible operators in this gauge theory but they are already very interesting. We know that the operators $O_i$ describe a gas of massless strings (when the number of fields in the operator is finite), D-branes (when the number of fields in the operator scales linearly with $N$) or ``geometry changing" objects (when the number of fields scales quadratically with $N$), which induce $\frac12$-BPS geometries found by Lin-Lunin-Maldacena \cite{Lin:2004nb} in the vicinity of the operator insertion.

We will discuss correlation functions of three protected operators \footnote{Here the representation $\mathcal{R}_i$ parametrizes the operator which is oriented along an R-charge direction parametrized by the six dimensional null vector $y_i$ and which is inserted at the four dimensional space-time location $x_i$.  $\chi_{\mathcal{R}}(\Phi)$ is the character of the field $\Phi$ in the $U(N)$ representation $\mathcal{R}$ and $ \mathcal{N}_{\mathcal{R}_i}$ are simple normalization factors. More details in the next section. These so-called Schur operators were introduced in \cite{Corley:2001zk}.}
\beq
\hat{Z}_3=\left\langle\prod_{i=1}^3  {O}_i \right\rangle  \,,\qquad   {O}_i \equiv \mathcal{N}_{\mathcal{R}_i}\, \chi_{\mathcal{R}_i}(y_i \cdot \phi(x_i))
\eeq
This object is at the same time trivial and extremely rich. 
On the one hand it is trivial: it is protected by supersymmetry, it has no coupling dependence and is obtained by simply Wick contracting the three operators at tree-level. On the other hand, preforming these simple Wick contractions is by no means trivial, especially for very large representations with each operator containing $\mathcal{O}(N^2)$ fields. It then becomes a rich combinatorial problem which leads to a beautiful matrix model problem describable through a rich world of $(1+1)D$ complex fluid dynamics of the sort shown in figure \ref{fig:asymmetricFlow}. Identifying these flows and studying its implications is the main purpose of this paper. We also managed to compute explicitly the 3-point functions for some particular  huge operators showing richness and complexity of these 1/2-BPS quantities. These are the unique such examples to our knowledge.  

We stress that this paper is about the CFT side. A beautiful question -- probably the most interesting one -- is how to translate the results herein obtained to the AdS bulk. We briefly speculate about some of this in the discussion, being fully aware of the fact that there is still a lot to understand here and in particular to relate in general the full fluid dynamics of CFT side  to the AdS side!

\section{Exact Description -- Matrix Models}  \label{matrixSec}

The three-point correlators are given by a simple kinematical factor uniquely fixed by conformal symmetry times a number $Z_{3}$ which only depends on the three representations~$\mathcal{R}_1$,~$\mathcal{R}_2$ and $\mathcal{R}_3$ of the three operators -- and of course on the number of colours $N$ -- and which is obtained  by simply Wick contracting the scalars between the operators. That is,
\beq
\hat{Z}_3= \texttt{kinematic} \times Z_3 \,, \qquad  \texttt{kinematic}  = \prod_{1\le i< j\le 3} \left( \frac{y_i\cdot y_j}{(x_i-x_j)^2}\right)^{\ell_{ij}}  \label{kin}
\eeq
where the bridge lengths $\ell_{ij}=(L_i+L_j-L_k)/2$ (where $k\neq i,j$) are the number of propagators connecting the various operators $\mathcal{O}_i$ which are made of $L_i$ fields each.\footnote{When some $\ell_{ij}=0$ there are no propagators between $\mathcal{O}_i$ and $\mathcal{O}_{j}$; instead since $L_k=L_{i}+L_j$ all fields in $\mathcal{O}_i$ and $\mathcal{O}_{j}$ connect to the bigger operator $\mathcal{O}_k$ and this correlator is called extremal. Extremal correlators have been studied intensively in the BMN times by the turn of the millennium, see for example \cite{Corley:2001zk,Kristjansen:2002bb}. Extremal correlators are way simpler than the generic non-extremal correlators which will occupy us here. In \cite{ZoharJaume} a very rich class of extremal correlators was studied in less supersymmetric theories. It would be fascinating to generalize those results and consider less extremal correlators there as well.  
} Each representation~$\mathcal{R}_i$ is given by a Young tableaux with~$L_i$ boxes. For example, we could have an operator with four fields given by  
\beq
\mathcal{R} = \ydiagram{2,2}
 \,, \qquad \chi_{R}(\Phi) = \tfrac{1}{8} \text{tr}(\Phi)^4+ \tfrac{1}{4} \text{tr}(\Phi^4)- \tfrac{1}{4} \text{tr}(\Phi^2)\text{tr}(\Phi)^2-\tfrac{1}{8}\text{tr}(\Phi^2)^2 \label{example22}
\eeq
and so on.\footnote{We recall known formulae relating the characters 
$\chi_\mathcal{R}(\Phi)$ 
to the eigenvalues~$\phi_j$ of~$\Phi$, as well as to multi-traces of powers of $\Phi$, 
\beq
\chi_{\mathcal{R}}(\Phi)= \frac{\det\limits_{i,j\le N} \phi_j^{h_i}}{\Delta(\phi)} \,,\qquad \qquad 
\chi_\mathcal{R}(\Phi)=\det_{i,j\le H} 
\oint \frac{dz}{z^{\lambda_j+i-j+1}} \exp \sum_{k=1}^\infty \frac{z^k}{k} \text{tr}(\Phi^k) \,.
\eeq
Here $\lambda_j$ is the number of boxes of the $j$-th of the Young-Tableau -- with H non-empty rows -- associated to the representation $\mathcal{R}$ while $h_j$ are the shifted row lengths defined in \ref{hDef} below. 
We encourage the unfamiliar reader to check these formulae on the  example (\ref{example22}). 
} The character basis is particularly nice since it forms an orthogonal basis of CFT operators \cite{Corley:2001zk}. The normalization constants above are chosen so that the basis is orthonormal, 
\beq
\langle   {O}_i    {O}_j \rangle =1\times \delta_{\mathcal{R}_i , \mathcal{R}_j}   \left( \frac{y_i\cdot y_j}{(x_i-x_j)^2}\right)^{2L_i} 
\eeq

Finally we have the dynamical factor $Z_3$ which is given by simply Wick contracting the fields between all operators $ {O}_i$ \textit{without} self-contractions within the operators themselves. We can thus count these Wick contractions by a matrix model where we code the fields of the three operator as a $N\times N$ matrix $M_i$ with a kinetic term such that the propagator between $M_i$ and $M_j$ is $\tfrac 1N$ for $i\neq j$ and $0$ for $i=j$. That leads us to $Z_3  =\mathcal{N} Z $ in terms of a simple~3-matrix model 
\begin{mdframed}[frametitlealignment=\centering,backgroundcolor=red!3, leftmargin=0cm, rightmargin=0cm, topline=false,
	bottomline=false, leftline=false, rightline=false] 
\beq
Z(\mathcal{R}_1,\mathcal{R}_2,\mathcal{R}_3) = \int \prod_{i=1}^3 d M_i \, \chi_{R_i}(M_i) \, e^{-N\,\text{tr}\Big(\frac12\sum\limits_i M_i^2-  \sum\limits_{i <j} M_i M_j  \Big) }  \label{M1M2M3}
\eeq
\end{mdframed}
This matrix model partition function $Z$ is the key player in this paper; it is our starting point. Similar matrix models solve beautiful combinatorial problems such as the  Potts model on  Random Graphs \cite{Kazakov:1987qg}, followed by its explicit solutions in \cite{Kostov:1988pe,Daul:1994qy,Eynard:1995nv,Eynard:1995zv}, etc. 

The normalization constant $\mathcal{N}$ is trivially given in terms of the general matrix model partition function by setting particular representations to become the trivial empty representation $\phi$ for which $\chi_\phi(M)=1$, 
\beq
\mathcal{N} = \sqrt{\frac{Z(\phi,\phi,\phi)}{\prod\limits_{i=1}^3 {Z(\mathcal{R}_i,\mathcal{R}_i,\phi)}}} \la{Ndef}
\eeq
This simple normalization nicely ensures the orthonormality property mentioned above. We are now going to manipulate (\ref{M1M2M3}) to arrive at various equivalent matrix model representations. At various intermediate steps we will find and drop simple factorized factors of the form~$\prod_{n=1}^3 f(\mathcal{R}_n,N)$. Dropping these normalization factors is totally fine since they all drop out anyway when multiplying the partition function $Z$ by its corresponding normalization factor~$\mathcal{N}$ in~(\ref{Ndef}) to obtain the physical structure constant $Z_3$.

\begin{figure}[t]
\begin{center}
\includegraphics[scale=0.5]{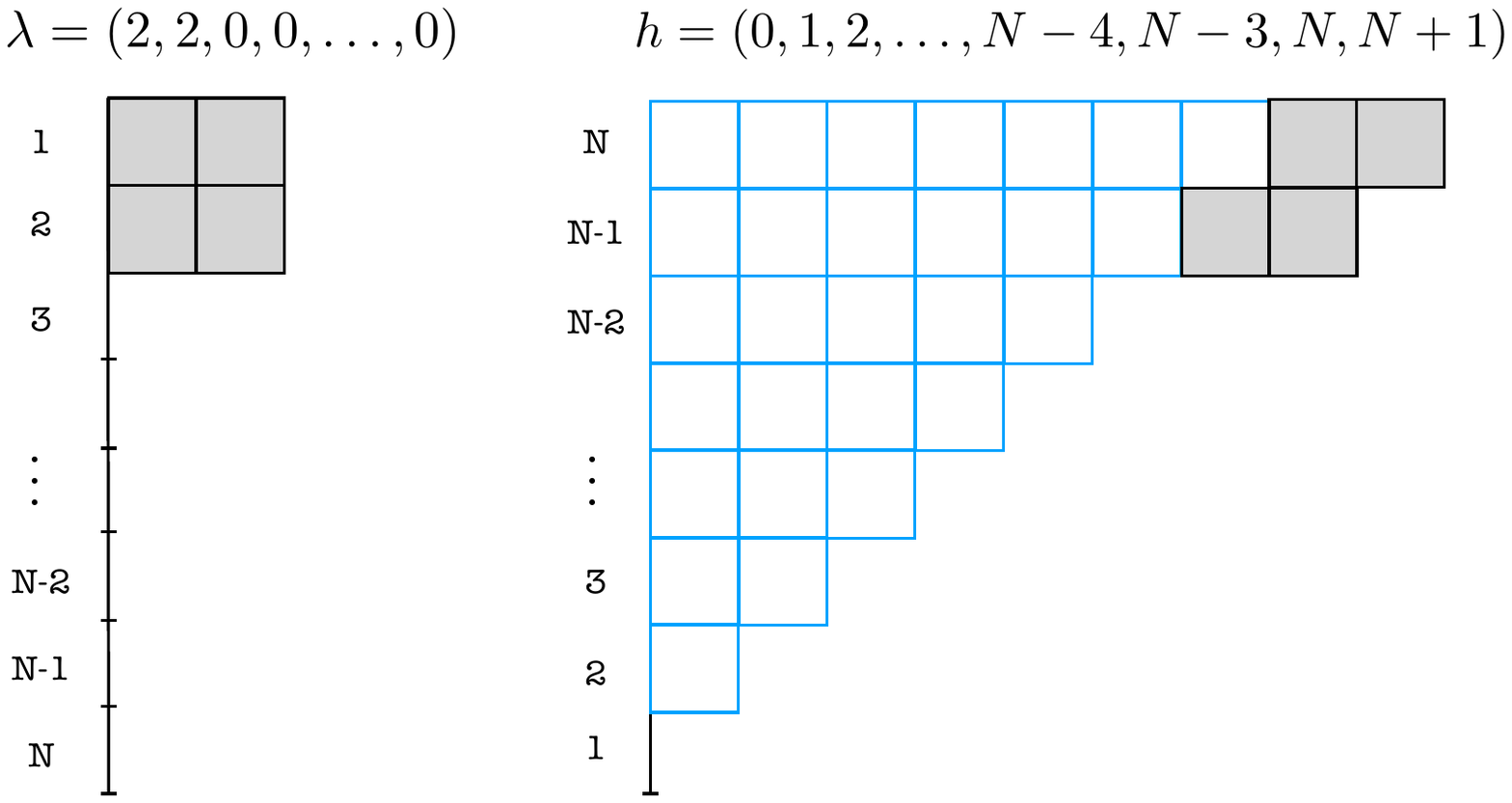}
\end{center}
\vspace{-3.5cm}
\caption{Each representation can be associated to a Young Tableau. On the left we represent a simple one discussed in (\ref{example22}) in the main text. Appearing everywhere will be the ``shifted highest weights" $h_j=\lambda_{N-j+1}+j-1$ which is the length of row $j$ of a shifted Young Tableau obtained from the original one by adding $j-1$ boxes to the $j$-th row. The empty Young Tableau with $\lambda=(0,0\dots,0)$ corresponds to a triangle shifted Young-Tableaux with $h=(0,1,2,3,\dots,N-1)$. 
 }  \label{lambdaH}
\end{figure}

The three matrix model in (\ref{M1M2M3}) can be reduced to a single matrix model!, albeit quite a non-trivial one. The simple idea is to introduce, following the trick of \cite{Kazakov:1987qg}, an auxiliary field $X$ to disentangle the~$M_i M_j $ interactions so that the dependences on the matrices $M_i$ totally factorize and can thus be easily integrated out. Details are in appendix \ref{3M1MA}. Then we end up with  a simple representation in terms of integration over the $N$ eigenvalues $x=(x_1,\dots,x_N)$ of the auxiliary matrix $X$, 
\begin{mdframed}[frametitlealignment=\centering,backgroundcolor=red!3, leftmargin=0cm, rightmargin=0cm, topline=false,
	bottomline=false, leftline=false, rightline=false] 
\begin{equation}
Z(\mathcal{R}_1,\mathcal{R}_2,\mathcal{R}_3) =\! \int \!d\mu(x) Q_1(x)Q_2(x)Q_3(x) \,, \qquad\!\!\!\!\! d\mu(x)=\prod_{i=1}^N dx_i\,  \Delta(x)^2\exp\Big(\!-\!\frac{N}{2} \sum\limits_{i=1}^N x_i^2\Big) 
 \label{singleM}   
\end{equation}
where $\Delta(x) \equiv  \prod_{i>j} (x_i-x_j)=\det_{i,j} x_i^{j-1}$ is the usual Vandermonde determinant and where, for each of the three operators, we have a wave function $Q$ given as a Slater determinant of Hermite polynomials~\footnote{We are abusing nomenclature a bit here -- of course, the usual Slater determinant is only the numerator of $Q_h(x)$ multiplied by exponential factors \la{slaterSubtletyFootnote}} as 
\beq
Q_n(x)=\frac{ 1 }{  \Delta(\hat x)} \, \det\limits_{1\le i,j \le N} H_{h_{n,j}}(\hat x_i)\,, \qquad \hat x= \sqrt{\frac{N}{2}} x\,,\label{Qdef}
\eeq
and where the information about the representation $\mathcal{R}_n$ is encoded in the very important shifted row lengths
\beq 
h_{n,j} = j-1+ (\text{number of boxes in row $N-j+1$ of the YT associated to $\mathcal{R}_n$})\,, \la{hDef}
\eeq 
see figure \ref{lambdaH}
\end{mdframed}
As a sanity check note that if we set one of the operators to be trivial, its $Q$-function trivializes to unity and we immediately recover the anticipated orthogonality for the remaining two point function. (Details in section \ref{2ptS}.)

\begin{figure}[t]
\begin{center}
\includegraphics[scale=0.6]{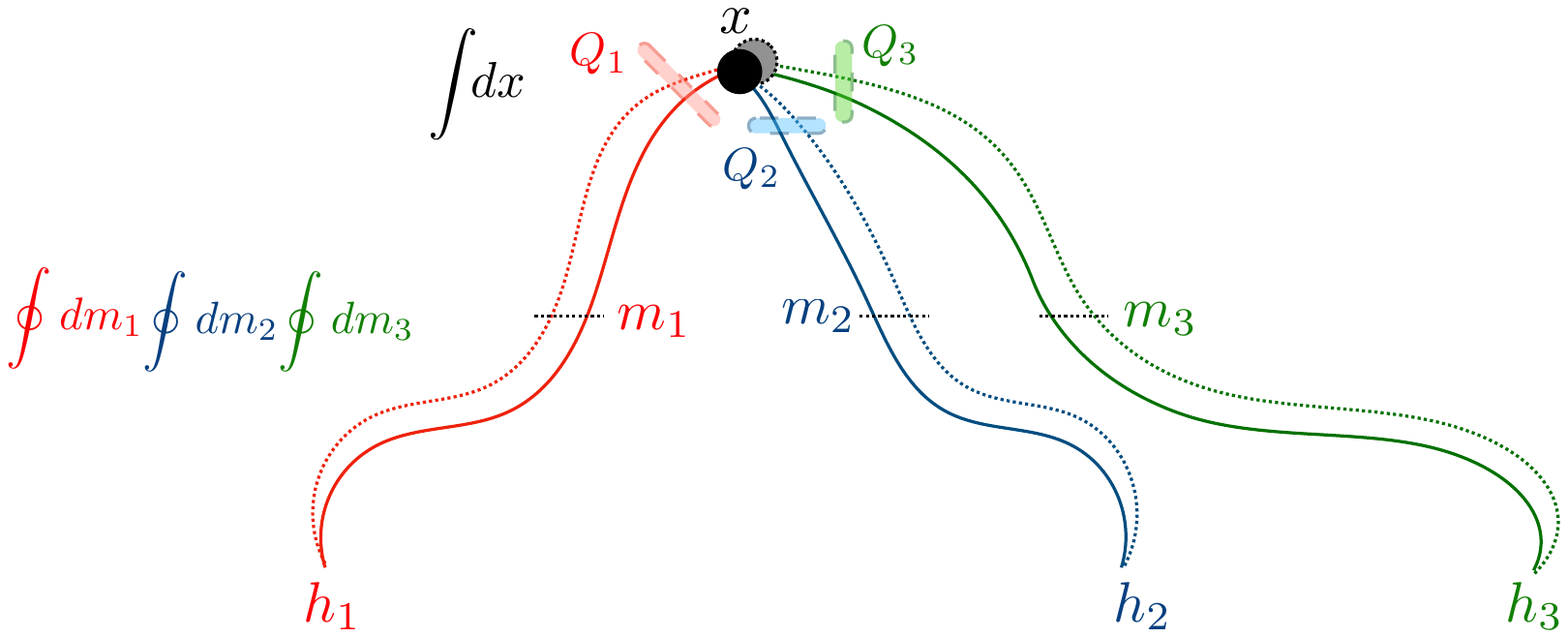}
\end{center}
\vspace{-6cm}
\caption{Graphical representation of the four-fold integral representation (\ref{fourInts}). The three sets of outer variables $h_n$ propagate into bulk variables $m_n$ which then propagate into a common point $x$. The points $m_n$ and $x$ are integrated over; in other words, these propagations are quantum. We can also think of the $m_n$ integrations as producing three quantum wave functions $Q_n(x)$ which are then glued together with a single last $x$ integration; that is precisely the representation (\ref{singleM}).
 }  \la{eChambers}
\end{figure}

For the purpose of computing correlators with a finite number of boxes -- or even for the purpose of computing large correlators with a finite number of rows -- we found the representation (\ref{singleM}) perfect. After all, this is an $N$ dimensional representation while the original matrix integral (\ref{M1M2M3}) was a $3N^2$ dimensional integral. If we use the generating function of Hermite polynomials we can trade each of them in (\ref{Qdef}) by an additional integral over an auxiliary variable leading to a very simple $4N$ dimensional representation (derivation details in appendix \ref{4intA})
\begin{mdframed}[frametitlealignment=\centering,backgroundcolor=red!3, leftmargin=0cm, rightmargin=0cm, topline=false,
	bottomline=false, leftline=false, rightline=false] 
\beqa
\!Z(\mathcal{R}_1,\mathcal{R}_2,\mathcal{R}_3) =\! \int \!d\mu(x) 
\prod_{n=1}^3 \oint d\mu(m_n) I(x,m_n)I(h_n,-\log m_n) \frac{\Delta(-\log m_n)}{\Delta(m_n)}\Delta(h_n)  \la{fourInts}
\eeqa
where (with $G$ being the Barnes G-function),
\beq
I(a,b)\equiv \frac{\det_{i,j} e^{N a_i b_i}}{\Delta(a)\Delta(b)} \times \frac{G(N+1)}{N^{N(N-1)/2}}\label{IZdef}
\eeq
can also be cast as an angular integral $I(a,b)= \int dU e^{N \text{tr}(A U^\dagger B U)}$ as shown by Harish-Chandra-Itzykson-Zuber.
\end{mdframed}

Using the Harish-Chandra-Itzykson-Zuber integral we could also have jumped directly from (\ref{M1M2M3}) to the representation (\ref{fourInts}) once we introduce the auxiliary matrix $X$ and recall that $\chi(M)$ can be written in terms of the $M$ eigenvalues as $\chi(M)= (\det_{i,j} m_i^{h_j}) /\Delta(m)$. In other words, once we introduce the auxiliary matrix $X$ to disentangle the $M_j$ interactions in (\ref{M1M2M3}), we end up with (\ref{singleM}) if we completely integrate out the matrices $M_i$ and we end up with (\ref{fourInts}) if we only integrate out the angular part of the matrices $M_i$. 

These matrix model representations are quantum expressions; they are exact, no approximation was taken. They are represented in figure \ref{eChambers}. 

\begin{figure}[t]
\begin{center}
\includegraphics[scale=0.6]{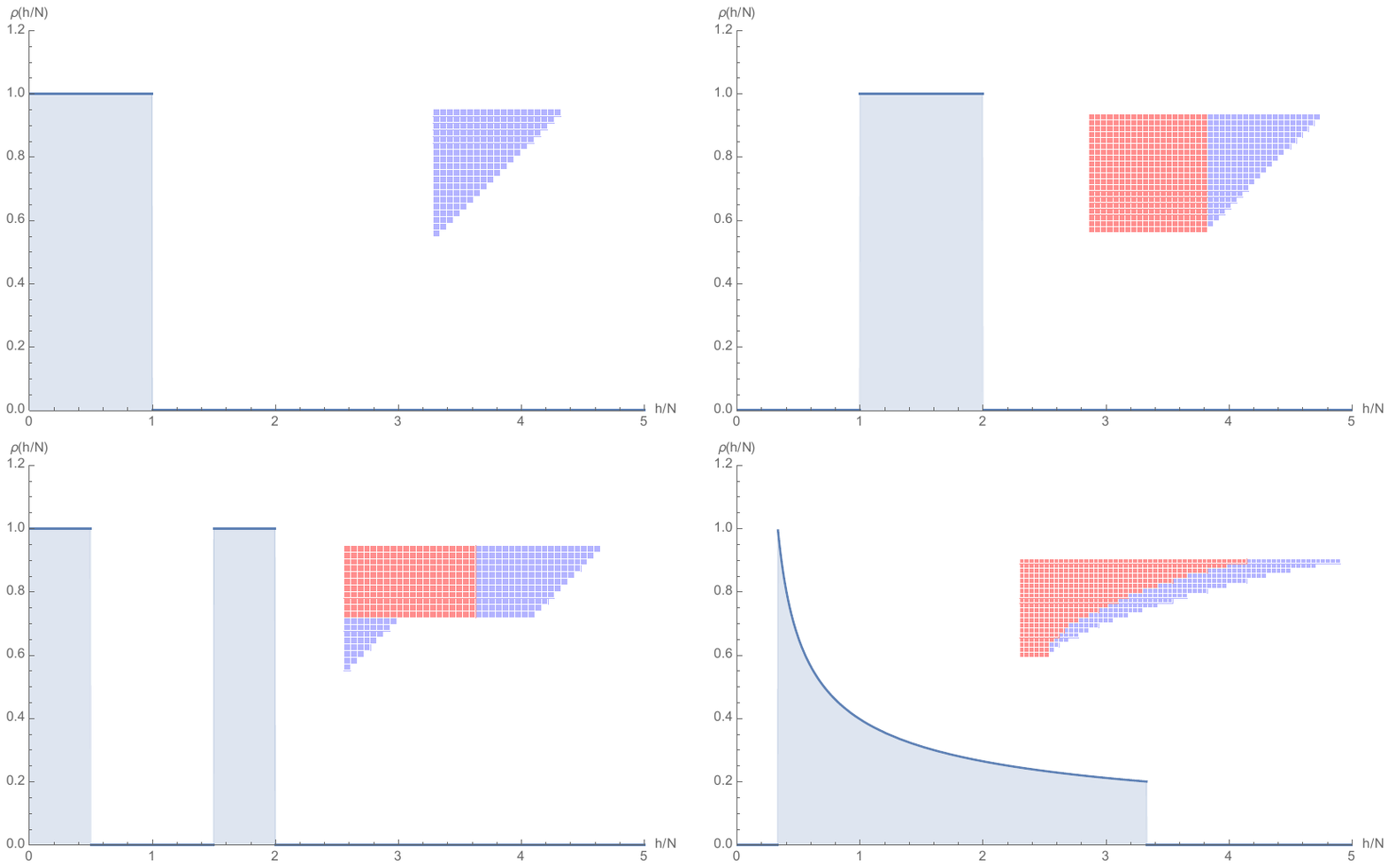}
\end{center}
\vspace{-2cm}
\caption{We consider here the large $N$ limit and very large representations with corresponding Young Tableaux with $\mathcal{O}(N^2)$ boxes. In the smooth classical limit we are considering here these can be nicely described by the density $\rho(h)$ of heights $h_j$ of the corresponding shifted Young Tableaux. 
 }  \label{Hhuge}
\end{figure}

We reach here the main focus of this paper: The limit of very large operators with $\mathcal{O}(N^2)$ fields with $N\to \infty$. These operators ought to correspond to backreacted geometries. For two point functions of huge operators, the map between the representation $\mathcal{R}$ and the corresponding geometry was suggested in a beautiful paper by Lin, Lunin and Maldacena \cite{Lin:2004nb}. It is not clear whether all large representations have a nice smooth classical dual but  at least for a typical map (probably corresponding to sufficiently smooth large YT's) the LLM proposal is applicable \cite{Balasubramanian:2005mg,deMelloKoch:2008hen,Berenstein:2004kk,Balasubramanian:2018yjq}. For the first three examples in figure \ref{Hhuge}, for instance, we would obtain  simply an annulus pattern of white and black regions describing the harmonic functions governing the ten dimensional dual geometry metric, as described in \cite{Lin:2004nb}.

Also, it is worth stressing that knowing the dual of a single operator does \textit{not} mean we have a nice gravity dual of a correlator of three such operators. After all, if all three operators are very heavy, all three  will simultaneously backreact on the geometry and the resulting picture will be a much richer geometry with three legs ending at the insertions of each of three operators -- close to which the metric ought to behave as expected from \cite{Lin:2004nb} up to the appropriate conformal transformations \cite{Abajian:2023jye,Abajian:2023bqv}. The physical dual picture will also strongly depend on any further relations between the operators. For instance, if all operators are huge but one is much smaller than the other two, and if we take these other two to be nearly identical, then the situation should simplify and become closer to the sort of probe heavy-heavy-light correlators people have extensively studied for strings (see e.g. \cite{Zarembo:2010rr,Costa:2010rz}) and branes (see e.g. \cite{Bissi:2011dc,deMelloKoch:2007rqf,Berenstein:2013md,Yang:2021kot,Jiang:2019zig}).

For the combinatorics analysis on the CFT side for very large operators, the two natural starting points for this analysis are the single matrix model representation (\ref{singleM}) and the four-matrix model representation (\ref{fourInts}). The first representation (\ref{singleM}) has a single integral but a more complicated integrand, with determinants of Hermite polynomials. We will consider it in detail in the next section where we will analyse some interesting cases of rectangular, trapezoidal and triangular YTs. The second has four integrals but the integrand is a very simple product of Itzykson-Zuber determinants (\ref{IZdef}) whose classical limit can be readily attacked borrowing some beautiful technology developed many years ago in \cite{Matytsin_1994,Das:1990kaa} in terms of fluid dynamics. This second approach is pursued next in section \ref{FluidSec}.

\section{Semi-Classics -- Slater Determinant Wave Functions} \label{SlaterSec}

In this section, we will consider a couple of particular cases of operators for which we can basically compute the three point functions -- either analytically or with very high precision numerics -- with ease. 

\begin{figure}
    \centering
    \includegraphics[width=\textwidth]{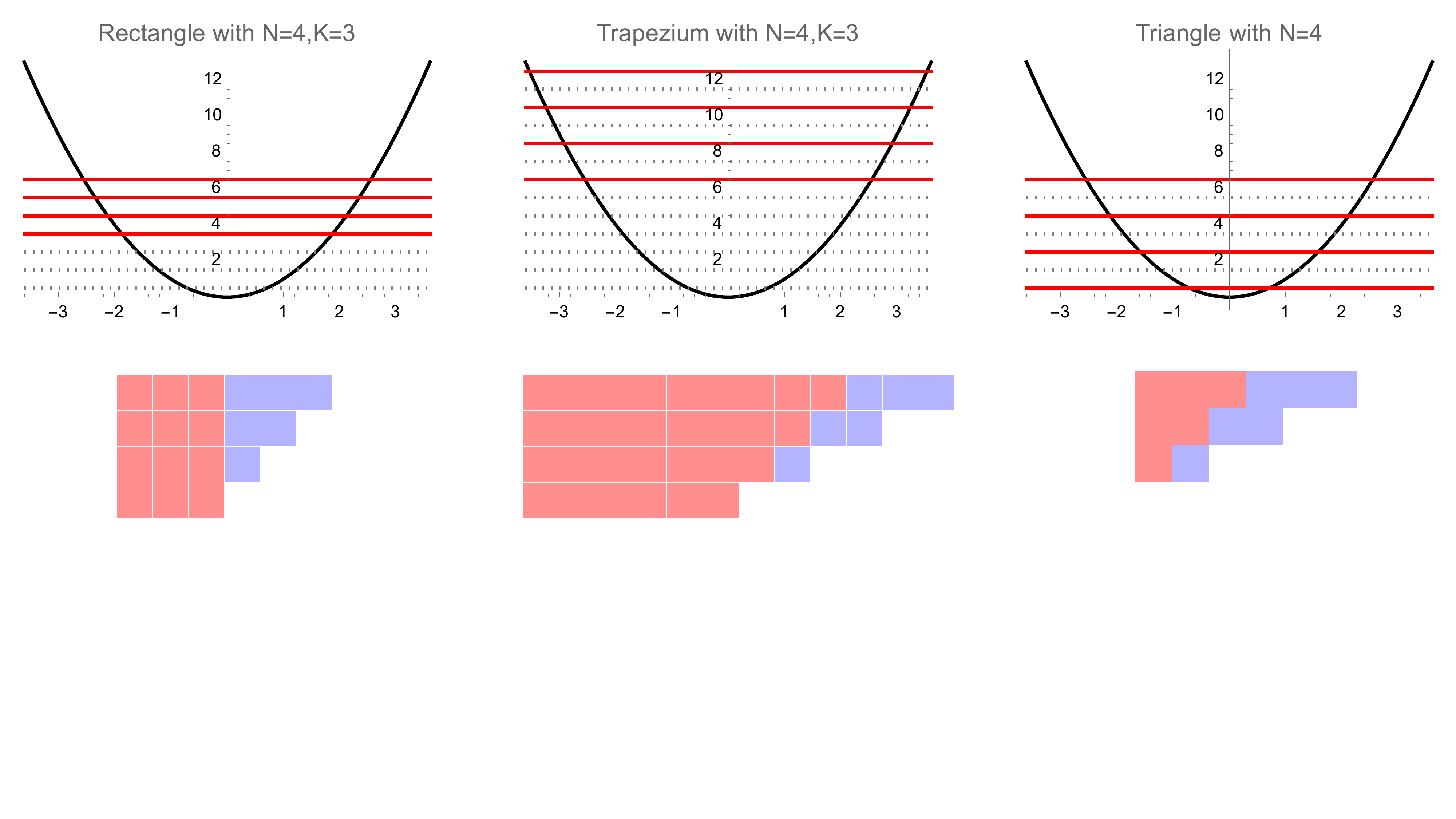}
    \vspace{-3.5cm}
    \caption{Rectangle and Trapezia can be described by systems of fermions in an harmonic potential. In both cases we have a gap: the first levels are empty. In the rectangle case all levels are filled after that gap while in the trapezia case we excite every other level after some gap. The triangle is a limiting case of the trapezia when the gap is sent to zero.}
    \label{oscillatorPictureAandB}
\end{figure}

These are the so-called \textit{rectangle} YT  characterized by shifted highest weights \(h_{j}=K+j-1\) and \textit{trapezium} YT with $h_j = 2K+2j-2$, see figure \ref{oscillatorPictureAandB}. We will refer to the trapezium with $K=0$ as the \textit{triangle}. 

For a \textit{rectangle} YT the function $Q_h(x)$ is a Slater determinant (upto some factors -- see footnote \ref{slaterSubtletyFootnote}) for $N$ fermionic oscillators with the first $K$ unfilled levels and the rest filled densely. The trapezium is similar except the first $2K$ levels are left empty and that excited odd levels are left empty as well, see figure \ref{oscillatorPictureAandB}. For such simple level distributions we can cast these Slater determinants in terms of a simple auxiliary matrix integral over an auxiliary $K \times K$ matrix $Y$ with eigenvalues $y_j$ as (dropping immaterial normalization factors),
\begin{eqnarray}\label{niceYT}
&&\!\!\!\!\!\!\!\!\!\!\!\!\!\!\!\!\!\!\!\!\mathcal{Q}_{K}^{\texttt{rectangle}}(x) = \frac{\underset{1\le k,j\le N  }{\det} H_{K+k-1}(\hat{x}_j)}{\Delta_{N}(\hat{x})}=
 \int \prod_{j=1}^K dy_j \, e^{-\frac N2 y_j^2} \prod_{\alpha=1}^N ( x_\alpha-y_j) \Delta_K(y)^2\la{slaterRect} \\
&&\!\!\!\!\!\!\!\!\!\!\!\!\!\!\!\!\!\!\!\!    \mathcal{Q}_K^{\texttt{trapezium}}(x) = \frac{\underset{1\le k,j\le N}{\det} H_{2K+2k-2}(\hat{x}_j)}{\Delta_{N}(\hat{x})}
    =\frac{\Delta(\hat{x}^2)}{\Delta(\hat x)}\times \int \prod_{j=1}^K dy_j \,e^{-\frac N2 y_j^2} \prod_{\alpha=1}^{N}(x_\alpha^2-y_j^2) \Delta_K(y^2)^2\la{slaterTrapz}
\eeqa
as derived in appendix \ref{slaterAp}. The first formula can be borrowed directly from \cite{Brezin_2000,Morozov:1994hh, Kimura:2021hph}. 

In particular, for the triangle, we get the most simple expression
\beqa
\mathcal{Q}^{\texttt{triangle}}(x) = \frac{\underset{1\le k,j\le N}{\det} H_{2(k-1)}(\hat{x}_j)}{\Delta_{N}(\hat{x})}
= \frac{\Delta(\hat{x}^2)}{\Delta(\hat x)} \,,\la{triangleQ}
\eeqa
as expected. Indeed, for any set of orthogonal polynomials $\pi_k(x)$, the Vandermonde determinant is given by $\Delta(x) = \det_{1\le i,k\le N} \pi_{i-1}(x_k)$ up to an overal constant. With $\pi_i(x)$ being a set of even polynomials, we get by the same token $\Delta(x^2)$. This is precisely what we have for the triangle. 

We could now plug such representations into (\ref{Qdef}) to obtain a simple representation of any three point correlator involving any combination of rectangles and trapezia as a simple eigenvalue integral over four set of variables: Three $y_k^{(n)}$ -- one for each $Q$ -- plus the $x_k\equiv y_k^{(0)}$ eigenvalues. In the large $N$ limit, saddle point equations will constraint all these varibles which we can depict as locations of four type of charges through a set of simple equations of the form 
\beq
F_{\color{blue} n}({\color{blue} y_k^{(n)}}) +  \sum_{{\color{red}(j,m)} \neq {
\color{blue}(k,n)} 
} F_{{\color{blue}n},{\color{red}m}}({\color{blue}y_k^{(n)}},{\color{red}y_j^{(m)}}) =0 \,.
\eeq
Solving such electrostacic equations -- for very large number of charges -- is very easy to do numerically. Sometimes it can also be done analyically as described momentarily. 

Let us consider two examples in some detail. We will consider the case of three identical rectangle operators as well as three triangle operators. For the triangle we have no auxiliary variables -- we simply have (\ref{triangleQ}) while for the three rectangle case we focus on the most symmetric solutions where all $y_k^{(1)}$, $y_k^{(2)}$ and $y_k^{(3)}$ are identical. For the rectangle we thus obtain the effective saddle point equations\footnote{In the continuum, sums like $\frac1N\sum_{j\neq k}^N \frac1{x_k-x_j}$ become principal part integrals $\dashint \frac{\rho(z)}{x-z}$. This principal part integral is nothing but the average of the resolvent $G(x)=\int \frac{\rho(z)}{x-z}$ on both sides of the cut where $x$ lives. Namely,~$\tfrac{1}{2}(G(x+i0)+G(x-i0))\equiv \sG(x)=\dashint \frac{\rho(z)}{x-z}$. The density is similarly related to the discontinuity of the resolvent across the cut, $\rho(x)=\tfrac{1}{2\pi i} (G(x-i0)-G(x+i0)) $.} 
\beqa
0 = -x_k + \frac2N\sum_{j\neq k}^N \frac1{x_k-x_j} + \frac 3N \sum_{j=1}^K \frac1{x_k - y_j}\,, \qquad k=1,2,\dots,N \la{SPr1}\\ 
0 = -y_k + \frac2N \sum_{j\neq k}^K \frac1{y_k-y_j} + \frac1N \sum_{j=1}^N \frac1{y_k - x_j}\,, \qquad k=1,2,\dots,K \la{SPr2}
\eeqa
while for the triangle YT they simply read 
\beq
0=-x_k  +  \frac1N \sum_{j\neq k}^N \left( \frac{2}{x_k-x_j}+\frac{3}{x_k+x_j}\right) \,, \qquad k=1,2,\dots,N \,,\label{triangleSP}
\eeq
the same as for the $O(-3)$  sigma model on random planar graphs~\cite{Kostov:1988fy}.

We first want to make the obvious but important point that these equations have many solutions! 
Take the rectangle equations for instance and focus on purely real solutions where $y_k, x_k$ are real. We can think of these real numbers as locations of charges that repel electrostatically. We can imagine starting with some random value for such charge locations and let them dynamically relax towards their electrostatic equilibrium configuration. In particular note that $y$ and $x$ particles repel each other with a force that blows up as they approach each other meaning that they can never cross. So if we start with an ordered configuration where $x$'s and $y$'s are sprinkled in a $\{x,x,y,y,y,x,y,\dots\}$ like pattern and let these particles relax to equilibrium they will lead to a solution of the saddle point equations where such ordering is preserved. That means we can have a single $x$ cut solution (if we start with all $y$'s to the left or right of all $x$'s), a two cut solution (if a fraction of the $x$'s is to the left of all the $y$'s and the remaining $x$'s are to their right), a three cut solution (if we start with three $x$ domains separated by two $y$ domains) etc.

\begin{figure}[t]
    \centering
    \includegraphics[width=\textwidth]{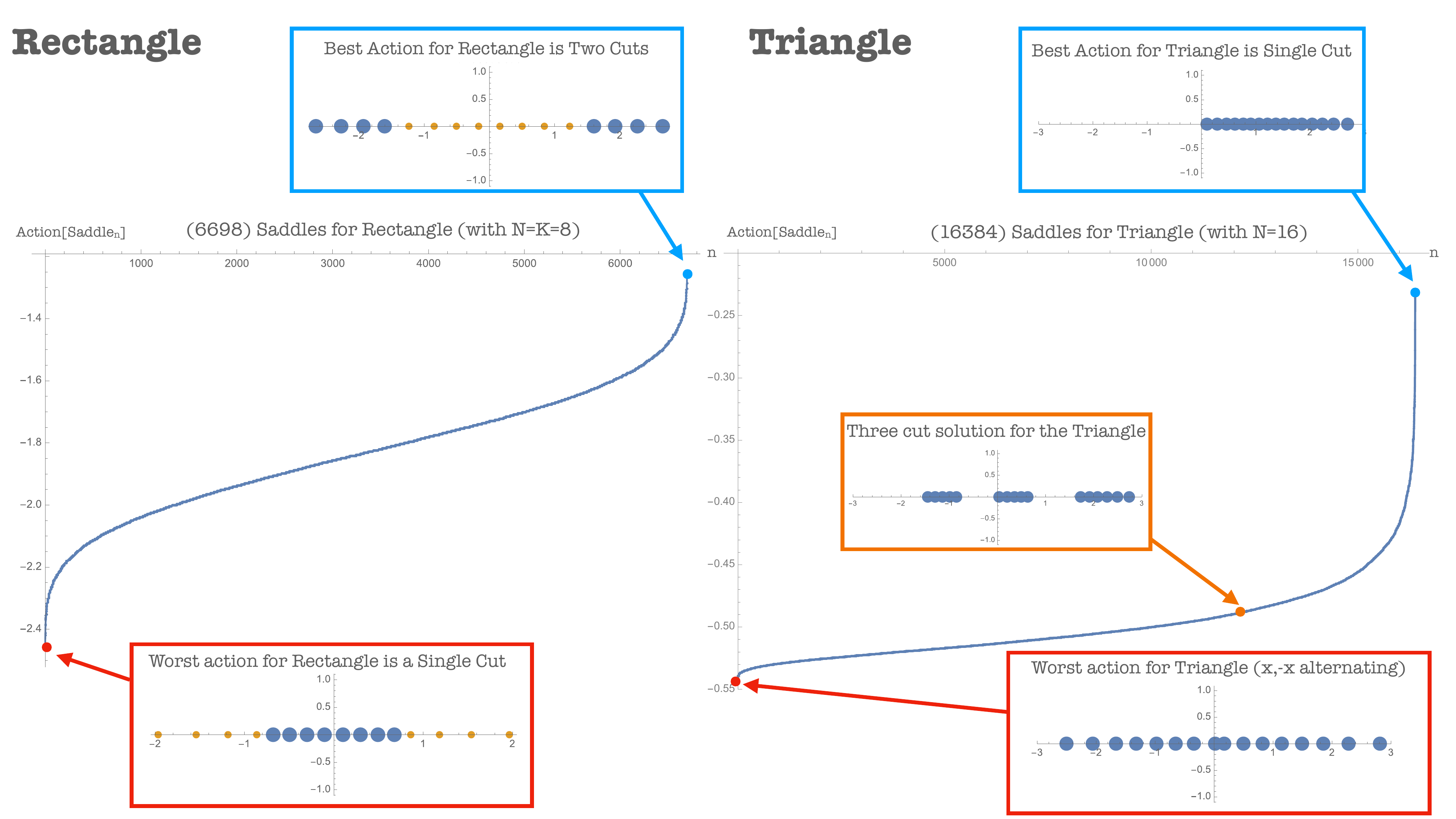}
    \caption{On the left we computed all the $x$ (blue) and $y$ (orange) real solutions to (\ref{SPr1}),(\ref{SPr2}) for some $N=K=8$ rectangle while on the right we depict the real $x$ eigenvalue solutions to (\ref{triangleSP}) for the triangle with $N=8$. Note that ever for such moderate values of $N=8$ there are thousands of saddle point equations in either case. We see that the least dominant contribution for the rectangle is the single cut solution in $x$ which for the triangle is actually the dominant solution! For the rectangle the dominant solution -- with the least negative action -- is the symmetric two cut solution. In both cases a rich set of multi-cut solutions show up as intermediate saddles as illustrated with a three cut solution on the right. }
    \label{manySPS}
\end{figure}

We have a similar story for the triangle even if  there are no $y$ particles. The reason is that a particle $x_k$ feels a particle $j$ through its location $x_j$ but also through its mirror image at $-x_j$. The presence of such images allows us to have multiple cut solutions as well despite the simplicity of the saddle point equations (\ref{triangleSP})! Imagine for instance starting with half of the $x$'s in $[-\Lambda,0]$ and the other half between $[+\Lambda,\infty]$ and letting them relax to equilibrium. The left most particle at $x_L$ close to $-\Lambda$ has an image at $-x_L$ close to $+\Lambda$ which will not allow for the right half of the $x$'s to move to the left and approach and merge with the left cut. So if we start with such a configuration with a big gap we will relax to a configuration with a big gap, in other words a two cut solution. We can easily cook up solutions with any number of cuts. To get the simplest one cut solution we can start with all $x$'s positive and relax towards equilibrium.\footnote{There is of course an equivalent solution where we start with all $x$'s negative since we have an obvious $X\to -X$ symmetry here.} 

In sum, for both the triangle and the rectangle we have in the large $N$ limit an infinite amount of saddle point solutions with an arbitrary number of cuts! This holds more generically for any non-trivial YTs. This is both a curse and a blessing. On the one hand it means we have a rich moduli of solutions, all of which we can study with large N methods -- including the fluid methods advocated later in this paper. They all carry interesting physics in the matrix integral as either the classical saddle or sub-leading instanton contributions which we might also want to analyse.\footnote{The leading and subleading actions will both be of order $N^2$ but they will differ by order $N$. As such $(e^{-S_\text{leading}}+e^{-S_\text{sub-leading}}+e^{-S_\text{sub-sub-leading}}+\dots)/e^{-S_\text{leading}} = 1+O(e^{-a N})$ as expected for instanton corrections. Note also that although we have exponentially many solutions with an entropy proportional to $N$, it cannot compete with leading action which is of order $N^2$.} On the other hand, if all we care is the saddle point solution corresponding to the lowest energy solution -- the one that will dominate when computing the structure constant of huge operators -- we need to work extra hard to find the right cut topology to focus on. We find, for example, that for the rectangles the symmetric two cut solution is the dominant solution while for the triangles (as well as for the trapezia) it is the single cut solution that dominates! This is illustrated in figure \ref{manySPS} where we scanned over the many thousand equilibrium configurations for some rectangles and triangles.\footnote{Trapezia are also given by a single cut. If we interpolate between these trapezia and the rectangles by changing the slope of $h$'s slowly -- say by setting $h_j=j-1+K+\alpha (j-1)$ with $\alpha$ between $0$ (rectangle) and $1$ (trapezia) -- we find a nice phase transition between the two -- at some intermediate $\alpha$. 
This conclusion arose of numerical explorations done together with Andrea Guerrieri. Would be fascinating to systematically explore such phase transtions not only for this case but more generically. Which Young-Tableaux's correlators are dominated by one cut saddle points? Which are dominated by two-cuts? Where is the transition between the two? Does it have physical implications?...}

\begin{figure}
    \centering
    \includegraphics[width=0.8\textwidth]{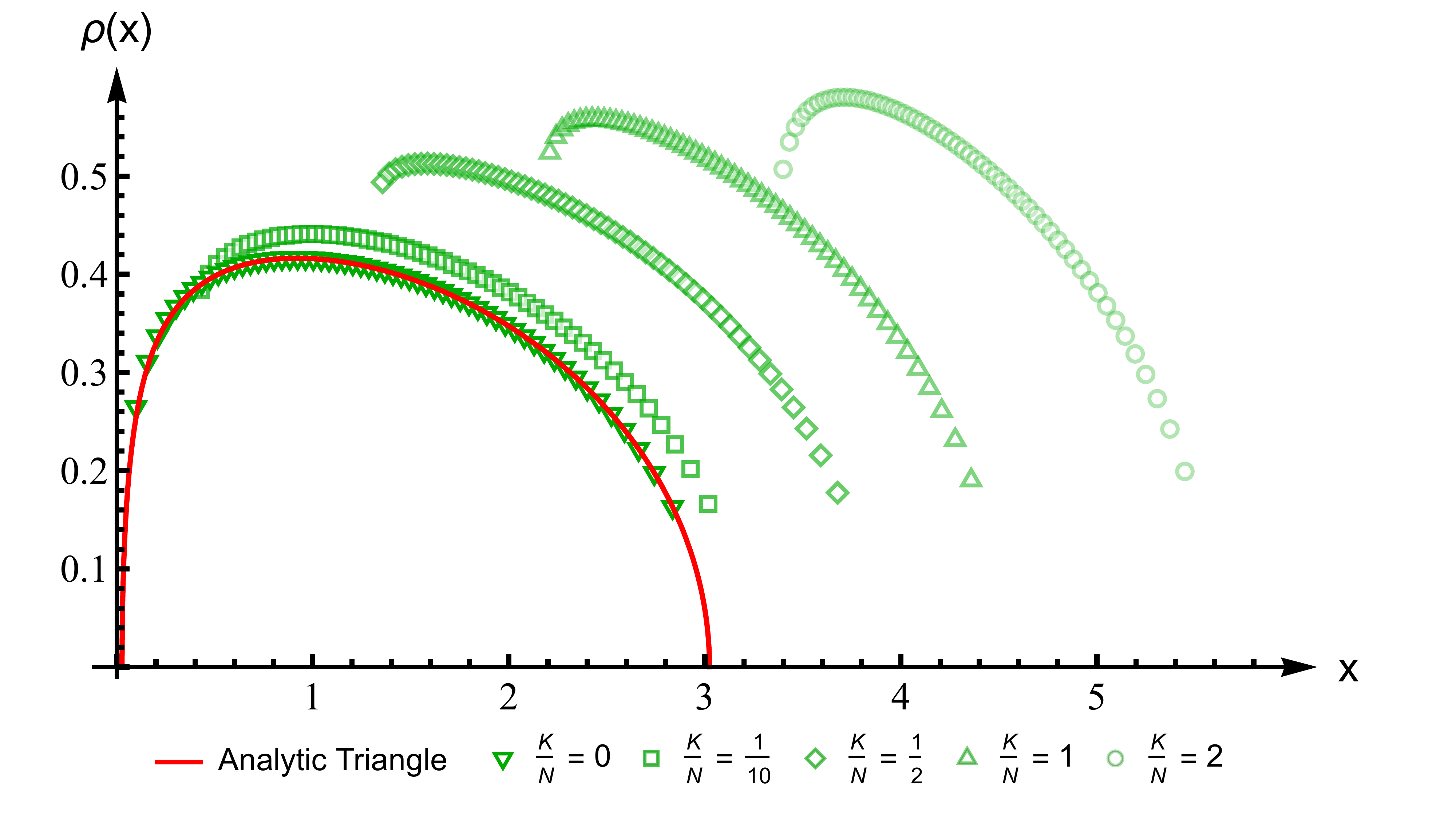}
    \caption{Densities (computed numerically for $N=60$) for several trapezia with varying $\gamma=K/N$ beautifully converge towards the triangle analytic solution as $\gamma\to 0$}
    \label{densityTriangleTrapz}
\end{figure}

For the case of three triangles, we were able to fully solve the problem in the continuum limit for the leading single cut solution. The SPE (\ref{triangleSP}) in the continuum limit becomes,
\beqa
    2\slashed G(x) - 3G(-x) = x
\eeqa
which takes the same form as the saddle point equations in the $O(n)$ model on random graphs \cite{Kostov:1988fy} for $n=-3$. This model was solved by Eynard and Kristjansen for any $n$ in terms of elliptic functions \cite{Eynard:1995zv}. Following them, we rederive a clean formula for the resolvent in appendix \ref{triangleAppendix} in terms of Jacobi Elliptic functions as 
\begin{eqnarray*}
    G(x(u)) &=& \frac{a\ sn(u,k)}{5}+\mathcal{N}\left(\frac{e^{-\frac{i \pi(1-\nu)}2 \left(\frac{u}{K}+1\right)} \theta \left(\frac{u-i(2-\nu )K'}{2 K}\right)}{\theta \left(\frac{u-i K'}{2 K}\right)}+\frac{e^{\frac{i \pi(1-\nu)}2 \left(\frac{u}{K}+1\right)} \theta \left(\frac{u+i(2-\nu )K'}{2 K}\right)}{\theta \left(\frac{u+i K'}{2 K}\right)}\right)\\
    x(u) &=& a\ sn(u,k)
\end{eqnarray*}
See appendix \ref{triangleAppendix} for details and values of the various parameters. Using this analytic form of the resolvent, we can compute the normalized action for three point function with arbitrary precision. We find
\beq \frac1{N^2}\log\left[Z(\Delta,\Delta,\Delta)\sqrt{\frac{Z(\phi,\phi,\phi)}{Z(\Delta,\Delta,\phi)^3}}\ \right] \approx 0.4465 \,.
\eeq

In figure \ref{densityTriangleTrapz} we depict the numerical density arising from numerically solving the saddle point equations for various trapezia with different $\gamma\equiv K/N$; when $\gamma\to 0$ the corresponding density beautifully agrees with the analytic density extracted from this resolvent. 

For the three point functions of trapezia and rectangles, we formulate the SPEs as a nice Riemann-Hilbert problem in appendix \ref{RectangleTrapeziaSPEap}. Their solution is more involved and will probably require higher than elliptic functions. We leave this task for future works.

\section{Semiclassics -- Fluids} \label{FluidSec}
In \cite{Matytsin_1994} (see also \cite{Das:1990kaa}) Matysin beautifully explained that at large $N$, the HCIZ integral admits a beautiful continuum limit description as 
\beq
\frac1{N^2} \log I(a,b) \simeq S_{\texttt{fluid}}[\rho_a,\rho_b] + S_{\texttt{bdy}[\rho_a,\rho_b]} 
\label{matytsinAction}
\eeq
where $\rho_a$ and $\rho_b$ are the eigenvalue densities for the matrices $A$ and $B$. The second term $S_{\texttt{bdy}}$, is an explicit function given by the following expression,
\beq
\!S_{\texttt{bdy}}[\rho_a,\rho_b]=\frac12\! \int\! dx\, x^2(\rho_a(x)+\rho_b(x))-\frac12 \!\int \!dx\, dy\, \log|x-y|\,(\rho_a(x)\rho_a(y)+\rho_b(x)\rho_b(y))-\frac34
\la{bdyAction}
\eeq
The remaining piece in (\ref{matytsinAction}), $S_\texttt{fluid}$, is the most important one since it carries all the dynamical information. It is the action of a $1\text{-}d$ inviscid fluid with equation of state 
\beq
P=-\frac{\pi^2}3 \rho^3 \label{Peq}
\eeq
and whose density $\rho(x,t)$ evolves in one time unit from 
\beq
\rho(x,t=0) = \rho_a(x) \qquad \text{into} \qquad \rho(x,t=1) = \rho_b(x)\la{boundaryConditions} \,.
\eeq
Explicitly, this fluid action reads
\beq
    S_{\texttt{fluid}}[\rho_a,\rho_b]=-\frac12 \int_0^1 dt\int dx\ \rho(x,t)\left(v(x,t)^2+\frac{\pi^2}3 \rho(x,t)^2\right)\la{fluidAction} \,.
\eeq
Importantly, note that this contribution depends on the full flow and not only on the densities at the end-points. One needs to solve for the density and velocity $\rho(x,t)$ and $v(x,t)$ at all times $t\in [0,1]$ to evaluate the action.

Extremizing the action (\ref{fluidAction}), we get the following Riemann-Hopf equation of motion,
\beq
    \frac{\partial f}{\partial t} + f\frac{\partial f}{\partial x}=0\la{fluidEOM}
\eeq
where,
\beq
    f(x,t)=v(x,t)+i\pi \rho(x,t) \label{fSplit}
\eeq
The real and imaginary parts of (\ref{fluidEOM}) give Euler equations for the one dimensional fluid with equation of state (\ref{Peq}). More generally, we can have complex flows, where the velocity and/or the support of the density are in the complex plane. In that case, the flow function $f(x,t)$ has a branch cut at the location of the fluid. The density and velocity, now identified as the discontinuity and average across a cut, satisfy the Euler equations.


\begin{figure}[t]
\begin{center}
\includegraphics[scale=0.6]{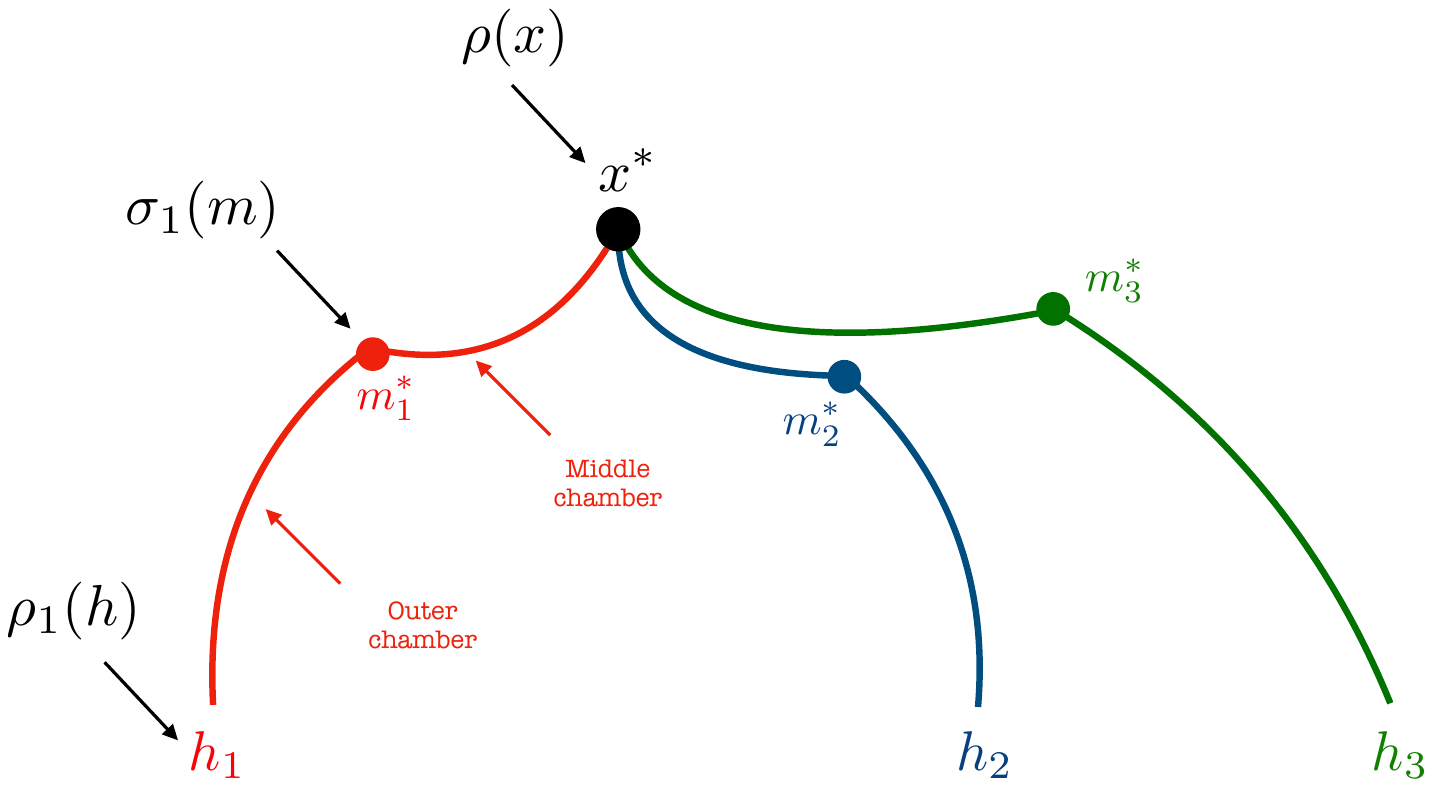}
\end{center}
\vspace{-5cm}
\caption{For very large operators and large $N$ all integrals become dominated by a leading trajectory. In other words, the quantum propagations of the previous figure become classical. This classical propagation is dominated by a fluid. Each leg of the diagram can be through as a fluid propagating in a \textit{chamber}. At the four dots -- which we dub as \textit{inner} junctions -- these fluids are glued together with some gluing conditions given by the saddle point equations for the four saddle point equations for the four variables~$x,m_1,m_2,m_3$. In total we have six chambers and four junctions.
 }  \la{cChambers}
\end{figure}

Importantly note that the equation has complex characteristics with eigenvalues $v\pm i\pi\rho$ and therefore is elliptic\footnote{Recall a system of $n$ first-order PDEs of the form $\frac{\partial \mathbf{u}}{\partial t}+\mathbf{A}(x,t) . \frac{\partial \mathbf{u}}{\partial x}=0$ is classified as hyperbolic if all eigenvalues of the matrix $\mathbf{A}(x,t)$ are real, elliptic if all eigenvalues are complex, and parabolic if $\mathbf{A}$ is not diagonalizable. In our case,
\beq
\partial_t\begin{pmatrix}
    v\\ \pi\rho
\end{pmatrix} + \begin{pmatrix}
    v&-\pi\rho\\ \pi\rho&v
\end{pmatrix} \partial_x\begin{pmatrix}
    v\\ \pi\rho
\end{pmatrix}=0
\eeq
and the $\mathbf{A}$ matrix has eigenvalues $v\pm i\pi\rho$. Note that with a more sensible equation of state like $P=+\frac{\pi^2}3 \rho^3$, we would get a hyperbolic system with eigenvalues $v\pm \pi\rho$, in line with our intuition for fluid flows.\la{hyperbolicFootnote}}. So, the boundary conditions (\ref{boundaryConditions}), where we fix the initial and final densities, leaving the velocities undetermined are indeed well-posed. The velocities both at the $a$ or $b$ end-points depend very non-trivially on the densities. If we instead provide hyperbolic boundary data -- i.e. specify initial density \textit{and} velocity, the solution will be unstable to perturbations.

Note that the velocities at the end point encode almost as much information as the full action itself. They are the derivatives of the action with respect to small variations of the initial or final densities:
\beqa
\partial_x\frac{\delta S_{\texttt{fluid}}[\rho_a,\rho_b]}{\delta \rho_a(x)} &=& v(x,t=0)\\
\partial_x\frac{\delta S_{\texttt{fluid}}[\rho_a,\rho_b]}{\delta \rho_b(x)} &=& -v(x,t=1)\la{velVariation}
\eeqa
We can now readily write the expected semi-classical result for the three point function partition function $Z$. It can be visualized as three legs, which start at some density $\rho(x)$ of the auxiliary matrix and end at the Young tableau densities $\eta_n(h)$. Each leg is further split into two flows, one going from $x^*\rightarrow m_n^*$ and the other from $-\log m_n^* \rightarrow h$. Given the YT densities $\eta_n(h)$, the densities of $x$ and $m_n$ are fixed by saddle point equations obtained by varying the corresponding matrices. We will denote the velocities in the first flow as $v_n(x,t)$ and in the second flow as $w_n(x,t)$. All in all,
\begin{mdframed}[backgroundcolor=blue!3, leftmargin=0cm, rightmargin=0cm, topline=false,
	bottomline=false, leftline=false, rightline=false] 
\vspace{-0.5cm}
\beqa
 && \!\!\!\!\!\!\!\!   \frac{1}{N^2} \log Z[\eta_1,\eta_2,\eta_3]\simeq 
 \sum_{n=1}^3  S_{\texttt{fluid}}[\rho,\sigma_n] +\sum_{n=1}^3 S_{\texttt{fluid}}[\tilde\sigma_n,\eta_n] +       \label{action3}\\
&&  
\qquad\qquad\qquad\qquad +\int\! dx\, x^2
    \Big(\rho(x)+\frac12\sum_{n=1}^3 \tilde \sigma_n(x)
    \Big) -\frac{1}{2} \!\int \!dx\!\int\! dy \log|x-y|\,
    \rho(x)\rho(y) 
\nn
\eeqa
where we used $\sigma_n(x)$ and $\tilde\sigma_n(x)=e^{-x} \sigma_n(e^{-x})$ to indicate the densities of $m_n$ and $-\log m_n$ respectively, at the junction $n$, see figure \ref{cChambers}.\footnote{We dropped simple normalization factors $\sum_n f[\eta_n]$ since they are not physical and drop out when building the physical three point functions, see discussion below (\ref{Ndef}).} 
Lastly, we have four gluing conditions which follow from the saddle point equations,
\beqa
\dashint dz\ \frac{\rho(z)}{x-z}&=&2x + \sum_{n=1}^3 v_n(x,0)\label{xVelGlue} \\
w_n(x,0)&=&-x - e^{-x}\ v_n(e^{-x},1)\,,\qquad n=1,2,3\label{mVelGlue}
\eeqa
\end{mdframed}

As a sanity check, let's count the degrees of freedom and constraints. We have six flows, each of which require two boundary conditions for a total of 12 degrees of freedom. At the meeting point of the three legs, we require that the densities match (2 constraints) and the velocities obey the gluing condition (\ref{xVelGlue}). We also have two additional gluing conditions per leg, one for velocity (\ref{mVelGlue}) and one for continuity of density. This leaves us with precisely three degrees of freedom, which are fixed by the YT densities $\eta_n$.

\subsection{Local and Global Problems}

The fluid equations
\beq
\partial_t v+v \partial_x v  =\pi^2 \rho \partial_x \rho \,, \qquad \partial_t \rho +\partial_x (\rho v)=0 \,. \la{fluidEOM2}
\eeq
following from Riemann-Hopf equation by splitting $f$ as in (\ref{fSplit}) define an \textit{integrable} model. Let us also mention that there is an equally important discretization of these fluid equations. The continuum fluid equations can be approximated as evolution equation of many discrete eigenvalues $x_i(t)$ under an inverse cube mutual attraction \cite{Bun:2014dha} known as the Calegero-Moser rational model,
\footnote{Let us recall briefly how this equation has anything to do with the fluid equations above. The reason is an anomaly. Indeed, for well separated $x_i$ and $x_j$ we can drop the interaction term because of the $1/N^2$ factor in front. For 
$x_{j=i+k} - x_i \simeq \frac{k}{N \rho(x_i)}-\frac{k^2\rho'(x_i)}{2N^2 \rho(x_i)^2}+\dots$  
we do get a contribution. When we plug this into the sum in (\ref{discrete}) the leading term leads to a $1/k^3$ sum which vanishes by parity while the subleading term contributes as 
  \beq
  3 \rho(x_i)\rho'(x_i) \sum_{k\neq 0} 1/k^{2} 
  =\pi^2\rho(x_i)\rho'(x_i) \,.
  \eeq 
  where we recognize precisely the right hand side of the first equation in (\ref{fluidEOM2}). For a nice review with more details see the lecture notes   \href{https://www.dam.brown.edu/people/menon/talks/cmsa.pdf}{https://www.dam.brown.edu/people/menon/talks/cmsa.pdf} by Menon. 
}  
\beq
\frac{d^2x_i}{d t^2}+\frac2{N^2}\sum_{j\neq i} \frac{1}{(x_i-x_j)^3}=0\,, \qquad i=1,\ldots N \label{discrete}
\eeq
Numerically, these coupled ODEs are easier to work with than Euler equations; we will come back to these below in later sections. As is well known, the Calogero-Moser model is also \textit{integrable}. See \cite{Polychronakos:2006nz} for a very nice review of such models, including rational and elliptic generalizations\footnote{It would be interesting to see if the large $N$ limit of these integrable generalizations (see \cite{Abanov:2008ft}) can be related to protected correlators in $\mathcal{N}=4$ SYM.}.

These models admit infinitely many conserved charges. 
In the continuum, we find a general large class of charges that to our knowledge were not known before. Namely,
\beq
 Q_{n,m} \equiv   \oint dx\int\limits^{f(x,t)} dz\ (x-tz)^n(x+(1-t)z)^m \label{newCharges}
\eeq
are conserved along any fluid flow.\footnote{To check that this is indeed a conserved charge the reader can show that upon using (\ref{fluidEOM}) its time derivative becomes an integral of a total derivative  $\displaystyle\oint dx\ \dfrac{\partial}{\partial x}\Big(-\!\! \displaystyle\int^{f(x,t)} \!\!\!\!\!\!\!\!\!\! dz\ z\,(x-tz)^n(x+(1-t)z)^m\Big)=0$. Note that we are not assuming any special analyticity properties for $f(x,t)$ here. The contour integral goes only \textit{around} the cut and can thus be expanded into a sum of real integrals \textit{on} the cut.}

Before reviewing the exact solvability of the fluid equations (\ref{fluidEOM}) or (\ref{fluidEOM2}) let us first stress that there are two natural problems to consider when dealing with such a fluid, in both its discrete or continuum formulation. 

A first type of problem is what we call an \textit{initial value} problem (IVP). Here we specify the initial density and velocity of the fluid at some $t=0$ and evolve the fluid from that point forward. In the discrete equations (\ref{discrete}), the IVP is the one where we specify the initial position $x_i(0)$ and velocity $x_i'(0)$ of each fluid bit and then simply evolve the differential equations  (\ref{discrete}) towards a future time $t=1$. Clearly, the IVP formulation of both the continuum and the discrete problems is local.

A second type of problem is what we call the \textit{boundary value} problem (BVP). Here we specify the initial density at some time $t=0$ and the final density at some time $t=1$ but we do not specify any velocity. This problem is \textit{global} since it is impossible to find the density and velocity at any time along the flow without solving the full flow. In the discrete equations (\ref{discrete}) the BVP is the one where we specify the initial and final positions $x_i(0)$ and $x_i(1)$ of the fluid bits. This can in principle be solved by global relaxation methods but that is not straightforward. 




In the continuum IVP problem we are given $f(x,0)=v(x,0)+i \pi \rho(x,0) \equiv F(x)$. Then we can write $f(x,t)$ at any future time in parametric form as\footnote{Note that we need to analytically continue the initial data $F(x)$ to the entire complex plane in order to use this formula \cite{Matytsin_1994}} 
\beq
f(x,t)=F(u) \,, \qquad x=u+t F(u) \,. \la{continuumSol}
\eeq
In the discrete IVP we are given an initial set of positions $\{x_1,\dots,x_N\}_{t=0} \equiv \mathbb{X}$ and velocities $\{\dot x_1,\dots,\dot x_N\}_{t=0} \equiv \mathbb{V}$. Then we can find $x_i(t)$ as
\beq
\{ x_i(t) \} = \text{eigenvalues of }\Big( \text{diag}(\mathbb{X})+ t P(\mathbb{X},\mathbb{V}) \Big) \,, \qquad P(\mathbb{X},\mathbb{V})_{ij} = \mathbb{V}_i \delta_{ij}-\frac{1}{N}\frac{1-\delta_{ij}}{\mathbb{X}_i-\mathbb{X}_j} \,. \la{discreteSol}
\eeq
As anticipated, the IVP is explicitly solvable! Checking (\ref{continuumSol}) is straightforward -- see appendix~\ref{directAp1}. The derivation of (\ref{discreteSol}) requires more ingenuity, see for example the nice lecture notes \href{https://www.dam.brown.edu/people/menon/talks/cmsa.pdf}{https://www.dam.brown.edu/people/menon/talks/cmsa.pdf} by Menon. 

Can we cast the integrable charges (\ref{newCharges}) in the discrete in terms of these matrices in (\ref{discreteSol})? In the three point function flows, this could allow us to directly relate moments of the Young Tableaux on one end to the moments of the eigenvalue matrices in the inner junctions hopefully allowing us to crack the BVP problem for the structure constants. We leave this to future work.

\section{The Two-Point Function} \label{2ptS}
Before moving to the three point function, let us warm up by looking at two point function,
\beq
Z(\mathcal{R}_1,\mathcal{R}_2,\phi)\,
\eeq
which also appears in the normalization factor (\ref{Ndef}). None of the results of this section will be new. After all, everything is known about the two point function for many decades already. The expert reader might want to jump this section on a first reading. We decided to include it here nonetheless for pedagogical purposes and to highlight some interesting aspects of some of the formulae above, both in the fluid classical limit as well as in the exact matrix model representations. 

When the third representation is trivial $Q_3(x)=2^{N(N-1)/2}$  the single matrix model partition function (\ref{singleM}) becomes simply
\beq
Z(\mathcal{R}_1,\mathcal{R}_2,\phi)=\textcolor{gray}{\frac{2^{N(N-1)/2}}{(N/2)^{N^2/2}}}\int \prod_{i=1}^N d{x}_i e^{- \sum_{i=1}^N {x}_i^2} \det_{1\le i,j \le N} H_{h_{1,j}}({x}_i) \det_{1\le i,j \le N} H_{h_{2,j}}({x}_i) \,. \nn
\eeq
Note that the Vandermonde factors cancelled out and we trivially rescaled the integration variables. In what follows, there are many irrelevant prefactors, that we color in gray. Now we use very standard matrix model manipulations. First, since the integrand is fully symmetric in any permutation of integration variables, we can replace one of the determinants by a single product, say $\prod_{i=1}^N H_{h_{1,i}}(x_i)$ and multiply the integral by $N!$. We thus obtain 
\beq
Z(\mathcal{R}_1,\mathcal{R}_2,\phi)=\textcolor{gray}{\frac{N!\ 2^{N(N-1)/2}}{(N/2)^{N^2/2}}} \,\det_{1\le i,j \le N}  \int_{-\infty}^{+\infty} dx  \,e^{-x^2} H_{h_{1,j}}(x)H_{h_{1,i}}(x) \,. \nn
\eeq
We are left with a simple determinant of one dimensional integrals! So far we used no property of the weights $h_j$ or the functions $H_n$. If now we recall that $H_n$ are orthogonal polynomials with respect to this Gaussian measure and that the $h_i$ are strictly monotonical we immediately get that the matrix reduces to a simple diagonal matrix and  
therefore 
\beq
Z(\mathcal{R}_1,\mathcal{R}_2,\phi)=\textcolor{gray}{\frac{N!\ 2^{N(N-1)/2}}{(N/2)^{N^2/2}}} \prod_{j=1}^N \delta_{h_{1,j},h_{2,j} } \prod_{j=1}^N  \int_{-\infty}^{+\infty} dx  \,e^{-x^2} H_{h_{1,j}}(x)^2 \nn
\eeq
Evaluating these integrals we obtain the final and well known exact result for the two point function normalization of two half BPS operators,
\beq
Z(\mathcal{R}_1,\mathcal{R}_2,\phi)= \delta_{\mathcal{R}_1,\mathcal{R}_2} \times \texttt{trivial}(N) \times \texttt{interesting}(h,N) \label{exact2pt}
\eeq
with 
$\texttt{trivial}(N)=$\textcolor{gray}{$N^{-N^2/2}2^{3N(N-1)/2}(2\pi)^{N/2} G(N+2)$}, with $G(N)$ being the Barnes gamma function and
\beq
\texttt{interesting}(h,N)=\prod_{j=1}^N \frac{\Gamma(h_j+1)2^{h_j-j+1}}{\Gamma(j)}  \,. \nn
\eeq
Note that \texttt{interesting} is manifestly normalized to unity for empty representations for which~$h_j=j-1$ and thus $\texttt{trivial}$ is nothing but the vacuum partition function 
\beq
\texttt{trivial}(N)=Z(\phi,\phi,\phi) \nn
\eeq
also appearing in (\ref{Ndef}). Very similar manipulations would allow us to derive all these results starting from the four-matrix model representation (\ref{fourInts}).\footnote{We would obtain the same \texttt{interesting} factor but a slightly different \texttt{trivial} factor. This is totally fine, see the discussion below (\ref{Ndef}).} Here, let us also note that the normalization factor (\ref{Ndef}) is given by
\beq
\mathcal{N} = \frac{(\texttt{trivial}(N))^{-1}}{\sqrt{\prod_{n=1}^3 \texttt{interesting}(h_n,N)}}\la{Nexpr}
\eeq
The most important features of the exact result~(\ref{exact2pt}) are
\begin{itemize}
\item \textbf{Orthogonality}. We see that the two point function requires the two representations to be the same, \beq
\mathcal{R}_1=\mathcal{R}_2=\mathcal{R}\,.\label{R1R2}
\eeq
This ensures that the characters are the best choice for the full orthonormal system of 1/2 BPS operators of $\mathcal{N}=4$ SYM.
\item \textbf{Row Factorization}. We see that \texttt{interesting} is a simple product of factors, one per row of the Young Tableau associated to  the representation $\mathcal{R}$. Because of this factorization, it is trivial to reconstruct this factor from its logarithmic derivative 
\beq
\partial_{h_i} \log \texttt{interesting}(h,N)=\log(2)+\frac{d}{dh_i} \log \Gamma(h_i+1) \nn
\eeq
Indeed, we can readily integrate back this relation imposing that $\texttt{interesting}=1$ for the vacuum to fix all integration constants. For large rows $h_i \gg 1$ relevant for the classical limit which we will turn to next, this simple relations reduces to 
\beq
\partial_{h_i} \log \texttt{interesting}(h,N) \simeq \log(h_i) \,. \label{hDer}
\eeq
\end{itemize}
Next we will discuss how to recover these two key results from the fluid description. 

\subsection{Fluids and the Two-Point Function}

\begin{figure}[t!]
    \centering
    \includegraphics[width=\textwidth]{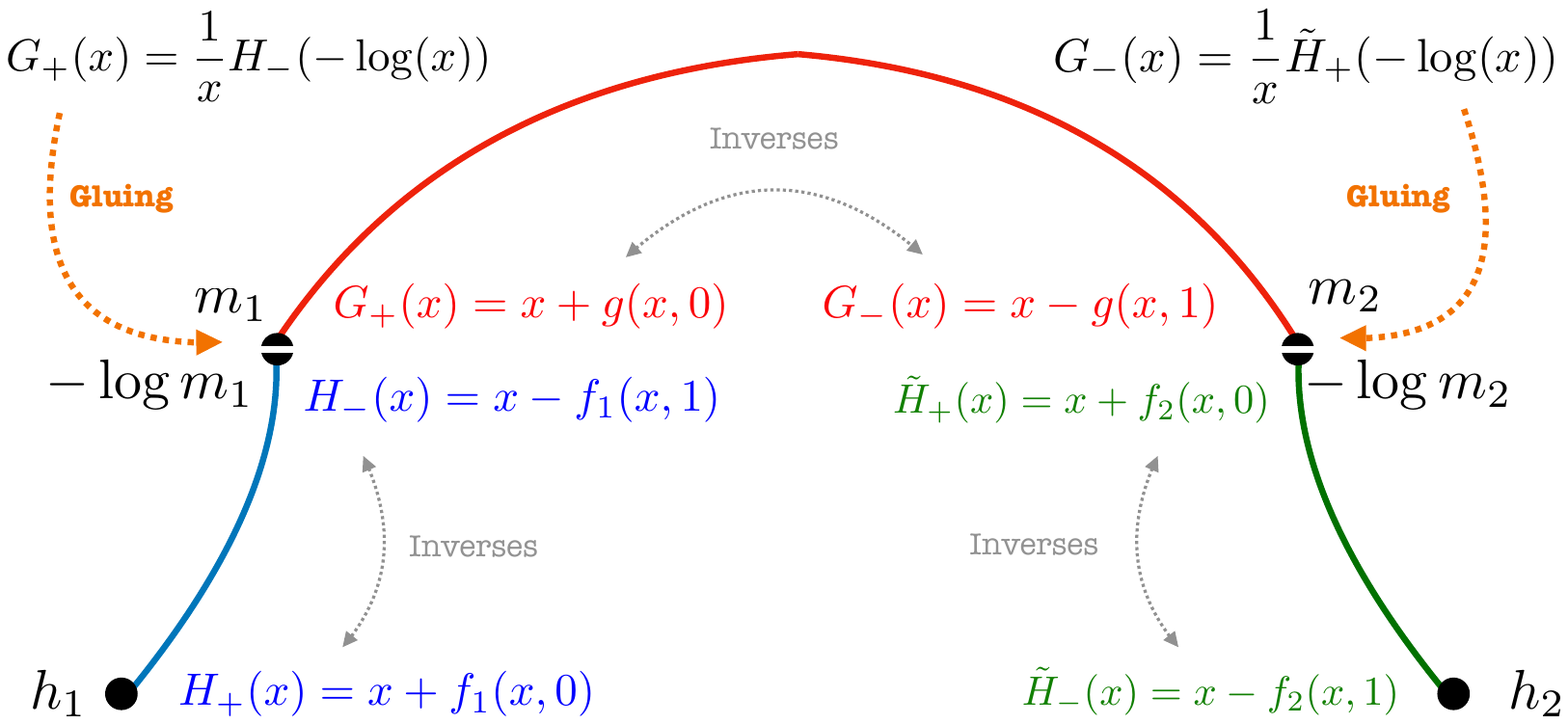}
    \vspace{-4.5cm}
\caption{The semi-classical limit consists of three flows, $(h_1\rightarrow -\log m_1)$, $(m_1\rightarrow m_2)$ and~$(-\log m_2\rightarrow h_2)$. The corresponding Hopf flow functions are denoted as $f_1(x,t)$, $g(x,t)$ and $f_2(x,t)$. Particularly important is what is stressed in the figure: 1) \textit{The functions $H_+$ and $H_-$ are inverse of each other, $H_+(H_-(x))=x$}, and so on for all end-point functions and 2) \textit{The gluing conditions relate $G_+$ and $G_-$ of the middle chamber with $H_-$ and $\tilde H_+$}. (These gluing conditions are slightly different from the three point function case due to the presence of extra exponentials in the measure there.) }
    \label{2ptFlows}
\end{figure}

For the fluids, let us work with a simpler two matrix integral. From appendix \ref{4intA}, we have
\beq
    \frac{Z(\mathcal{R}_1,\mathcal{R}_2,\phi)}{\prod_{n,i} \Gamma(h_{n,i}+1)} = \oint \prod_{n=1}^2\left[\prod_i dm_{n,i}\ \Delta(m_n)\Delta(-\log m_n)\Delta(h_n) I(-\log m_n, h)\right] I(m_1,m_2)  
\la{twoPt3Chamber}
\eeq
We have three HCIZ integrals and thus three fluid flows with two gluing conditions at two inner junctions, see figure~\ref{2ptFlows}. We will now extensively use the various definitions and identities in this figure together with the important conserved charges introduced above in (\ref{newCharges}).
At $t=0,1$, these simplify to
\beq
    Q_{n,m} = \frac1{m+1}\oint \frac{dx}{2\pi i}\ x^n (x+f(x,0))^{m+1} = \frac{-1}{n+1}\oint \frac{dx}{2\pi i}\ x^m (x-f(x,1))^{n+1}\la{PnmBdy}
\eeq
In particular, note that for $m=0$ this implies
\beqa
    \oint dx\ x^n H_+(x) &=& \frac1{n+1}\oint dx\ H_-(x)^{n+1}\\
    \oint dx\ x^n \tilde{H}_-(x) &=& -\frac1{n+1}\oint dx\ \tilde H_+(x)^{n+1}
\eeqa
The LHS of the above equations are nothing but moments of Young tableaux distributions of~$\mathcal{R}_1$ and $\mathcal{R}_2$ respectively. Plugging in the gluing conditions and changing integration variables, we get
\beqa
    \oint dx\ x^n H_+(x) &=& -\frac{1}{n+1}\oint dy\ y^n\, G_+(y)^{n+1}\\
    \oint dx\ x^n \tilde H_-(x) &=& \frac{1}{n+1}\oint dy\ y^n\, G_-(y)^{n+1}
\eeqa
Using again (\ref{PnmBdy}) but now with $m=n$, we immediately see that all moments of the two YTs are equal to each other. So, the Young tableaux must be identical and we rederived (\ref{R1R2}) in the fluid language.\footnote{For a real distribution of YT weights with finite support, this follows from the uniqueness of the Hamburger moment problem with these assumptions}

Next let us show how (\ref{hDer}) comes about in the fluid language. Differentiating (\ref{twoPt3Chamber}) by the YT weights, 
\beqa
    \frac1{N^2} \partial_{h_i} \log Z \approx \frac1{N^2} \partial_h \frac{\delta \log Z}{\delta \eta(h)} = 2\log h + 2h + v_1(h,0) - v_2(h,1) 
\eeqa
where $\eta(h)$ is the density of both the Young Tableaux, and $v_1(x,t)$ and $v_2(x,t)$ are the velocities in the first and last chambers. Showing (\ref{hDer}) is therefore equivalent to proving that at the saddle point
\beq
    v_1(h,0) - v_2(h,1) = -2h - \log h\la{twoPtVelocityIdentity} \,.
\eeq
This last condition follows straightforwardly from a simple identity 
\beqa
    H_+(u) + \tilde H_-(u) = -\log(u) \label{simpleId}
\eeqa
which follows from the various identities and definition in figure \ref{2ptFlows} through simple manipulations.\footnote{Here is one derivation: we can use that $G_+$ and $G_-$ are inverses of each other by plugging $x=G_-(z)$ in the leftmost gluing condition in figure \ref{2ptFlows} to get \beqa
    z &=& \frac{H_-(-\log G_-(z))}{G_-(z)} = \frac{H_-\left(-\log\left(\frac{\tilde H_+(-\log z)}z\right)\right)}{\frac{\tilde H_+(-\log z)}z}
\eeqa
Now, with $-\log z = \tilde H_-(u)$ and noting that $\tilde H_+$ and $\tilde H_-$ are again inverses of each other, we get
\beqa
    H_-\Bigl(-\log(u) - \tilde H_-(u)\Bigr)= u\,,
\eeqa
which immediately leads to (\ref{simpleId}) once we act with $H_+$ on both hand sides and use -- again -- the inverse property for these end-point functions $H_\pm$.
}
Indeed, when we plug the definition of the end-point functions in terms of densities and velocities in (\ref{simpleId}), the former cancel (since they are identical on both end-points as we just proved) while the velocity differences reduce precisely to (\ref{twoPtVelocityIdentity}) as desired. The fact that the flow  to the resulting Young tableau depends only on the difference of two velocities, and not on each of them individually, should be related to the zero-mode in the saddle point for this two-point function producing the functional delta-function equalling the two Young tableaux. This phenomenon should be similar to the one described for the normalization integral in~\cite{Alexandrov:2002fh}. 

Although we only discussed orthogonality and the velocity relations in the continuum, they even hold for the discrete flows. Starting with a distribution of $\{h_i\}$ and \textit{any} initial velocity, we get the same distribution $\{h_i\}$ at the end of the third chamber! Moreover, the final velocities are such that (\ref{twoPtVelocityIdentity}) holds. In figure \ref{fig:ortho}, we show three different flows starting from the positions but different velocities and they beautifully reach the same final positions even though the intermediate flows look quite different. We confess that we would like to have a more solid analytic understanding of this sort of miraculous discrete picture. 
\begin{figure}
    \centering
    \includegraphics[width=\textwidth]{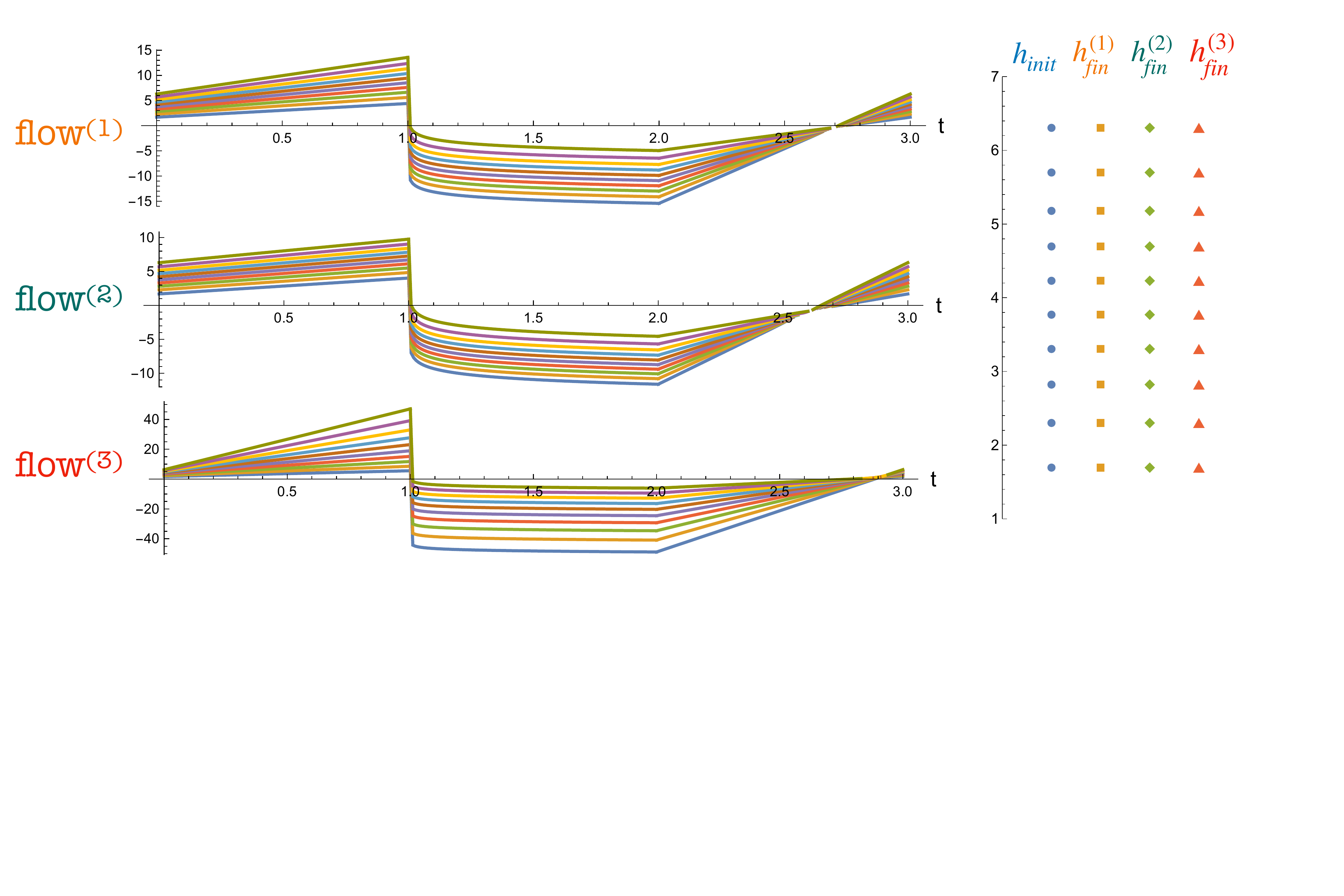}
    \vspace{-4cm}
    \caption{For a given distribution of YT weights, choosing any initial velocity leads us to the same final YT distribution. On the left, we show three different flows with different initial velocities but same initial positions. In the middle chamber, we plot logarithm of the eigenvalues in order to have nice continuous flows. The flows look quite different at intermediate times, but end up at the same positions as seen in the right panel.}
    \label{fig:ortho}
\end{figure}

\section{Huge-Huge-Huge Correlators -- The General Case} \label{FluidSec2}
In appendix \ref{characterExample} we considered the large $N$ limit of characters as a perfect playground for the various fluid techniques discussed above. The strategy employed in that appendix for the computation of the large $N$ limit of general characters nicely transposes to the case of interest -- the problem of computation of structure constants of {\it\ huge}  $\frac12$-BPS operators in $\mathcal{N}=4$ SYM. Following that appendix we proceed in two steps
\begin{itemize}
\item First we explain how some particular simple operators can be described very efficiently through some very simple fluid flows. We call these example our \textit{seeds}. These seeds come in different kinds, depending on their topological properties. We will have $n\to m$ cut structures (plus corresponding filling fractions) where an $n$ cut distribution in $x$ flow to an $m$ cut distribution in $h$; the simplest topology which we will study in greatest detail will be the $1\to 1$ cut evolution. 
\item Finally we describe  general operators in a given topological sector by starting with these seed solutions and adiabatially deforming them to whatever Young tableaux we might be interested in. This is done by exploiting the underlying integrability of the (1+1)d hydrodynamical problem. 
\end{itemize}

\subsection{Seeds: Semi-Circles}

Recall the main challenge of the fluid approach: The direct evolution of a fluid is trivial but the boundary value formulation -- which is what we need -- is very non-trivial. 

Let us illustrate how to take advantage of the power of the direct evolution by considering first the simplest case of three identical operators. We can start with an arbitrary $x_i$ distribution $\mathbb{X}=\{x_1,\dots x_N\}_{t =0}$ in the middle chamber. Automatically we know the initial set of velocities $\mathbb{V}=\{v_1,\dots v_N\}_{t =0}$ from the regularity condition (\ref{xVelGlue}) which in the discrete would read (since we are assuming identical operators $(v_1)_i=(v_2)_i=(v_3)_i=v_i$)
\beq
v_i=\frac{1}{3}\Big(-2x_i+\frac{1}{N} \sum_{j\neq i} \frac{1}{x_i-x_j} \Big) \,.\la{xGlueSymm}
\eeq
But then, if we know both the initial positions $\mathbb{X}$ and velocities $\mathbb{V}$ of the fluid bits we can trivially evolve the fluid from the center using (\ref{discreteSol}) all the way to end points where we can read of the final $\mathbb{H}=\{h_1,\dots,h_N\}$ distribution.\footnote{At the $m_j^*$ junctions we need to glue the fluid as illustrated in figure \ref{cChambers}. In the discrete formulation the gluing conditions (\ref{mVelGlue}) simply read 
\beq
(-\log x_i)_{t=1 \text{ in the inner leg}}=(x_i)_{t=0 \text{ in the outer leg}} \,, \qquad (\log x_i- x_i v_i )_{t=1 \text{ in the inner leg}}=(v_i)_{t=0 \text{ in the outer leg}} \,, \nn
\eeq
allowing us to get the initial positions and velocities in the beginning of the outer legs from the final positions and velocities of the inner legs. Then we use again (\ref{discreteSol}). } Moreover, since we now have the full fluid evolution we can simply plug it in the fluid action and obtain in this way the complete large $N$ limit of such huge-huge-huge correlators with such $\mathbb{H}$ distribution. 

\begin{figure}[t]
    \centering
    \qquad\includegraphics[width=\linewidth]{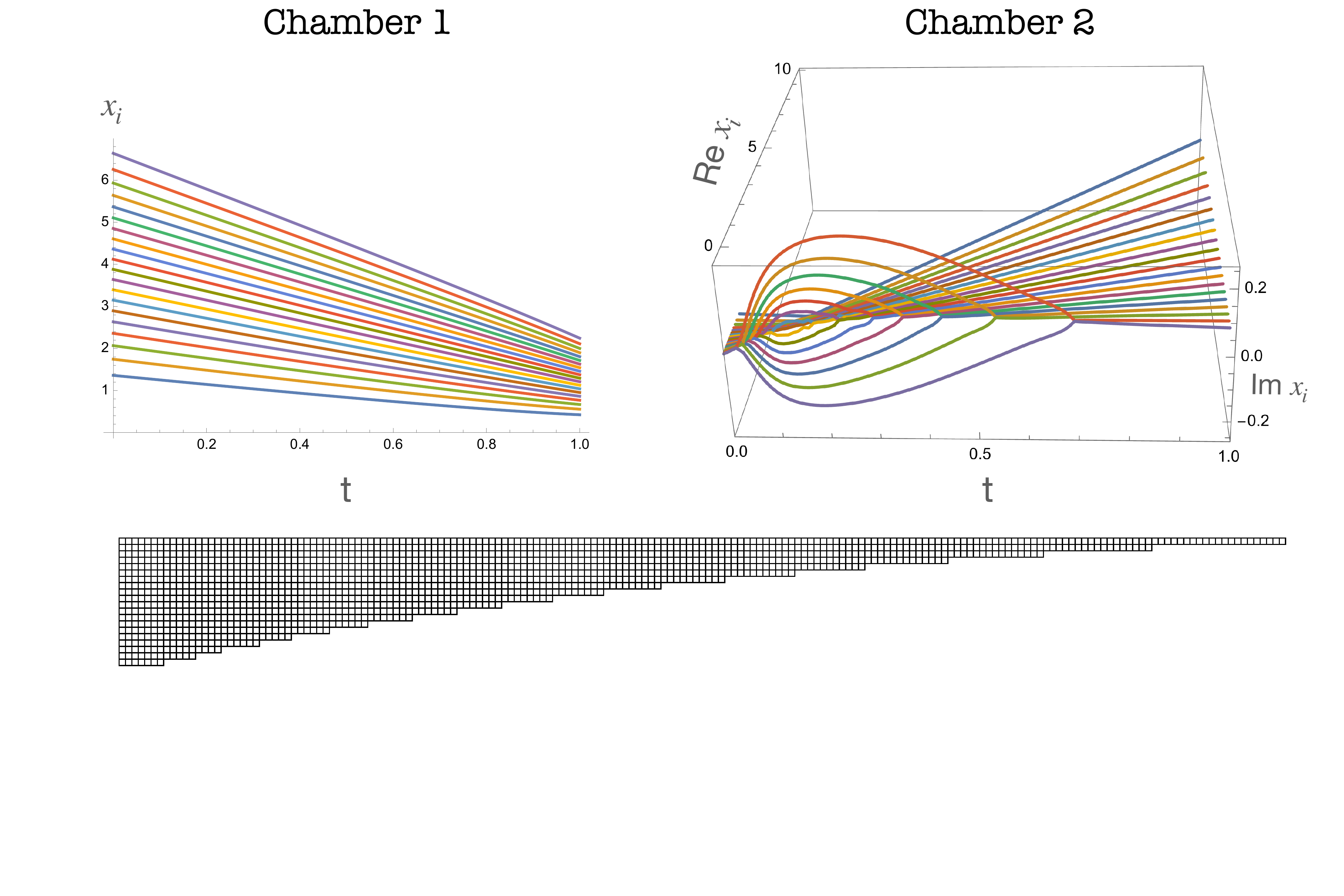}
    \vspace{-3cm}
    \caption{Three point function flows, starting with Hermite roots at $\mathbb{X}$ and equal initial velocities. For this plot, we chose $a(0)=3$ and $b(0)=4$. The eigenvalues at the end of the first chamber are also Hermite roots. In the second chamber, the flow develops shock waves and becomes complex. At the end of the chamber, we obtain the sparse Young tableau, shown at the bottom}
    \label{hermiteSeedFlow}
\end{figure}

Straightforward as this is, there is an obvious drawback of this method: Given some initial guess for $\mathbb{X}$ we have no control over what $\mathbb{H}$ we will get! 

A priori, we could even get un-physical $\mathbb{H}$'s at the end of the flow: It could have an un-physical support for example or be given by a density which would go above $1$ which is forbidden for a shifted Young-Tableaux \cite{Douglas:1993iia}.\footnote{Recall that by definition (\ref{hDef}) we have that $h_j$ are monotonic and $h_j > j-1$. In the large $N$ limit this immediately implies that the $\rho$ density is bounded $\rho(h)\le 1$. In other words, the shifted young tableaux clearly always has slope smaller or equal to $1$ and this slope is nothing but the $h$ density in the continuum limit.} Luckily we never found such unphysical examples. Even if we did, this would not be a big deal since these seed examples will be used as starting points and will then be deformed to any desired physical $\mathbb{H}$ distribution.

The simplest possible distribution of eigenvalues $x_i$ is the Wigner semi-circle distribution. In the continuum, it is known that a semi-circle distribution with linear initial velocity flows to another semi-circle \cite{Bun:2014dha}. In fact, in this case, the roots of the Hermite polynomials turn out to be \textit{exact} solutions of the \textit{discrete} flow equations for any $N$! This is probably a well known result and is a consequence of \cite{calogeroHermite, Agarwal:2019omb} and related references but we could not find it explicitly stated in the literature. Explicitly, the statement is that $z_i(t)$ satisfying
\beqa
    H_N\left(\sqrt{2N}\ \frac{z_i(t) - b(t)}{a(t)}\right) &=& 0 \label{HermiteRoots}
\eeqa
with simple $a(t)$ and $b(t)$ given in appendix \ref{Hap} are an exact solution to the flow equations (\ref{discrete}). As explained in the appendix, this gives us a nice analytical solution for the first chamber. After gluing to the second chamber however, the eigenvalues are no longer Hermite roots. This does not scare us -- we can solve the system numerically, using (\ref{discreteSol}) and we find legitimate YTs. For instance, with initial positions as Hermite roots given with an initial radius $a(0)=3$ and center $b(0)=4$, we obtain the valid Young tableau shown in figure \ref{hermiteSeedFlow}. 

Note that this flow is much richer than the character flows in section \ref{characterExample}. In the second chamber, the eigenvalues collide, leading to formation of ``shock waves" and the flow becomes complex. At later times, due to the attractive interaction, the complex conjugate pairs collide and become real once again. 

\subsection{General Young Tableaux Flows Through Deformation} \label{defSec}
\begin{figure}[t]
    \centering
    \includegraphics[width=\textwidth]{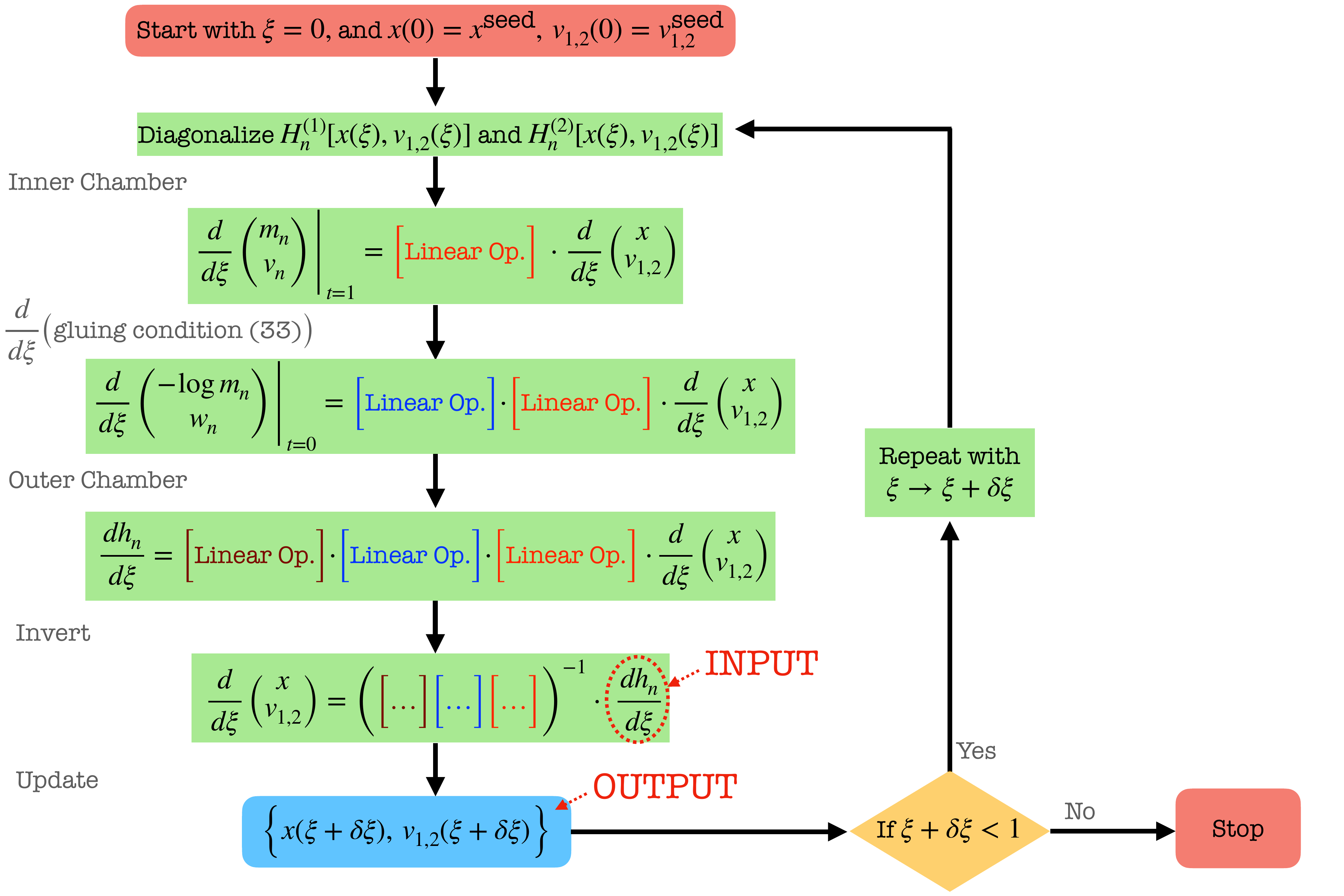}
    \caption{A depiction of the algorithm to deform three point function flows. Here, $H_n^{(1)}$ is defined in \eqref{HPDef} and $H_n^{(2)}$ is defined similarly for the second chamber. We input the deformation rates $h_n'(\xi)$ and this procedure computes for us the new initial positions and velocities at $\xi=1$.}
    \label{fig:flowChart3pt}
\end{figure}
In this subsection, we will describe how to deform the seed flow of previous section in order to solve any one cut to one cut flows. The basic principle is the same as for the character example in section \ref{deformationChiSec} but with the added complication of extra flow chambers. See figure \ref{fig:flowChart3pt} for a quick summary.

As discussed above, once the initial conditions are known, the problem is solved. So the question we need to address is -- how do $\mathbb{X}$ and $\mathbb{V}_n$ change when the YTs are deformed. With $\xi$ as a deformation parameter, let $h_i^{\texttt{seed}}$ be the YT of the seed flow at $\xi=0$. At $\xi=1$, let $h_n^{\texttt{target}}$ for $n=1\ldots3$ be the three target YT distributions. As in the character example, we need to pick some YT deformation rates such that,
\beqa
\int_0^1 d\xi\, \frac{dh_{n,i}}{d\xi}&=&h_{n,i}^{\texttt{target}} - h_i^\texttt{seed}
\eeqa

\begin{figure}[t]
    \centering
    \includegraphics[width=\textwidth]{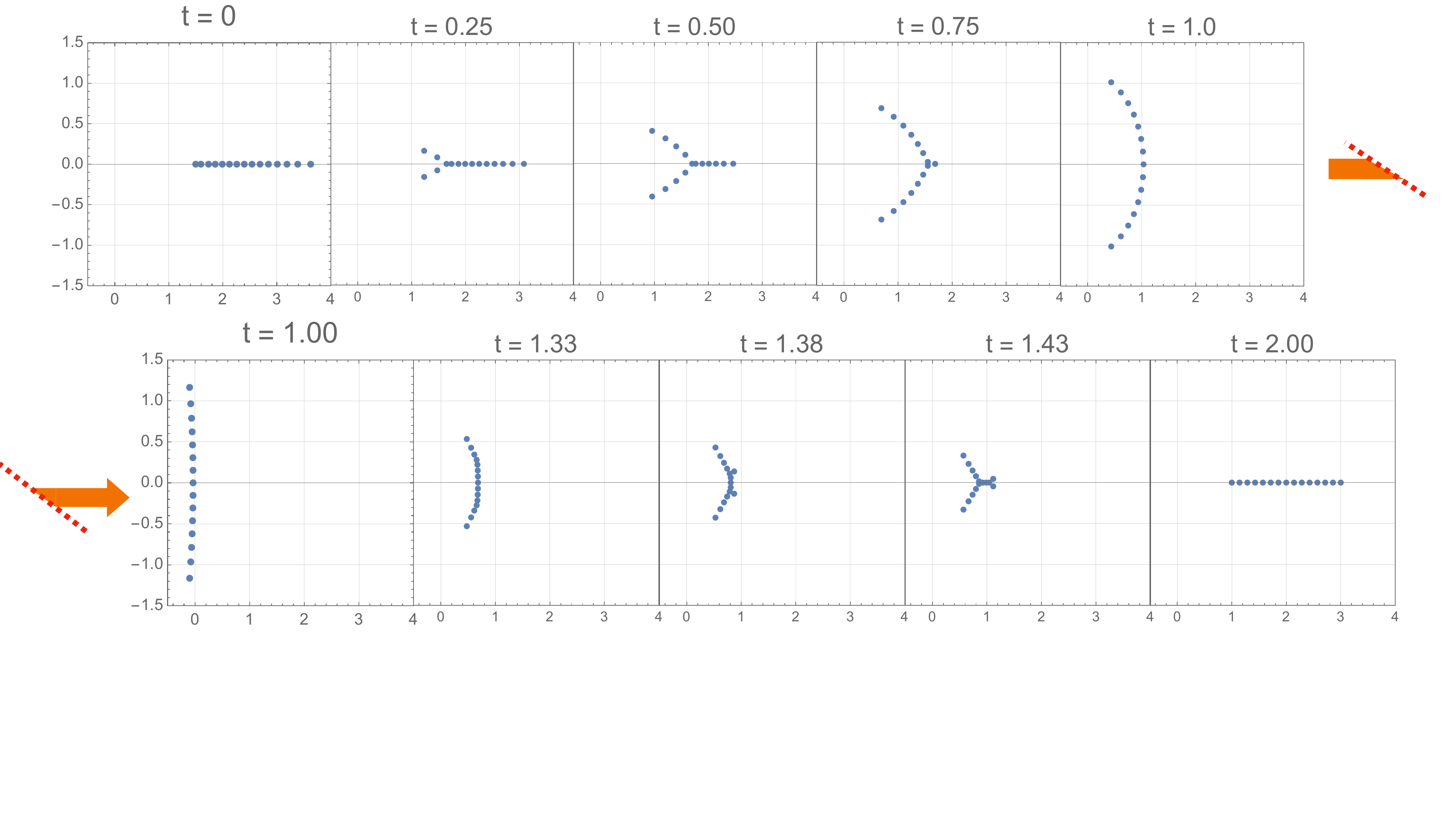}
    \vspace{-2 cm}
    \caption{Discrete flow for three equal trapezia with $\frac KN=1$, obtained by deforming the flow in figure \ref{hermiteSeedFlow}. The two rows correspond to the flows in the two chambers, which are glued at $t=1$. At $t=0$ in the first chamber, the eigenvalues are denser near the left branch point. The collisions that follow open up a shock wave which propagates to the right. After gluing to the second chamber, this complex flow collapses back to the real axis due to more collisions.}
    \label{fig:EqualTrapzFlow}
\end{figure}

As we deform, we need to ensure that the gluing conditions are satisfied. The $x$ gluing condition (\ref{xVelGlue}) can be immediately solved by requiring that
\beqa
    v_{3,i} &=& -v_{1,i}-v_{2,i} - 2x_i + \frac1N \sum_{j\neq i} \frac1{x_i-x_j}
\eeqa
We now need to solve for $\dfrac{dv_{1,2}}{d\xi} \text{ and } \dfrac{dx}{d\xi}$. Let us define 
\beq
    P_n\equiv P(x,v_n)\ ,\quad\text{ and }\quad H^{(1)}_n\equiv \text{diag}(x) + P(x,v_n)\,.\la{HPDef}
\eeq
Using eigenvalue perturbation theory in \eqref{discreteSol}, we obtain the following variations at the endpoint of the first chamber in terms of the variations at the inlet,
\beqa
    \frac{d m_{n,i}}{d\xi} &=& \bra{i} \frac{dH^{(1)}_n}{d\xi}\ket{i}\la{FirstThreeptDeformation}\\
    \frac{d v_{n,i}}{d\xi} &=& \bra{i}\frac{dP_n}{d\xi}\ket{i} + \sum_{j\neq i}\frac{\bra{i}P_n\ket{j}\bra{j}\frac{dH^{(1)}_n}{d\xi}\ket{i}+\bra{j}P_n\ket{i}\bra{i}\frac{dH^{(1)}_n}{d\xi}\ket{j}}{m_{n,i}-m_{n,j}}
\eeqa
where $\ket{i}$ are the eigenvectors of $H^{(1)}_n$. The gluing conditions (\ref{mVelGlue}) can be differentiated to give,
\beqa
    \frac{d w_{n,i}}{d\xi}&=& \left(\frac1{m_{n,i}}-v_{n,i}\right)\frac{dm_{n,i}}{d\xi} - m_{n,i} \frac{dv_{n,i}}{d\xi}
\eeqa
Finally, using eigenvalue perturbation theory again, we have
\beqa
    \frac{dh_{n,i}}{d\xi} &=& \bra{\tilde i}\left(\text{diag}\left(-\frac{d \log m_n}{d\xi}\right) + \frac{dP(-\log m_n,w_n)}{d\xi}\right)\ket{\tilde i}\la{LastThreeptDeformation}
\eeqa
where $\ket{\tilde i}$ are the eigenvectors at the end of the second chamber. The equations (\ref{FirstThreeptDeformation}-\ref{LastThreeptDeformation}) are linear in the derivatives and are trivial to solve numerically. Doing so, we obtain the deformation rates $\dfrac{dv_{1,2}}{d\xi}$ and $\dfrac{dx}{d\xi}$ in terms of $\dfrac{dh_n}{d\xi}$. We then integrate to go from the seed flow to the target flow. The deformation algorithm is summarized in figure \ref{fig:flowChart3pt}

The result of this procedure for the case when $h^{\texttt{target}}$ is a trapezium with $\frac KN=1$ for all three legs is shown in figure \ref{fig:EqualTrapzFlow}. The time slices show how in regions with high particle density, collisions occur leading to shock waves and complex flows, which beautifully collapse back in the second chamber to give the real distribution corresponding to the trapezia.

Note that the algorithm developed in this section allows for the three external operators to be all distinct as exemplified in figure \ref{fig:asymmetricFlow}.

\subsection{Zippers in Discrete Flows}
At this point, the reader might wonder why we do not start with the $x_i$ distributions following from the nice Slater determinants of section \ref{SlaterSec} corresponding to trapezia and rectangles. There is a small obstruction to this proposal, namely the potential formation of so-called \textit{zippers}.

We open here a small detour for a discussion of such exotic objects. To our knowledge such objects were first encoutered in \cite{KolyaPedro}. Consider a resolvent $$G(z)\equiv \frac{1}{N} \sum_{j=1}^N \frac{1}{z-z_j}$$ which converges in the $N\to \infty$ continuuum limit to $$G(z)=\frac{1}{z}\,.$$ What $z_j$ distribution yields such simple resolvent? One obvious possibility is to put all roots at the origin, $z_j=0$, but there are infintiely many other choices! Any distribution of $z_j=R e^{2\pi j/N}$ along a circle yields the same resolvent outside such circle. Indeed, 
\beq
G(z) = \frac{1}{N} \sum_{j=1}^N \frac{1}{z-R e^{2\pi j/N}} \simeq \oint\limits_{|w|=R} \frac{dw}{2\pi i} \frac{1}{w(z-w)} = \left\{ \begin{array}{ll}
1/z &,\,\,\, |z|>R\\ \\
0 &,\,\,\, |z|<R
\end{array}
\right. \label{imitation}
\eeq
We see that there is a closed contour -- in this case a simple circle -- outside which the function is exactly what we want. If we were to analytically continue the function from this outside region we would never see the inner region and we would simply conclude that $G(z)=1/z$. Note that this is in clear contradistinction with usual cuts which are not closed contours. There we can do monodromies around branchpoints, slide to different sheets and so on. Note also that the two sheets where $G(z) \simeq 1/z$ and $G(z) \simeq 0$ are connected at finite $N$ in the sense that we can indeed connect $|z|>R$ where $G(z) \simeq 1/z$ to $|z|<R$ where $G(z) \simeq 0$ passing through the middle of the roots $z_j$ located in the $|z|=R$ circle. When we take $N\to \infty$, however, they become disconnected. If we start outside and analytically continue inwards we get the function $G(z)=1/z$ everywhere while if we start inside and analytically continue outwards we get the function $G(z)=0$ everywhere. 

Such closed contours arose in \cite{KolyaPedro} when studying the classical limit of Bethe root distributions describing classical strings in $AdS_5\times S^5$ following the algebraic curve formalism of \cite{Beisert:2005di,Beisert:2005bm}. There again such zippers would create closed regions. Inside these regions different sheets of a large Riemann surface would be swapped, see figures 7 and 8 therein. The Riemann surface itself, would remain the same under this shuffling. And while the inside and outside of the zippers would be connected for finite number of Bethe roots, they could never be detected in the continuum limit for the same reason we described above.  

In sum: Zippers are closed contours which can be physical -- and describe Bethe root distributions rendering the spectrum of energy of classical strings -- or numerical artifacts of a discretization of a continuum problem whose solution involves some kind of Riemann surface. In either case the claim is that because zippers are closed contours -- without any branchpoints -- when they appear they do not affect the continuum limit Riemann surfaces. They might at most re-shuffle the various sheets a little bit. 

\begin{figure}[t]
    \centering
    \includegraphics[width=0.9\textwidth]{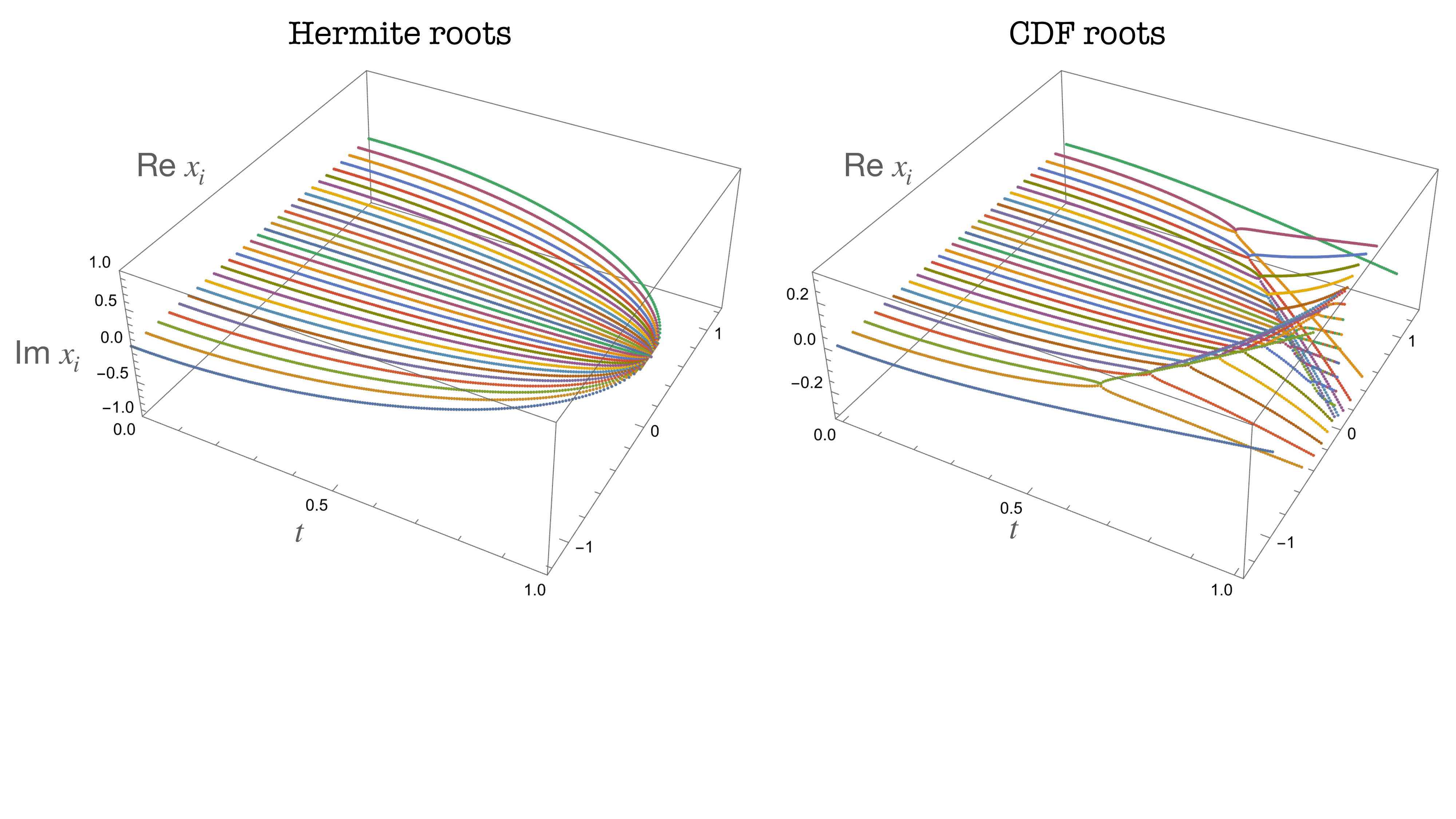}\vspace{-2cm}
    \caption{Starting from two nearly identical distributions at time $t=0$, the flow evolution can lead to drastically different looking distrutions at $t=1$. In the continuum limit we claim that they are the same however and that what we are seeing here is the dynamical formation of so-called \textit{zippers} described in the text, see also the next figure.}
    \label{fig:flowsH}
\end{figure}

Let us now see how such objects would arise for our fluid problems. 
Consider for instance a single chamber flow with the initial positions sampled from a centered semi-circle distribution with radius $a(0)=\sqrt{2}$ and zero initial velocity. Let us consider two different initial conditions,
\begin{itemize}
    \item Hermite roots $x_i$, obeying $H_N\left(\sqrt N x_i\right) = 0$
    \item CDF roots $y_i$, obeying $\dfrac iN = \int\limits_{-\sqrt 2}^{\displaystyle y_i} dz\ \texttt{prob}(z)$
\end{itemize}
where $\texttt{prob}(z) = \frac1\pi \sqrt{2-z^2}$ is the probability distribution of the semi-circle distribution. At large $N$, clearly both $x_i$'s and $y_i$'s have the same distribution. However, as we shall see their flows look rather different. 

\begin{figure}[t]
    \centering
    \includegraphics[width=0.9\textwidth]{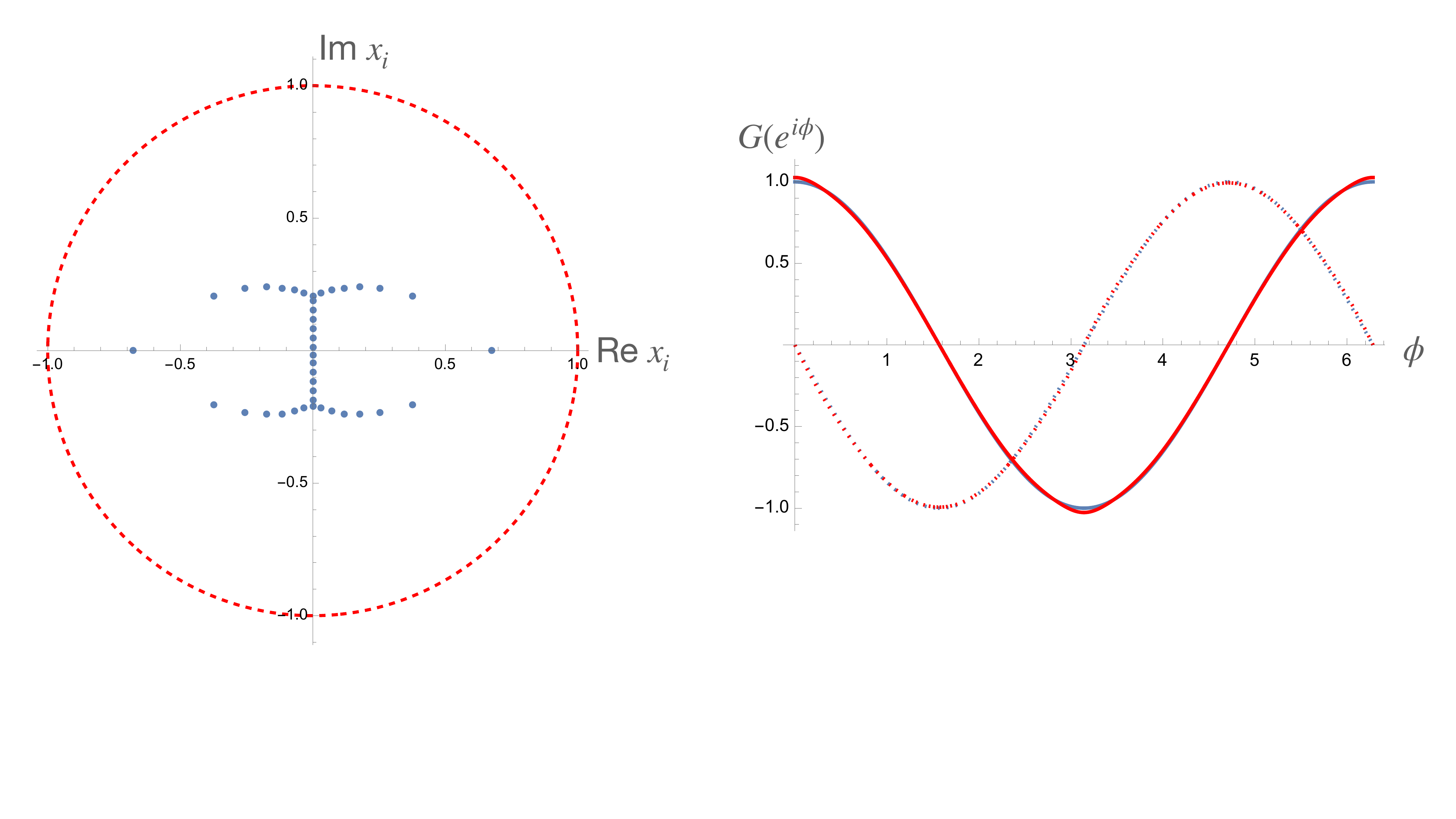}\vspace{-2cm}
    \caption{Hermite roots at $t=0$ evolve to a distribution where they all collapse to the origin at $t=1$ as illustrated on the left in figure \ref{fig:Hmoments}. However, using the cumulative distribution function as starting points leads to a very different pattern of $x_i$'s depicted in here on the left. The claim is that these roots are forming a zipper (with more number of roots, they will form nice closed contours) and that in the continuum limit they lead to precisely the same moments as the collapsed distribution. (Indeed, compare this distribution to figure 9 in \cite{KolyaPedro}.) On the right we test this hypothesis by plotting (in red) the (real and imaginary parts of) resolvent of these Bethe roots along a circle encircling them. We see that the resolved is perfectly indistinguishable from the simple $G(z)=1/z$ resolvent (plotted in blue and hardly visible as it is right below the red curve!).}
    \label{fig:Hmoments}
\end{figure}

For the $x_i$'s, as discussed earlier, the flows are analytically solvable and they correspond to Hermite roots with $a(t) = \sqrt{2-2t^2}$ and $b(t)=0$ in (\ref{HermiteRoots}). In particular, at $t=1$, we see that the flow becomes singular -- all eigenvalues collapse to the origin. The flow for the CDF roots can be computed numerically and is shown in figure \ref{fig:flowsH}. At first glance, there seems to be a big problem because the flows look nothing alike. On closer examination, it turns out that they have the same moments, as shown in figure \ref{fig:Hmoments}

One should not be too surprised with this phenomenon of small perturbations of initial conditions leading to big differences in downstream flows. After all, in the continuum limit, this amounts to imposing hyperbolic initial conditions for an elliptic PDE (see footnote \ref{hyperbolicFootnote}). However these errors are still under control thanks to integrability. In particular, recall that the conserved charges (\ref{newCharges}) reduce to moments of eigenvalue density at the end points for $m=0$ (\ref{PnmBdy}). Therefore, these must agree given that both Hermite and CDF flows at large $N$, have the same density and velocity distributions. Said differently, as described above, the fluid problem above turns into a Riemann Hilbert problem describing in the continuum a rich Riemann surface for some functions $G_\pm(z)$. Here we find zippers re-shuffling some sheets of this surface in some isolated closed regions but leaving the full Riemann surface invariant as it ought to.

\begin{figure}[t]
    \centering
    \includegraphics[width=\textwidth]{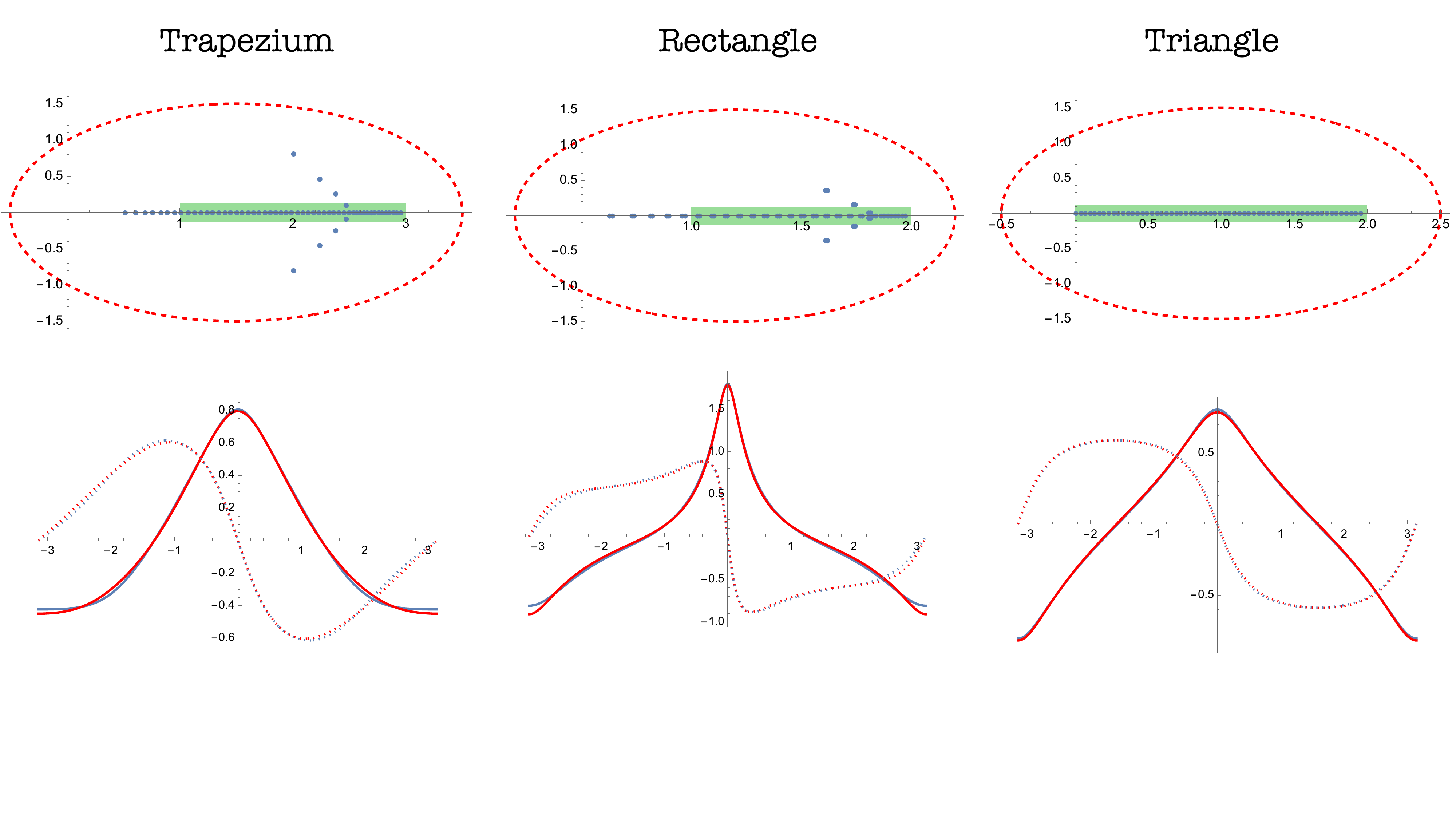}
    \vspace{-2cm}
    \caption{Zippers seem to be in play also for the full two chamber flow for trapezia and rectangles. For some reason -- maybe luck -- there are no zippers for the triangle. On the top we depict in blue the distribution for $h_i$ obtained by evolving the saddle point solutions for $x_i$ described in section \ref{SlaterSec} in the middle chamber all the way to the outside operators. The green line indicates where we would expect the $h_i$'s to lie based on the YT's that they ought to describe. Clearly, only for the triangle is the final $h_i$ distribution in perfect agreement with this expectation. However, when we plot the resolvents along the red dashed ellipses we see that the blue roots perfectly agree with the analytic expectations as depicted in the bottom panels! 
    }
    \label{fig:zippers3}
\end{figure}

Starting with the solution of SPEs the Slater determinant representations for rectangles and trapezia, we also find such zippers, as seen in figure \ref{fig:zippers3}. This is unfortunate because at the end point of these flows, we find degenerate eigenvalues. This greatly slows down the eigenvalue perturbation theory methods we use. On the other hand, as also illustrated in that figure, while the roots can sometimes look a bit weird they lead to perfect resolvents in the continuum limit. For the trapezium, for instance, we see that the funny arrow shaped distribution\footnote{Somehow these arrow distributions do not really look like closed zipper contours but probably they are and we would see that more clearly as we increase the number of roots. At least that is what the perfect match in the bottom panels of figure \ref{fig:zippers3} indicate!} of roots is indeed in perfect agreement with the expectation
\beq
G(z) = \int_{1}^3 \frac{dw}{2\pi i} \frac{1/2}{z-w} = \frac{1}{2} \log \frac{z-1}{z-3}
\eeq
for a trapezium with $\gamma=1$. (Recall that for a trapezium the density is $1/2$ and the support goes from $\gamma$ to $\gamma+2$.)

For the case of three triangles, the final distribution is perfectly real. So the triangle three point flow is also a great starting point for the deformation games of  section \ref{defSec}. In fact for the numerics, we often found that triangle flow is a better starting point than Hermite flow. 

Let us end with a simple observation: Such zippers usually arise when $N$ is large  and are usually absent with $N$ is not too large (say when $N\simeq 10$). This is quite natural if we recall the very first example we opened this section with. It is only for large $N$ that the uniform distribution of roots around a unit circle mimics well the continuum resolvent as shown in~(\ref{imitation}). For finite $N$ the only way to get a precise $G(z)=1/z$ would be to collapse all roots to the origin. And then the zippers would be gone. For our fluids, we often found it very instructive to start with a reasonable but not too large $N$ to obtain a nice full flow without zippers where the final $h_i$ distribution can be very cleanly identified. Then when we increase $N$ the final $h_i$ distribution starts becoming more exotic looking but we then know not to fear such exotic shapes; they are nothing but some harmless zippers being formed.


\section{Two Nice Examples: Very Long YTs and Graviton Gases} \label{simpleSec}

\subsection{Very Long Young Tableaux}\la{longYTsec}
It was shown in~\cite{Abajian:2023jye} that for three symmetric representations, that is for three operators with $h^{(i)}_{j}=j-1+L_{i}\delta_{j,1}$, the structure constant can be analytically computed for any $N$ in terms of a simple hypergeometric function 
\begin{equation}
\label{symC}   
Z_\texttt{all symmetric} = \frac{
 _3F_2(-\ell_{12},-\ell_{13},-\ell_{23};1,1{- N}-\frac{L_1+L_2+L_3}{2} ;1)({ N})_{\frac{L_1+L_2+L_3}{2}} 
}{  (-)^{\frac{L_1+L_2+L_3}{2} } \sqrt{(- N-L_1+1)_{L_1} (- N-L_2+1)_{L_2} (- N-L_3+1)_{L_3}}}   \,.
\end{equation}
It can be also derived from our matrix model representations. In this section we will be mostly interested in reproducing its asymptotics for very large operators $ L_1\sim L_2\sim L_3\gg N$. In this regime, since $L\gg N$ it is easy to see that the leading asymptotics of this result are $N$ independent. We can effectively drop $N$ or simply set $N$ to any number such as $N=1$. In \cite{Abajian:2023jye} this was dubbed as an \textit{Abelianization} of the structure constants in this very long row limit. We will suggest an extension of this result below, see figure \ref{longYT}.

\begin{figure}[t]
    \centering
    \includegraphics[width=\textwidth]{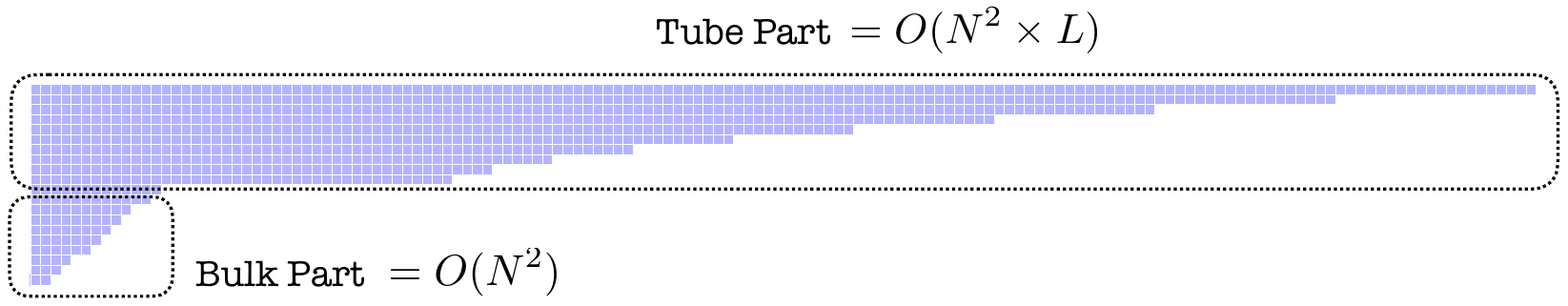}
    \vspace{-8.5cm}
    \caption{YT where the first rows are much longer than the rest. We can introduce a parameter $L \gg 1$ so that $h_j \sim L$ for the first $K$ rows and $h_j \sim 1$ for the remaining $N-K$ ones. Then the contribution to the three point function action of these first rows -- which we call the \textit{tube part} -- will give a leading contribution, of order $L$ and the remaining ones -- which we call the \textit{bulk part} -- will be subleading. Moreover, because of this decoupling the three point function asymptotics will be $N$ independent and we can effectively replace $N$ by any value greater or equal than $K$.
    }
    \label{longYT}
\end{figure}

As pointed out in \cite{Abajian:2023jye}, the exact result (\ref{symC}) has the same asymptotics as 
\begin{align}
Z_{\rm all\,\, symmetric}\simeq \frac{2^{(L_1+L_2+L_3)/2}\sqrt{\pi}\,\,L_1!\,L_2!\,L_3!}{(\frac{L_1+L_2-L_3}{2})!(\frac{L_1+L_3-L_2}{2})!(\frac{L_3+L_2-L_1}{2})!} 
 \label{rhs}
 \end{align}
which we can trivially rederive from our results once we use the $N$ independence alluded to above. Indeed, setting $N=1$ in (\ref{singleM}) we would end up with a simple integral of three Hermite polynomials but that indeed precisely yields the right hand side of (\ref{rhs})!, 
\begin{align}
I= \int\,  dx \,e^{- \,x^2}\, H_{L_1}(x)H_{L_2}(x)H_{L_3}(x)
=\frac{2^{(L_1+L_2+L_3)/2}\sqrt{\pi}\,\,L_1!\,L_2!\,L_3!}{(\frac{L_1+L_2-L_3}{2})!(\frac{L_1+L_3-L_2}{2})!(\frac{L_3+L_2-L_1}{2})!} \,.
 \end{align}
It is instructive to derive the large $L$ limit of this expression. The easiest way is to take the result on the right hand side and use Stirling. The instructive exercise is to do it directly through simple manipulations of the integral on the left hand side. Using the Hermite generating function
\beq
H_h(x) = h! \oint \frac{dm}{2\pi i} e^{-m^2+2mx-(h+1)\log(m)} \la{hermiteGenFunc}
\eeq
we have 
\beq
I=\frac{L_1! L_2! L_3!}{(2\pi i)^3}\oint dm_1 \oint dm_2 \oint dm_3 \int dx e^{-S_\text{eff}(m_1,m_2,m_3,x)}
\eeq
which we can immediately estimate through saddle point which is located at 
\beqa
m_i = \frac{1}{2} \sqrt{\frac{(L_{i}+L_{i+1}-L_{i-1})(L_{i-1}+L_{i}-L_{i+1})}{L_{i-1}+L_{i+1}-L_i}}\,,\qquad x=-\frac{1}{2} \frac{\sum\limits_{i=1}^3 (L_i^2-2 L_{i-1}L_{i+1}) }{\prod\limits_{i=1}^3\sqrt{L_{i-1}+L_{i+1}-L_i}} \,.\la{threeHermiteSPE}
\eeqa
Here $i$ is defined modulo $3$ so that $L_{-1}\equiv L_3$ and $L_4=L_1$. Plugging this saddle into the effective action $S_{\text{eff}}$, and normalizing by two Hermite integrals, yields precisely the desired asymptotics,
\beq
    \log I \simeq \underbrace{\strut -\sum_{i=1}^3 \left(\frac{L_{i+1}+L_{i-1}-L_{i}}2\right) \log\left(\frac{L_{i+1}+L_{i-1}-L_{i}}2\right) - \frac12 L_i \log L_i}_{\equiv\ \texttt{tube}(L_1,L_2,L_3)}\la{tubeDef}
\eeq

We will now show how this result can be extended to YTs with many long rows. The details of the derivation are in appendix \ref{apLongYT}. We consider a representation such that the corresponding YT has a very long set of rows as illustrated in figure \ref{longYT}. We can formally do it by scaling the first $K$ rows by a large parameter $L$. Then the structure constant semi-classics still isolates a leading contribution which we call the \textit{tube} part 
\beq
\lim_{N,L\to \infty} \frac{1}{N^2 L} \Big( \log C \simeq N^2 \times L \times \texttt{Tube Part}+ N^2 \times L^0 \times  \texttt{Bulk Part}+\dots \Big)=\texttt{Tube Part} \la{doubleLimit}\,,
\eeq
which will be the subject of the study of this section. This tube contribution arises from the contribution of the first long rows which decouple from the remaining ones. In practice that means we can effectively drop the smaller rows. That effectively corresponds to replacing $N$ by $K$ and simply taking the single $L\to \infty$ limit -- instead of the double limit in (\ref{doubleLimit}) -- generalizing the abelianization proposal in \cite{Abajian:2023jye}. 

Consider a single chamber HCIZ flow. Let the first $K$ eigenvalues of $X$ be very large, obeying $|x_i+t v_i| \gg \frac1{|x_j-x_k|}$ for $i<K$ and $j\neq k$. The flow is then given by the eigenvalues of a block diagonal matrix (\ref{discreteSol}),
\beqa
\{x_{n,i}(t)\} &\approx& \text{eigenvalues of }\begin{bNiceArray}{c|c|c|c}[margin]
x_1+t v_{n,1}&&&\Block{3-1}{0}\\

& \Ddots \\
& & x_K+t v_{n,K} \\
\hline
\Block{1-3}{0} &&&  \text{diag}(\mathbb{X}^{\texttt{bulk}}) + t P(\mathbb{X}^{\texttt{bulk}},\mathbb{V}_n^{\texttt{bulk}})
\end{bNiceArray}\nn
\eeqa
where $\mathbb{X}^{\texttt{bulk}} = \text{diag}(x_{K+1} \ldots x_N)$ and $\mathbb{V}^{\texttt{bulk}}_n$ is defined similarly. From this we see that the first $K$ eigenvalues decouple from the bulk part and evolve freely. 

In the limit where $K$ rows of the Young tableaux very long, these eigenvalues decouple from the bulk in both the first and second chamber (see appendix \ref{apLongYT}). So, we have two free evolutions, glued together by (\ref{mVelGlue}). The eigenvalue positions at the end of the second chamber, $\{y_{n,i}\}$ is the set of Young tableaux weights $\{h_{n,i}\}$. However, they need not be ordered -- in general, we have $y_{n,i} = h_{n,\pi(i)}$ for some permutation $\pi$. The flow for these large eigenvalues is shown in figure \ref{fig:longYTFlow}.
\begin{figure}[t]
    \centering
    \includegraphics[width=\textwidth]{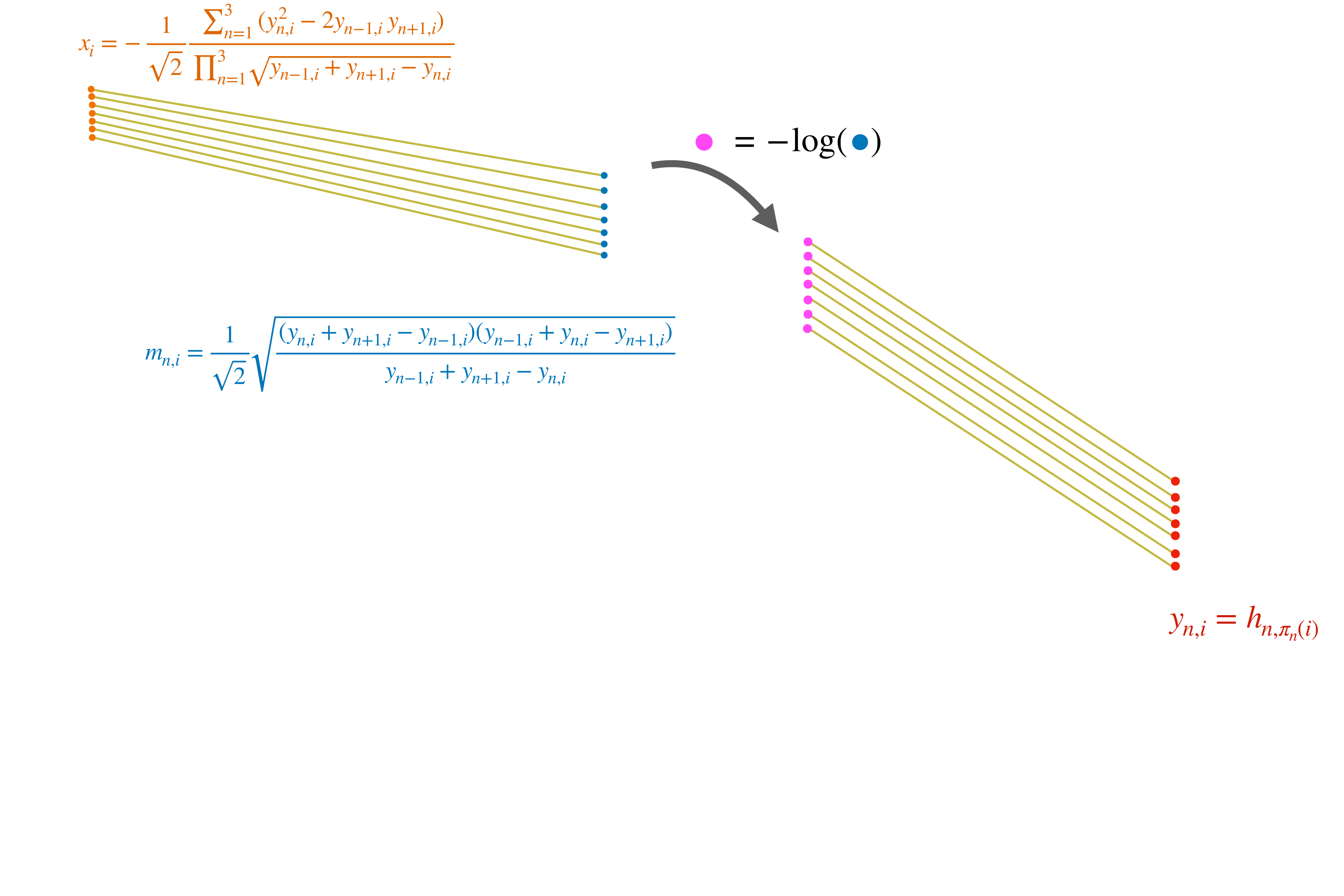}
    \vspace{-3.5cm}
    \caption{A schematic depiction of flows for YTs with rows much bigger than $N$. As usual, we have two chambers, $(x\rightarrow m)$ and $(-\log m\rightarrow y)$. The flows are labeled by permutations of the YT weights, $\pi_n$. In the limit under consideration, the eigenvalues evolve freely, as indicated by the straight lines. The saddle point locations of $x_i$ and $m_{n,i}$ are shown in the figure. Note that -- not coincidentally -- these have the very same functional dependence as in the case of integral of three Hermites (\ref{threeHermiteSPE}).}
    \label{fig:longYTFlow}
\end{figure}

Plugging in the flow from figure \ref{fig:longYTFlow}, we obtain the following leading contribution in the action for the three point function,
\beqa
\frac1{N^2L} \log Z \simeq \max_{\pi,\pi'\in S_K}\left[\frac1{NL}\sum_i \texttt{tube}(h_{1,i}, h_{2,\pi(i)}, h_{3,\pi'(i)})\right]\la{maxActionTubes}
\eeqa
where $\texttt{tube}(h_a,h_b,h_c)$ is the function defined in (\ref{tubeDef}). As alluded to above, this formula shows that the structure constants are Abelianized in this limit. We first break up the Young tableaux into its rows. The three point function is then a simple product of $N$ terms, each of which is a fully symmetric three point function (\ref{rhs}) formed by picking one row from each tableau, see figure \ref{fig:abelianization}. We then maximize over ways of pairing up the rows to get the leading contribution.
\begin{figure}[t]
    \centering
    \includegraphics[width=\textwidth]{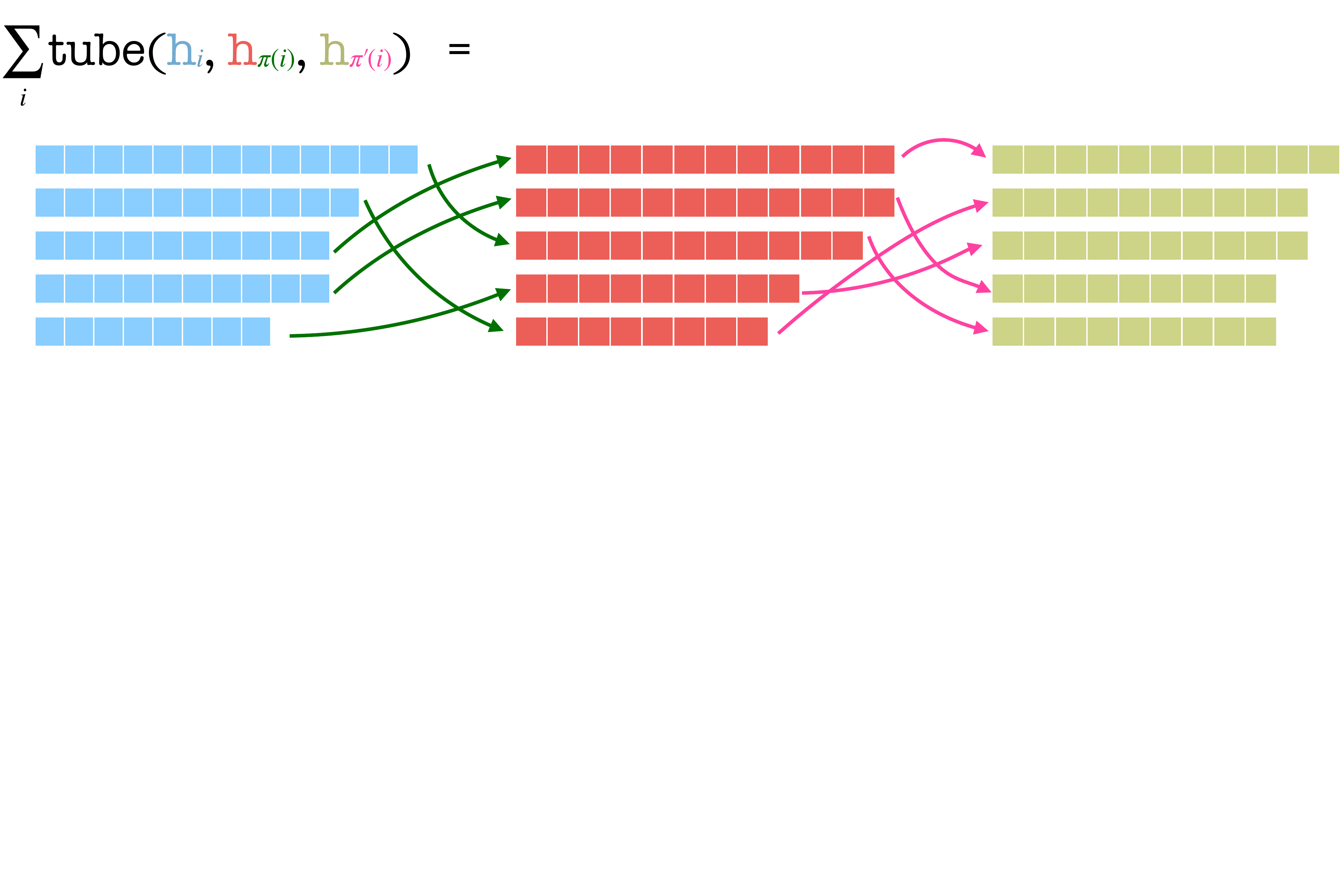}
    \vspace{-6.5cm}
    \caption{The saddle point contributions in the tube limit are given by a sum of terms, each of which is a $\texttt{tube}$ formed by picking one row from each YT. The action is obtained by maximizing over all possible pairings, shown here as green and pink arrows.}
    \label{fig:abelianization}
\end{figure}

\subsection{Graviton Gas Example}
We want to consider here a simple class of operators which we dub the \textit{graviton gas}. (In \cite{Paul:2022piq} they were introduced with the more sober name of \textit{maximal-trace operators}.)
They are defined as starting with a $20'$ operator, dual to a single graviton, and raising it to a big power. Explicitly, we have three graviton gas operators (which are also $1/2$ BPS), 
\beq
\texttt{gas}_i=\mathcal{N}_i \,(\Tr(y_i\cdot \phi(x_i))^2)^{L_i/2}  \label{gas}
\eeq
each with dimension $L_i$ so that its correlator will be still of the same form (\ref{kin}). This three point function is then simply given by a partition function like (\ref{M1M2M3}) with the characters replaced by simple $\text{Tr}(M_i^2)^{L_i/2}$ factors. Then this partition function can be trivially computed by considering a trivial vacuum partition function with no insertions but a modified kinetic action obtained by replacing $M_i^2 \to \alpha_i M_i^2$. Then we can generate the operator insertions by simply taking derivatives with respect to $\alpha_i$ of the gaussian partition function and then setting $\alpha_i=1$. Carrying this out leads to a nice simple representation for the three point function very reminiscent of the three symmetric operator case discussed in the previous subsection, namely, 
\beqa
&&\!\!\!\!\!\!\!\!\!\!
\<\texttt{gas}_1\texttt{gas}_2\texttt{gas}_3\>=\texttt{kinetic} \times  \label{gas3F2}\\ && \!\!\!\!\!\!\! \times \Big(\mathbb{C}_\texttt{gas}=
\frac{\Gamma\big(\frac{N^2}{2}+\sum\limits_{i=1}^3\frac{L_i}{2}\big)\sqrt{  \Gamma\!\left(\frac{N^2}{2}\right)\prod\limits_{i=1}^3 L_i! } }{\ell_{13}!\ell_{23}!\ell_{12}!
\sqrt{\prod\limits_{i=1}^3
\Gamma\!\left(\frac{N^2}{2}+L_i\right) 
}} \times {}_3F_2(-\ell_{23},-\ell_{13},-\ell_{12};\tfrac{1}{2}
   ,1-\tfrac{N^2}{2}-\sum\limits_{i=1}^3\tfrac{L_i}{2};1) \nn   \Big)\,.
\eeqa
This result is exact, it holds for any $L_1,L_2,L_3,N$.  Nice simplifications take place when these are large. If we set $L_i = \alpha_i N^2$ we obtain two big simplifications when $\alpha_i$ is very large or very small, namely
\beqa
\log \mathbb{C}_\texttt{gas} \simeq \left\{
\begin{array}{ll}
\texttt{tube}(\tfrac{1}{2}L_1,\tfrac{1}{2}L_2,\tfrac{1}{2}L_3) & \alpha_i \ll 1 \\ \\
\texttt{tube}(L_1,L_2,L_3) & \alpha_i \gg 1 
\end{array}\right. \,. \label{AbLimits}
\eeqa
Tubes again! When the operators are huge, with many more fields than the number of colours squares, the matrix nature of these operators can effectively be discarded as in \cite{Abajian:2023jye}. We can as well replace the gas operators (\ref{gas}) by effective abelian 
 operators 
\beq
\texttt{gas}_i^{\text{effective, }\alpha \gg 1}  \simeq  \phi(x_i)^{L_i} 
\eeq
without any matrix structure so that once we Wick contract three of these we readily obtain an expression as (\ref{rhs}) for counting which operators from each field contract with the other with $e^{\texttt{tube}(L_1,L_2,L_3)}$ asymptotics. Each of the $L_i/2$ gravitons of each operator effectively breaks into its two scalar constituients in this limit. The $\a_i\gg 1$ is therefore a sort of \textit{deconfining} limit. The $\a_i\ll 1$ would be in this sense a \textit{confining} limit. When the operators are large but much smaller than $N^2$ the trace of two scalars that makes the graviton does not want to be broken as that would be suppressed at large $N$ lacking the huge entropy contribution of the previous \textit{deconfining} limit. The gravitons therefore preserve this dimmer configuration of two fields and we can now replace the gas operators by a gas of $L_i/2$ effective dimer operators $\Phi(x)$ -- $\Phi$ is basically the $20'$ operator which is raised to the $L_i/2$ power in (\ref{gas}) -- which do not break apart in this low entropy \textit{confining} limit so that  
\beq
\texttt{gas}_i^{\text{effective, }\a \ll 1}  \simeq  \Phi(x_i)^{L_i/2} 
\eeq
This is why the asymptotics in this limit are exactly as before with a simple $L_i \to L_i/2$ replacement. In this limit we again simply count how many fields from operator $i$ connect to operator $j$ but now these fields are the full graviton and not its individual constituents. 

For intermediate $\a_i$ we have
\beq
\log \mathbb{C}_\texttt{gas} =N^2\, \mathbb{S}_\text{effective}(\a_1,\a_2,\a_3) \,, \label{Seff3}
\eeq
which interpolates between these two abelian limits.
It would be interesting to find its gravity dual. A proposal for the dual geometry created by similar gas operators was recently proposed in \cite{Giusto:2024trt}. Their geometry would hold and reproduce our correlators when $L_1=L_2=N^2 \a \gg 1$ and $L_3=L=O(1)$ so that we have a single background geometry with a small probe on top and 
\beq
\mathbb{C}_\texttt{gas}=(N^2 \a)^L \times (\texttt{Simple polynomial of degree L in $\a$})
\eeq
where these polynomials can be readily obtained from the exact result (\ref{gas3F2}).
One can also replace $\texttt{gas}_3$ by any operator with a finite number of fields and one should be able to obtain a match with their supergravity computation. (A beautiful pioneering paper studying VEVs of operators in these sort of supersymmetric states and matching between gauge theory and gravity is \cite{Skenderis:2007yb}). Another  nice recent paper with several HHL computations is \cite{Holguin:2023orq}.) It would be instructive to extend the result \cite{Giusto:2024trt} -- perhaps perturbatively -- to include backreaction of the third operator eventually reproducing the full three legged geometry (\ref{Seff3}). How would the two simplifying abelian limits (\ref{AbLimits}) matifest themselves geometrically? 

In \cite{Paul:2022piq,Paul:2023rka,Brown:2023why} four point HHLL type integrated correlators arising from supersymmetric localization were considered; as before, these large charge results should be interpreted as small probes (now two probes instead of one) proving the large background generated by the huge operators. It would be very interesting if these integrated correlators could also be extended to describe configurations with more than two huge operators.

Conversely, one could take all results of this paper valid for three large operators and study the limit where one of them is much smaller than the other two. This would be a slightly improved HHL limit, in the sense that, since we now  have the exact matrix model representations, we would have full control over the backreaction induced by the third operator. We should be able to study not only diagonal configurations (the two huge operators being the same) but also off-diagonal configurations (such that the action of the third operator backreated on one of the huge operators transforming it slightly). As emphasized in \cite{Escobedo:2011xw,Yang:2021kot} such backreaction effects -- which are often  overlooked in the literature -- can often have a major physical consequences.

\section{Conclusions}

We studied in this work three point correlation functions of $1/2$ BPS operators in $\mathcal{N}=4$ SYM. For generic -- i.e. non-extremal correlators -- such correlators are given by nice combinatorial problems which can be cast as simple matrix models. We proposed two such matrix model representations in section \ref{matrixSec}. One uses a single matrix with an involved integrand -- see (\ref{singleM}) -- and the other involved four matrices and a much simpler integrand -- see (\ref{fourInts}). Each of them has its own advantages.

In some simple cases -- when the correlators become extremal, or when the external operators are particularly simple (see e.g. section \ref{simpleSec}) or when one operator becomes the identity and the three point function reduces to a two point function (see section \ref{2ptS}) -- these matrix models can be analytically evaluated yielding exact finite $N$ results. 

The main focus of this paper is the study of these correlators in the limit of a large number of colours and when the BPS operators are \textit{huge}, with a very large, $\mathcal{O}(N^2)$ conformal dimension. It is important to stress that this is \textit{not} a trivial planar limit where only planar diagrams contribute to the correlation function! On the contrary, since the operators are huge, the number of possible diagrams contributing to these correlators grows exponentially. Because of this entropic enhancement, graphs of all genera matter. In the matrix model language we are studying these matrix models at large $N$ in the presence of very heavy sources which shift the vacuum planar saddle to a totally different one.  

When these operators are simple enough, these sources can be simplified and the single matrix representation (\ref{singleM}) is the most powerful. As we recall in section \ref{SlaterSec}, these sources can be seen as Slater determinants, multi-fermion wave functions in a simple harmonic well. Simple sources are those for which these fermions occupy a simple enough pattern of energy levels such that the semi-classical limit of these wave functions is easily accessible. Examples of the simple sources are the fermion fillings of harmonic potential corresponding to empty, triangular, rectangular or trapezium-like Young tableaux. For general huge operators, the Slater determinants related to arbitrary fillings of harmonic potential appear to be quite complex objects, certainly worth studying further.

Meanwhile, we found the four matrix approach more convenient for our purposes.  We explore in this work a beautiful connection between simple determinants of eigenvalues arising in all sorts of matrix model manipulations and one dimensional inviscid (1+1)D fluids obeying the Riemann-Hopf  hydrodynamical equations described in section \ref{FluidSec}. Such duality allows us to write down equations describing three fluids, one for each operator, which start their lives with some initial density profile. Different external huge operators are given by different Young Tableaux which code these different initial densities. Then the four matrix model is evaluated in this semi-classical limit by evolving these external densities towards a middle point where these fluids meet and where a proper regularity condition ought to be imposed, see figure \ref{fig:asymmetricFlow} for a nice example.\footnote{In the bulk of the paper we often considered three identical external operators for simplicity. The generalization to the case of different operators is straightforward as described in section \ref{defSec} and -- indeed --  figure \ref{fig:asymmetricFlow} is a simple example of one such asymmetric flows governing the three point function of three distinct operators.} This is a beautiful problem. It turns out that this one dimensional fluid is integrable! (This is a very different incarnation of \textit{integrability}    as compared to the usual planar integrability of free strings governing the renormalization properties of finite operators in ${\cal N}=4$ SYM at large $N$~\cite{Beisert:2010jr,Gromov:2013pga}; it will be fascinating to see how much of it survives as we extend this geometrical regime of huge operators further, see also discussion below.) This integrability means that some fluid problems can be trivially solved. It is straightforward, for example, to evolve the fluid forward given some initial density \textit{and} velocity in what is called the \textit{initial value problem}. Note that our problem, instead, is cast as a \textit{boundary value problem} where we are given partial boundary conditions at various ends instead (like when we study the flight of a projectile knowing its initial and final locations and being ignorant of the initial or final velocities). In contradistinction to the direct problem, solving this boundary value problem turns out to be a formidable challenge. 

The solution comes from combining the two approaches just described. We use simple operators with simple sources to solve these matrix models. We then translate these solutions into simple fluid flows. Finally, to study more complicated operators we use these simple flows as starting points and adiabatically change the operators towards the operators we are ultimately interested in. This strategy is put in practice in section \ref{FluidSec2}. 

The study of  (1+1)D free fermion gas by means of fluid dynamics is a well known approach applied to various physical systems. For example, it is employed for the study of shape and polymer formation phenomena~\cite{pallister2022limit,pallister2024phase} 
and references therein.  We hope our methods can be applied to such problems. 

How and why do these fluids show up on the dual side? Here we might want to detail what we mean by the dual side. We could either mean standard holography where we have type IIB supergravity in ten dimensions and some complicated three legged geometry describing these three point correlators (a la  \cite{Abajian:2023bqv,Abajian:2023jye}) \textit{or} we could mean twisted holography \cite{Costello:2018zrm,Costello:2016mgj} with its six dimensional truncated dual which should also fully capture these supersymmetric obserbables. In any case, be it in ten dimensional type IIB supergravity describing the geometrical limit of IIB superstrings (usual holography) or in the six dimensional BCOV theory describing the geometrical limit of the topological B model (twisted holography), it would be fascinating to see how these fluid equations emerge. Are they a mini version of Einstein equations for instance, so that we are deriving the gravity dual from the CFT? That would be conceptually very satisfactory. A small step in this direction could be to re-interpret some of the features of the two-point function geometry -- encoded in the nice geometries of Lin-Lunin-Maldacena \cite{Lin:2004nb}. There, the fluid density at the end points -- that is the Young-Tableaux density $\eta(h)$ -- shows up as coding black and white regions parametrizing an harmonic function which plays a prominent role in the ten dimensional metric. It would be interesting to see whether the fermionic coherent states found at the end of our ``inner chamber" (that is, the $m_n$ eigenvalues conjugate to the shifted weights $h_n$) are already enough to specify the boundary conditions for the geometries \cite{Balasubramanian:2005mg, Balasubramanian:2007zt,deMelloKoch:2008hen}; the flows for those would be simply given by figure \ref{cChambers} with the three outer legs amputated. What is the role of the fluid velocity $v(h)$? From the fluid analysis of section \ref{2ptS}, we expect it to be simply given by $-2h-\log(h)$; looking for a nice physical quantity in these 10d geometries with such simple functional dependence might be enlightening. In terms of twisted holography, we could try to bridge the gap between the huge (of order $N^2$) operators considered here (dual to geometries) and the very large (or order $N$) operators studied by Budzik and Gaiotto \cite{Budzik:2021fyh} (dual to D-branes)  to try to uplift the latter to full geometry backreacting operators; fluids should then  arise in the twisted dual. 

One of the interesting elements that must arise geometrically when delving on the holographic dual side, is the proliferation of geometric phase transitions. Indeed, our matrix model \eqref{singleM} 
has a rich parameter space labeled by three Young tableaux. Such matrix models usually have a complicated phase space, with plenty of possibilities of having multiple patterns of phase transitions in the large $N$  limit (usually of 3rd order, like in various random matrix models related to statistical mechanics on planar graphs, reviewed for example in~\cite{Kazakov:1988ch}). Undoubtedly, we should have such phase transitions in our matrix integral and they should be dual to interesting gravitational transitions. 
Since 3rd order phase transitions are very difficult to detect numerically, it would be great to generalize the exact result of the type of three triangular diagrams described in section~\ref{SlaterSec} to more general Young tableaux, such as trapezium and rectangular tableaux, and solve the corresponding integral equations given in appendix~\ref{RectangleTrapeziaSPEap} to develop more intuition. It is also desirable to work out a formalism for analysing this matrix integral and the Slater determinants entering there based on algebraic curves, similarly to the one well known for the two-matrix models~(see e.g.~\cite{Kazakov:2002yh,Alexandrov:2002fh,Kazakov:2004du}) or for characters of big Young tableaux (see e.g. \cite{Kazakov:1995ae,Kazakov:1996zm,Kazakov:1995gm,Kazakov:2021uio}).

It would also be very nice to extend and generalize our results from the CFT perspective. There are many such possible directions. Even for the three point case we discussed in this paper -- and for the sort of operators we considered here -- many questions are still open. 

As explained in the text, in the large N limit we have many saddle points. Which one dominates can be sometimes figured out on a case by case analysis but a general understanding of this rich phase space and of the various potential phase transitions between these various saddles is wide open. In the continuum limit these various saddles can be distinguished by their topology. In fluid language we can have disjoint domains of fluids at an initial time merging and slipping until they reach a final configuration with another number of disjoint domains. The simplest case where a single lump of fluid evolves to another simply connected domain is what we call the $1\to 1$ cut topology and was the example studied in most detail in this paper. Would be fascinating to develop a more efficient technology to deal with all these more general cases. In principle this should be possible through algebraic geometric methods. After all these fluids can be cast as rich Riemann-Hilbert problems as explained in appendix \ref{doubleInversionAp}. Another promising approach is to use the conserved charges in the continuum to ``integrate out" the $M_n$ matrices and relate the moments of the YT to the initial conditions at $X$. Doing so, we get
\beq
    \sum_{r=0}^{n+1}C_{n,r} \int\limits_{\text{supp }\rho} \!\!\frac{dx}{2\pi i}
    x^{n-r} \Big(G_+(x-i0)^{n+r+1} -G_+(x+i0)^{n+r+1} \Big)
    =\! \int \! dh\ h^n \,\eta(h)\, , \label{ChargesEq}
\eeq
where $C_{n,r}=\frac{(-1)^r (r-n-1) \binom{n+1}{r}}{(n+1) (n+r+1)}$.
For identical operators, for instance,
\beq
G_+(x\pm i0 )=\frac{x}{3}\pm i \pi \rho(x) +\,\frac{1}{3}\, \dashint \frac{\rho(y)}{x-y}\,. \nn
\eeq
We see that (\ref{ChargesEq}) yields a trivial map from the density of eigenvalues in the middle chamber to the physical YT density,  $\rho(x) \to \eta(h)$. All we would need is to find a clever way to get the inverse map $\eta(h)\to \rho(x)$. Of course, this inverse map cannot be single valued -- as we have seen there is a large moduli space of solutions for $\rho(x)$. One idea to numerically explore this very rich non-single valued map would be to make an ansatz for $G_+(x)$ with several free parameters which we would fix by imposing (\ref{ChargesEq}). We hope to explore these ideas in future works.

What about generalizations? We could consider four point functions for example. Here, we can do two things. One is to pick polarizations $y_i$ to each operator tuned to its location $x_i$ such that the four point function of these supersymmetric operators still preserves the same amount of supersymmetry as the three point function. This tuning was proposed in \cite{Drukker:2009sf,Beem:2013sza} where it was explained that these supersymmetric correlators are independent of the operator locations as well as of the coupling of the theory; of course, they will still have a very non-trivial dependence on $N$ and on the details of the various operators. Such supersymmetric correlators can also be described purely in terms of twisted holography. And they should have nice fluid descriptions. We will have more chambers and more gluing junctions but the general picture carries over. It would be nice to work it out in detail. 

The other thing we could do is to pick $y_i$ uncorrelated to the locations $x_i$. This would no longer be described by twisted holography. We would need full fledged holography to describe such non-supersymmetric correlators. Even at tree level we would encounter matrix models which we would not know how to solve. Instead of (\ref{M1M2M3}) we would end up with a similar expression with \textit{four} matrices -- one for each operator -- but now the kinetic term would be generic so that the problem becomes unsolvable with current matrix model techniques.\footnote{A related toy model -- still unsolvable with current known techniques -- was studied in \cite{Bargheer:2019kxb} precisely when studying some simple set of four point correlation functions in this gauge theory, see equation~(2.6) therein. Already in the quarter century old matrix model review \cite{Kazakov:2000aq} it is pointed out -- see e.g. first paragraph in page 4 -- such general matrix models with several matrices are typically not solvable when the kinetic terms involves couplings between matrices forming \textit{closed} rings of $M_i M_{i+1}$ couplings since when integrating out the angular part link-by-link one is left with a final integration when closing the link. This remains an unsolved problem today.
} It would be remarkable if we could develop new techniques to overcome this obstacle. Of course, even if we could, the tree level result would in this case be just the beginning. We would then need to incorporate loops. In a dream scenario, tree level would be described by a yet to be unveiled integrable fluid system corrected in a controllable way at each loop order. The re-summation to strong coupling should then be a formidable exciting problem as it ought to unveil black holes and chaos in the four point function OPE decomposition. 

It all still feels a bit too far from BHs. Half BPS operators are dual to beautiful geometries but they are not dual to black holes. In the CFT this is easy to understand; each operator is made of a single big matrix multiplied many times inside many traces. One can not generate the huge number of operators we need to describe black holes with huge entropy with such simple operators made out of a single matrix. We suspect that with two matrices the situation might be already radically different with a huge associated entropy and we look forward to exploring this possiblity further. Of course, the picture of simple integrable fluids should break down. What would be the simplest toy model representing interesting features of the resulting modification when BHs are at play? Maybe some viscous fluid or a chaotic mixture of two fluids? Such modifications would more naturally exhibit the sort of chaos and universality we usually associate to black holes.

\section*{Acknowledgments} 
We thank J.~Abajian, A.~Abanov, P.~Anempodistov, F.~Aprile, D.~Gaiotto, A.~Guerrieri, S.~Komatsu, I.~Kostov, A.~Milekhin, S.~Pasterski and P.~Wiegmann for numerous enlightening discussions.  We thank Chat-GPT for help with one of the paragraphs of this paper. Research at Perimeter Institute is supported in part by 
the Government of Canada through the Department of Innovation, Science, and Economic Development Canada and by the Province of Ontario through the Ministry of Colleges
and Universities. 
PV is supported in part by Discovery Grants from 
the Natural Sciences and Engineering Research Council of Canada, 
and by the Simons Foundation through the ``Nonperturbative Bootstrap'' collaboration (488661). V.K. thanks
the Perimeter Institute for Theoretical Physics and the Simons Center for Geometry and
Physics, where a part of this work has been done.
This work was additionally supported by   
FAPESP Foundation through the grants 2016/01343-7, 2017/03303-1, 2020/16337-8. 

\section*{Appendices}
\appendix

\section{From 3-Matrix Model to Single-Matrix Model} \label{3M1MA}
In this appendix, we will derive the single matrix integral representation of the three point function (\ref{singleM}). The starting point is the three matrix integral which we reproduce from the main text,
\beq
Z_3(\mathcal{R}_1,\mathcal{R}_2,\mathcal{R}_3) = \int \prod_{i=1}^3 d M_i \, \chi_{R_i}(M_i) \, e^{-\frac N2\text{tr}\Big(\sum\limits_i M_i^2- 2\sum\limits_{i <j} M_i M_j  \Big) } 
\eeq
First, following the recipe of~\cite{Kazakov:1987qg} for the solution of Potts model on planar graphs,  let us introduce an auxiliary $N\times N$  Hermitian matrix $X$. The integral can be rewritten as,
\beqa
Z_3(\mathcal{R}_1,\mathcal{R}_2,\mathcal{R}_3) &=& \frac{\int d X\,\prod_{i=1}^3 d M_i \, \chi_{R_i}(M_i) \, e^{-\frac{N}2\text{tr}\Big(\left(X-\sum\limits_i M_i\right)^2+\sum\limits_i M_i^2-2 \sum\limits_{i <j} M_i M_j  \Big) }}{\int d X e^{-\frac N2 \tr X^2}}\nn \\
&=& \frac{\int d X\, e^{-\frac N2\tr X^2}\prod_{i=1}^3 \left(\int d M_i\, e^{-N\, \text{tr}(M_i^2-X M_i)} \chi_{R_i}(M_i)\right)}{\int d X e^{-\frac N2 \tr X^2}}\label{threeMatrixManipulation} 
\eeqa
Now, consider the $M_i$ integral in the numerator. Plugging in the definition of the character, we have
\beqa
\texttt{integral}&=&\int d M\, e^{-N \text{tr}(M^2-X M)}\ \frac{\det\limits_{i,j\le N} m_i^{h_j}}{\Delta(m)}\\
&=& \text{vol}\left(\frac{U(N)}{U(1)^N\times S_N}\right) \ \int \prod\limits_i d m_i\, dU\, e^{-N \text{tr}(M^2-X M)} \Delta(m)\det\limits_{i,j\le N} m_i^{h_j}\nn
\eeqa
where in the second line, we changed variables to the eigenvalues of the matrix $M$. The jacobian for this transformation gives us $d M=\text{vol}\left(\frac{U(N)}{U(1)^N\times S_N}\right)\prod_i dm_i\, dU\, \Delta(m)^2$ where $U\in U(N)/U(1)^N$ and $dU$ is normalized such that $\int dU=1$ (note that we need the coset to not overcount the diagonal unitaries. Similarly, we divide by the volume of $S_N$ to account for permutations of the eigenvalues $m_i$). Using the HCIZ formula (\ref{IZdef}) and that $\text{vol }U(N)=\frac{(2\pi)^{N(N+1)/2}}{G(N+1)}$, we have
\beq
\texttt{integral}= \frac1{N!}\left(\frac{2\pi}{N}\right)^{\frac{N(N-1)}2} \int \prod\limits_i d m_i\, e^{-N \text{tr}(M^2)} \frac{\det\limits_{i,j} e^{N m_i x_j}}{\Delta(x)}\det\limits_{i,j} m_i^{h_j}
\eeq
We have here an integral over a product of two determinants that are fully antisymmetric in $m_i$. Clearly,
\beqa
\texttt{integral}&=& \left(\frac{2\pi}{N}\right)^{\frac{N(N-1)}2}\frac{1}{\Delta(x)} \det_{i,j}\int dm\, e^{-N (m^2 - m x_i)}m^{h_j}
\eeqa
This integral can be evaluated in terms of Hermite polynomials using the following identity,
\beqa
\int dm\, e^{-N(m^2-mx)}m^h &=&  \frac{2i\sqrt\pi}{(2i\sqrt{N})^{h+1}} \times e^{\frac {x^2}4} H_h\left(i\frac{\sqrt{N}x}2\right)
\eeqa
So, we have 
\beq
\texttt{integral}=\underbrace{\frac{(2i\sqrt{\pi})^{N}}{(2i\sqrt{N})^{\sum_i (h_i+1)}}\left(\frac{i\pi}{\sqrt N}\right)^{\frac{N(N-1)}2}}_{\texttt{prefactor}(h)}\times\frac{e^{\frac N4\tr X^2}}{\Delta\left(\frac{i\sqrt{N} x}2\right)} \det_{i,j} H_{h_i}\left(\frac{i\sqrt N x_j}2\right)
\eeq
Plugging this expression back into (\ref{threeMatrixManipulation}) and rescaling the $X\rightarrow -i\sqrt{2} X$ in the numerator, we get
\begin{eqnarray*}
Z_3(\mathcal{R}_1,\mathcal{R}_2,\mathcal{R}_3)&=&  \left(\frac{N}\pi\right)^{N^2/2}\prod_{n=1}^3\texttt{prefactor}(h_n)\times \int d X\, e^{-\frac N2\tr X^2}\prod_{n=1}^3 \frac{\det\limits_{i,j} H_{h_{n,i}}\left(\sqrt{\frac{N}2 }x_j\right)}{\Delta\left(\sqrt{\frac N2}x\right)}
\end{eqnarray*}
As explained below eq.\,(\ref{Ndef}), the prefactors multiplying the integral will drop after normalization. Therefore, we obtain the following equivalent expression which gives the same normalized three point function,
\beq
Z(\mathcal{R}_1,\mathcal{R}_2,\mathcal{R}_3)=\int d\mu(x)\, Q_1(x)Q_2(x)Q_3(x)
\eeq
where $d\mu(x)$ and $Q_i(x)$ are given by (\ref{singleM}) and (\ref{Qdef}) respectively. 
\section{A Simple Four Matrix Model} \label{4intA}
In this appendix, we will derive the four matrix integral form of the three point function~(\ref{fourInts}). Let us start with the one matrix integral,
\beqa
Z(\mathcal{R}_1,\mathcal{R}_2,\mathcal{R}_3) &=& \int d\mu(x)\, Q_1(x)Q_2(x)Q_3(x)\nn
\eeqa
Now, in the Slater determinants $Q_n(x)$, let us plug in the generating function (\ref{hermiteGenFunc}). 
\beqa
    Z(\mathcal{R}_1,\mathcal{R}_2,\mathcal{R}_3) &=& \prod_{n,i} \Gamma(h_{n,i}+1)\int d\mu(x) \prod_{n=1}^3 \frac{\det\limits_{i,j} \oint \frac{dm}{2\pi i m} m^{-h_{n,i}} e^{-\frac N2 m^2 + Nm x_j}}{\Delta(\hat x)}
\eeqa
Dropping the prefactors, which do not contribute after normalization, we have
\beqa
    Z(\mathcal{R}_1,\mathcal{R}_2,\mathcal{R}_3) &=& \int d\mu(x)\ \prod_{n=1}^3\oint \frac{d\mu(m_n)}{\Delta(m_n)^2}\frac{\det_{i,j} e^{N m_{n,i} x_j}}{\Delta(x)} \det_{i,j} e^{-\log m_{n,j}(h_{n,i}+1)}
\eeqa
Finally, multiplying and diving by appropriate Vandermonde determinants, we get the four matrix representation that we are after, 
\beqa
    Z(\mathcal{R}_1,\mathcal{R}_2,\mathcal{R}_3) &=& \int d\mu(x) \prod_{n=1}^3\oint d\mu(m_n)\frac{\Delta(-\log m_n)}{\Delta(m_n)}\Delta(h_n) I(x,m_n)I(-\log m_n,h_n)\nn
\eeqa
For the two point function, using the same manipulations, we arrive at
\beq
    \frac{Z(\mathcal{R}_1,\mathcal{R}_2,\phi)}{\prod_{n,i} \Gamma(h_{n,i}+1)} = \int d\mu(x) \prod_{n=1}^2 \oint d\mu(m_n) \frac{\Delta(-\log m_n)}{\Delta(m_n)}\Delta(h_n) I(x,m_n)I(-\log m_n,h_n)\la{twoPtWithX}
\eeq
Now, recall that we can compose two HCIZ integrals as,
\beq
    \int d\mu(x) I(x,m_1)I(x,m_2) = \int dU\ e^{\frac N2 \text{tr}(M_1+U M_2 U^\dagger)^2}
\eeq
Integrating over $x$ in (\ref{twoPtWithX}) using this composition of HCIZs, we obtain
\beq
    \frac{Z(\mathcal{R}_1,\mathcal{R}_2,\phi)}{\prod_{n,i} \Gamma(h_{n,i}+1)} = \oint \prod_{n=1}^2\left[\prod_i dm_{n,i}\ \Delta(m_n)\Delta(-\log m_n)\Delta(h_n) I(-\log m_n, h)\right] I(m_1,m_2)  
\eeq
\section{Derivation of Slater Determinant Formulae} \la{slaterAp}
In this appendix, we will prove the formulae for Slater determinants of rectangles and trapezia (\ref{slaterRect},\ref{slaterTrapz}) used in the main text. 

\subsection{Rectangles}
Consider first the case of rectangles for which the result is well known \cite{Brezin_2000,Morozov:1994hh,Kimura:2021hph}. We rederive it here for completeness. We would like to show that
\beq  \mathcal{Q}^{\texttt{rectangle}}_K(x) = \frac{\det\limits_{i,j} H_{K+i-1}(x_j)}{\Delta(x)} \propto \int d\nu(y) \prod_{\substack{i=1\ldots N\\j=1\ldots K}}(x_i-y_j)
\eeq
where we introduced for compactness a new measure,
\beq
    \nu(y) = \left(\prod_{i=1}^K dy_i\ e^{-\frac N2 y_i^2}\right) \Delta_K(y)^2
\eeq
We will use orthogonality of two points functions (\ref{exact2pt}) to show this. For some arbitrary YT weights $\{\tilde h_i\}$, consider
\beqa
    \mathcal{I} &=& \int \prod_i dx_i\ e^{-x_i^2}\Delta(x)\mathcal{Q}^{\texttt{rectangle}}_K(x) \det_{i,j} H_{\tilde h_i}(x_j)\\
    &=& \int dx\, dy\ e^{-\sum_i(x_i^2+y_i^2)} \Delta(x)\Delta(y)^2\prod_{i,j}(x_i-y_j) \det_{i,j} H_{\tilde h_i}(x_j)
\eeqa
Defining $z = (x_1,\ldots,x_N,y_1,\ldots,y_K)$, we have
\beqa
    \mathcal{I} &=& \int dz\ e^{-\sum_i z_i^2} \Delta(z) \Bigl(\Delta(y)\det_{i,j} H_{\tilde h_i}(x_j)\Bigr)
\eeqa
Recall that the Vandermonde $\Delta(z) \propto \det\limits_{1<i,j<N+K} H_{i-1}(z_j)$. After antisymmetrizing the $x$'s and $y$'s, the term $\Bigl(\Delta(y) \det_{i,j} H_{\tilde h_i}(x_j)\Bigr)$ is also a wavefunction for for $N+K$ fermions in a harmonic potential. The first $K$ levels are fully filled and the rest of the fermions are in levels $\tilde h_i$. Therefore, by orthogonality, the integral would vanish unless $\tilde h_i = K+i-1$. Keeping track of proportionality constants, we obtain the following result,
\beq
    \mathcal{Q}^{\texttt{rectangle}}_K(x) = \frac{\det\limits_{i,j} H_{K+i-1}(x_j)}{\Delta(x)} = \frac{2^{\frac{1}{2} N (2 K+N-1)}}{\int d\mu(y)}\int d\mu(y) \prod_{\substack{i=1\ldots N\\j=1\ldots K}}(x_i-y_j)
\eeq
\subsection{Trapezia}
Now, let us prove the Slater determinant formula for trapezia,
\beq
    \mathcal{Q}^{\texttt{trapezium}}_K(x)=\frac{\det\limits_{i,j} H_{2(K+i-1)}(x_j)}{\Delta(x)} \propto \frac{\Delta(x^2)}{\Delta(x)} \int d\nu(y)\frac{\Delta(y^2)^2}{\Delta(y)^2}\prod_{\substack{i=1\ldots N\\j=1\ldots K}} (x_i^2-y_j^2)
\eeq
For $K=0$, the YT reduces to a triangle. In this case, the determinant of Hermites is fixed by antisymmetry and matching the degree. We get,
\beq
    \det_{i,j} H_{2(i-1)}(x_j) = 2^{N(N-1)}\Delta(x^2)\la{hermiteEven}
\eeq
Dividing by $\Delta(x)$, we get the Slater determinant formula for triangles. For general $K$, for some YT weights $\{\tilde h_i\}$, consider the integral
\beqa
    \mathcal{I} &=& \int \prod_i dx_i\ e^{-x_i^2}\Delta(x)\mathcal{Q}^{\texttt{trapezium}}_K(x) \det_{i,j} H_{\tilde h_i}(x_j)\\
    &=& \int dx\, dy\ e^{-\sum_i(x_i^2+y_i^2)} \Delta(x^2)\Delta(y^2)^2\prod_{i,j}(x_i^2-y_j^2) \det_{i,j} H_{\tilde h_i}(x_j)
\eeqa
Once again, defining $z = (x_1,\ldots,x_N,y_1,\ldots,y_K)$, we have
\beqa
    \mathcal{I} &=& \int dz\ e^{-\sum_i z_i^2} \Delta(z^2) \Bigl(\Delta(y^2)\det_{i,j} H_{\tilde h_i}(x_j)\Bigr)
\eeqa
Using the formula (\ref{hermiteEven}), we see that $\Delta(z^2)$ is the wavefunction for $N+K$ fermions filling the lowest lying even levels. The term in the parenthesis, $\Bigl(\Delta(y^2) \det_{i,j} H_{\tilde h_i}(x_j)\Bigr)$ is a wavefunction with first $K$ even levels filled and with the other $N$ fermions in levels $\tilde h_i$. Therefore, by orthogonality we require $\tilde h_i = 2(K+i-1)$. Taking care with the prefactors, we obtain the following result,
\beqa
   \mathcal{Q}^{\texttt{trapezium}}_K(x)&=&\frac{\det\limits_{i,j} H_{2(K+i-1)}(x_j)}{\Delta(x)}\\
   &=&\frac{2^{N(2K+N-1)}}{\int d\mu(y) \Delta(y^2)^2\Delta(y)^{-2}} \frac{\Delta(x^2)}{\Delta(x)} \int d\nu(y) \frac{\Delta(y^2)^2}{\Delta(y)^2}\prod_{\substack{i=1\ldots N\\j=1\ldots K}} (x_i^2-y_j^2)\nn
\eeqa

\section{Analytic Solution for Three Triangles}\la{triangleAppendix}
The SPE for the resolvent $G(x)$ is given by the following Riemann Hilbert problem for $n=-3$,
\begin{eqnarray}
    2\slashed{G}(x) + n G(-x) = x\la{SPETriangle}
\end{eqnarray}
Numerics with the discrete SPEs tell us that the dominant saddle has a single cut on the first sheet. The SPE (\ref{SPETriangle}) then tells us that if we tunnel through the cut on the first sheet, we see another cut on the second sheet that is reflected around the origin. Going through this second cut, we see a third cut and so on. 

In order to solve the SPE, we will uniformize this infinite sheeted Riemann surface. We do this using elliptic functions\footnote{We use \texttt{Mathematica}'s conventions. In particular, $sn(u,k) = \texttt{JacobiSN}[u,k^2]$, $K=\texttt{EllipticK}[k^2]$ and $K' = \texttt{EllipticK}[1-k^2]$}, following Eynard and Kristjansen's solution of the $O(n)$ model on random graphs\cite{Eynard:1995zv} (see also~\cite{Kostov:2006ry}),
\beqa
    x(u) &=& a\ sn(u,k)\\
    k&=& \frac ab
\eeqa
where $a$ and $b$ are the square root branch points of the resolvent in the $x$ variable. This maps the cut $x$ plane into the rectangle $[-K,K]\times[-i K',iK']$ as shown in figure \ref{ellipticMap}. We have,
\begin{itemize}
    \item $x(K)=a$ and $x(-K)=-a$
    \item $x(K+i K')=x(K-iK')=b$ and $x(-K-iK')=x(-K+iK')=-b$
    \item The segment $(a\pm i\epsilon,b\pm i\eps)$ is mapped to $(K,K\pm i K')$
\end{itemize}
\begin{figure}
    \centering
    \includegraphics[width=0.8\textwidth]{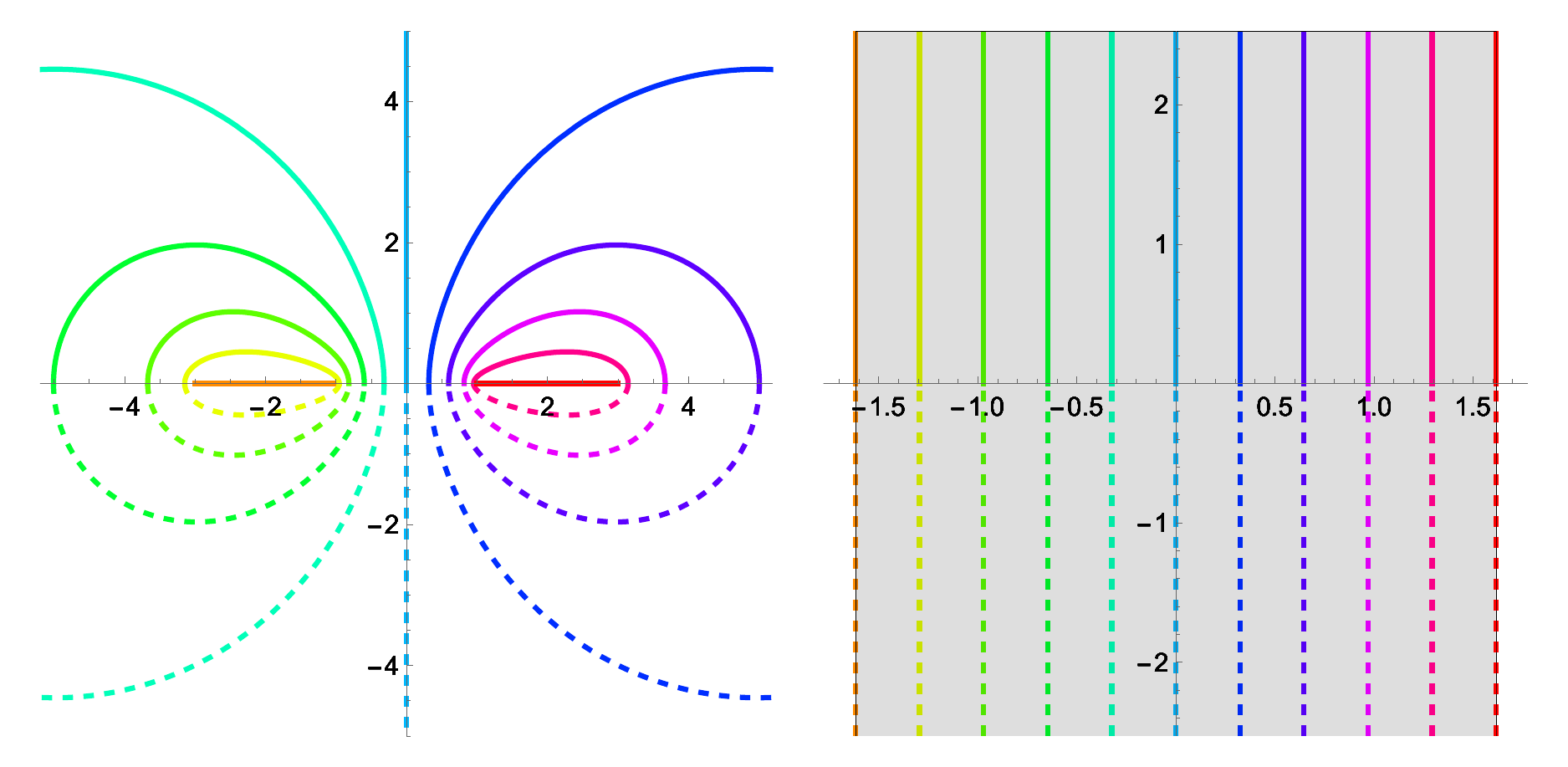}
    \caption{A depiction of the map $x=a\ sn(u,k^2)$. On the left, we have the $x$ variable and on the right, the shaded region is the fundamental rectangle $[-K,K]\times [-K',K']$. The right edge of the rectangle is mapped to the right cut in $x$ and left edge to left cut}
    \label{ellipticMap}
\end{figure}
Now, we can split $G(x)=g(x)+H(x)$ into the regular part ($g(x) = \frac{x}{2-n}$) and non-analytic part, $H(x)$ which satisfies a homogenous SPE,
\begin{eqnarray}
    2\slashed{H}(x)+n H(-x) = 0\nn
\end{eqnarray}
and has the large $x$ asymptotics,
\beq
H(x) \sim \frac{x}{n-2} + \frac 1x + \ldots\la{triangleHAsymptotics}
\eeq
In the $u$ variable, with slight abuse of notation, the SPE reads,
\begin{eqnarray}
    H(K+u)+H(K-u)+nH(-K+u)&=&0\,, \qquad u\in [-iK',iK']
\end{eqnarray}
Similarly, given that there is no cut for $x\in [-b,-a]$, we have,
\begin{eqnarray}
    H(-K+u)&=&H(-K-u)\,, \qquad u\in [-iK',iK']
\end{eqnarray}
The previous two equations together with analyticity of $H(u)$ imply
\begin{eqnarray}
    H(u+2K)+H(u-2K)+nH(u)&=&0\label{homoSPE}\\
    H(-u-2K)&=&H(u)\label{cond1}\\
    H(u+2iK')&=&H(u)\label{cond2}
\end{eqnarray}

Following Eynard-Kristjansen let us define two new functions $H_\pm(u)$ such that,
\begin{eqnarray}
    H(u)&=&e^{-\frac12 i(1-\nu)\pi} H_+(u) + e^{\frac12 i(1-\nu)\pi} H_-(u)
\end{eqnarray}
where $\nu =\frac1\pi\cos^{-1}(\frac n2)$. We require that these functions obey the following properties,
\begin{eqnarray}
    H_{\pm}(u+2K)&=&e^{\mp  i(1-\nu)\pi} H_{\pm}(u)\nn\\
    H_\pm(u+2iK')&=&H_\pm(u)\label{HPcond}\\
    H_-(u)&=&H_+(-u)\nn
\end{eqnarray}
We can check easily that if $H_\pm(u)$ has the above properties, the conditions (\ref{homoSPE}-\ref{cond2}) are automatically satisfied. Finally, in order for $H(u)$ to have the right asymptotics (\ref{triangleHAsymptotics}), we need $H_\pm(u)$ to have a simple pole at $u=iK'$ (i.e. at $x=\infty$). 

To write down the analytic expression for the resolvent, our basic building block will be the first elliptic theta function $\theta(z)$ ($=\texttt{EllipticTheta}[1,\pi z, \texttt{EllipticNomeQ}[k^2]]$ in \texttt{Mathematica}). It has a single simple zero in the fundamental domain $[-K,K]\times[-iK',iK']$ at $z=0$. Recall that it has the quasi-periodicity properties,
\begin{eqnarray}
    \theta\left(z + i\frac{K'}{K}\right) &=& -\theta(z) e^{-i\pi\left(2z+i\frac{K'}{K}\right)}\\
    \theta(z+1)&=&-\theta(z)
\end{eqnarray}
The function $H_+(u)$ which obeys (\ref{HPcond}) and has a single pole at $x=\infty$ is unique upto an overall normalization factor,
\begin{eqnarray}
    H_+(u) &=& \mathcal{N}\ \frac{\theta(\frac{u-i (2-\nu)K'}{2K})}{\theta(\frac{u-iK'}{2K})} e^{-i\pi(1-\nu)\frac{u}{2K}}
\end{eqnarray}
The resolvent is given by,
\begin{eqnarray}
    G(u) = \frac{a\ sn(u,k)}{5}+\mathcal{N}\left(\frac{e^{-\frac{i \pi(1-\nu)}2 \left(\frac{u}{K}+1\right)} \theta \left(\frac{u-i(2-\nu )K'}{2 K}\right)}{\theta \left(\frac{u-i K'}{2 K}\right)}+\frac{e^{\frac{i \pi(1-\nu)}2 \left(\frac{u}{K}+1\right)} \theta \left(\frac{u+i(2-\nu )K'}{2 K}\right)}{\theta \left(\frac{u+i K'}{2 K}\right)}\right)
\end{eqnarray}
There are three unknown parameters in the above solution: $a, k\text{ and } \mathcal{N}$. These are fixed by requiring that the resolvent has the asymptotics $G(x)\sim 0\, x^1 + 0\, x^0 + 1\, x^{-1} + \mathcal{O}(x^{-2})$. For $n=-3$, which is the case we are interested in, we have 
\beqa
    a&\approx& 0.0271840\\
    k&\approx& 0.0089906\\
    \mathcal{N} &\approx& 0.1852505 + 0.6047044\, i
\eeqa

\section{Simplifying the SPE for Rectangles and Trapezia}\label{RectangleTrapeziaSPEap}
In this appendix, we will cast the SPEs for rectangles and trapezia in a simple form. We explain that they cannot be solved using the same methods as the triangle case.
\subsection{SPE for Rectangles}
Let us first consider the case of three rectangles. The SPEs (\ref{SPr1},\ref{SPr2}) in the continuum limit become
\beqa
    -y + 2\gamma \dashint dz\,\frac{\sigma(z)}{y-z} + \int dz\, \frac{\rho(z)}{y-z} &=& 0\\
    -x + 2\dashint dz\, \frac{\rho(z)}{x-z} + 3\gamma \int dz\, \frac{\sigma(z)}{x-z} &=& 0
\eeqa
where $\rho(z)$ and $\sigma(z)$ are the densities of the $x$ and $y$ eigenvalues respectively. We also introduced a new parameter $\gamma = \frac KN$ which is kept finite in the large $N$ limit. Consider first the SPE for $y$. Let $g(y)$ and $G(x)$ be the resolvent for the $y$'s and $x$'s respectively. We have,
\beqa
    g(y) &=& -\frac1{2\pi i\gamma} \int_\alpha^\beta dz\, \frac{z-G(z)}{y-z} \sqrt{\frac{(y-\alpha)(y-\beta)}{(z-\alpha)(z-\beta)}}\\
    &=&\frac1{2\gamma} (y-\sqrt{(y-\alpha)(y-\beta)})+\frac1{2\gamma}\int dz \frac{\rho(z)}{y-z}\left(-1+\sqrt{\frac{(y-\alpha)(y-\beta)}{(z-\alpha)(z-\beta)}}\right)
\eeqa
As seen in the section \ref{SlaterSec}, the dominant saddle for the rectangles case has two cuts in $G(x)$ and one cut in $g(y)$, say in the interval $(\alpha,\beta)$. Now, lets introduce a new variable $z$,
\begin{eqnarray}
    y(z)&=& \frac{\alpha+\beta}2 + \frac{\alpha-\beta}4 \left(z+\frac1z\right)\\
    g(y(z))&=& \frac{\alpha+\beta}{4\gamma}+\frac{\alpha-\beta}{4\gamma z} -\frac{F(-1)}{\gamma(\alpha-\beta)}\frac{z-1}{z+1}-\frac{F(1)}{\gamma(\alpha-\beta)} \frac{z+1}{z-1}+\frac{4 F\left(\frac{1}{z}\right)}{\gamma(z-1) (z+1) (\alpha -\beta)}\nn
\end{eqnarray}
where $F(z)=\int dt \frac{r(t)}{z-t}$ and $r(z)dz=\rho(x)dx$. So, for $z\gg 1$, $F(z)\approx \frac1z$. 
Now, imposing the asymptotics $g(m)\sim \frac1m$, we get the conditions
\begin{eqnarray}
    F(1)&=&\frac{\beta -\alpha }{4 \gamma }+\frac{1}{16} (\alpha -\beta ) (3 \alpha +\beta )\\
    F(-1)&=&\frac{\alpha-\beta}{4 \gamma }+\frac{1}{16} (\alpha -\beta )(\alpha +3 \beta )
\end{eqnarray}
Now, the SPE for $X$ is 
\begin{eqnarray}
    -x + 2\dashint dy \frac{\rho(y)}{x-y} + \frac3\gamma g(x)=0
\end{eqnarray}
Lets plug in $g$ and change variables as before. After substituting $F(\pm 1)$, we get
\begin{eqnarray}
    z\ ReF(z)+ \left(\frac{3F\left(\frac1z\right)}{2z} - \frac{\Re F\left(\frac1z\right)}{z}\right)=\frac{1}{16} \left(\frac{z^2}{2}+\frac{1}{z^2}\right) (\alpha -\beta )^2+\frac{1}{8} \left(\frac{z}{2}+\frac{1}{z}\right) \left(\alpha ^2-\beta ^2\right)-\frac{3(\alpha -\beta )}{8 \gamma }\nn
\end{eqnarray}
Again from the numerics in figure \ref{manySPS}, we see that the cuts in $g$ and $G$ touch but do not overlap. This means that in the $z$ variable the $X$ cut is in the region $|z|\ge1$, such that $\Re F(1/z)=F(1/z)$ for $z$ on the cut. To obtain a homogenous equation lets define a new function 
\begin{eqnarray}
    f(z)&=& z F(z) - \frac{(\alpha-\beta)^2}{16 z^2} - \frac{\alpha^2-\beta^2}{8z}+\frac{1}{4\gamma}
\end{eqnarray}

It has the following asymptotics for $z\gg 1$,
\begin{eqnarray}
    f(z) \sim 1+\frac{1}{4\gamma} + \left(\frac{\beta^2-\alpha^2}{8z} + \int dt\ r(t) \frac{t}z \right) + ...
\end{eqnarray}
Also, plugging in $z=\pm 1$, we see that $f(\pm 1)=0$. The SPE in terms of $f(z)$ is simply,
\begin{eqnarray}
    \Re f(z) + \frac12 f\left(\frac1z\right)=0
\end{eqnarray}
This equation is very similar to the case of triangles -- in fact if we change variables to $w=\frac{z+1}{z-1}$, we get the $O(N)$ model SPE, $\Re f(z(w))+\frac12 f(z(-w))=0$. 

However, we did not yet manage to solve this simple looking equation. The reason being overlapping branch points. Recall from figure \ref{manySPS} that for both the one- and two-cut solutions the $x$ and $y$ cuts touch. In terms of $f(z)$, this implies that the cut in the first sheet is at $z\in(1,z^*)$ and it touches the reflected cut $z\in(\frac1{z^*},1)$ for some $z^*$. This prevents us from naively using the elliptic parametrization $w(u)= A\ sn(u,k)$ because we have $A=0$.

\subsection{SPE for Trapezia}
For three trapezia, the SPEs in the continuum limit are
\beqa
    -y + 2 y\left(2\gamma \dashint dz\, \frac{\sigma(z)}{y^2-z^2} + \int dz\, \frac{ \rho(z)}{y^2-z^2} \right)&=& 0\\
    -x + \dashint dz\, \rho(z)\left(\frac{2}{x-z}+\frac{3}{x+z}\right) +3\gamma \int dz\, \frac{2x \sigma(z)}{x^2-z^2} &=& 0
\eeqa
As before, $\rho(z)$ and $\sigma(z)$ are densities of $x$ and $y$ eigenvalues, and $\gamma = \frac KN$. Following the same steps as the rectangles case, for $g(y)\equiv \int dz \frac{\sigma(z)}{y^2-z^2}$, we have
\beqa
    g(y(z))&=& -\frac{1}{4 \gamma} + \frac{2z}{\gamma\alpha^2(z^4-1)} \left(F\left(-\frac1z\right)-F\left(\frac1z\right)\right) +\frac{z^2+1}{2\gamma\alpha^2(z^2-1)}(F(1)-F(-1))\nn\\
    &~&\quad +\ i\frac{z^2-1}{2\gamma\alpha ^2 \left(z^2+1\right)}\left(F(-i)-F(i)\right)
\eeqa
where we used the map $y(z)=-\frac\alpha2\left(z+\frac1z\right)$ (we choose a symmetric cut for $g(y)$ between $(-\alpha,\alpha)$ because that's what the numerics indicate) and defined $F(z)\equiv \int dt \frac{r(t)}{z-t}$, with $r(z)dz = \rho(x)dx$. From the asymptotics $g(y)\sim \frac1{y^2}$, we get 
\beqa
    F(i)&=&F(-i)-\frac{1}{4} i \alpha ^2 (2 \gamma -1)\\
    F(1)&=&F(-1)+\frac{1}{4} \alpha ^2 (2 \gamma +1)
\eeqa
Plugging the resolvent of $y$ into the SPE for $x$, 
\beqa
   z^2 \text{ Re }F(z)-\frac{3}{2} z^2 F(-z)-\frac{5}{2} F\left(\frac{1}{z}\right)+3 F\left(-\frac{1}{z}\right) = -\frac{\alpha ^2 \left(z^4-6 \gamma  z^2-4\right)}{4 z}
\eeqa
To simplify further, let us define
\beq
    f(z) \equiv z^2 F(z)+\frac{13 \alpha ^2 z^3}{64}+\frac{1}{2} \alpha ^2 \gamma  z+\frac{3 \alpha ^2}{64 z}
\eeq
Then, we the SPE becomes the following homogenous equation,
\beq
    \text{Re}f(z) -\frac{3 f(-z)}{2} + 3z^2 f\left(-\frac{1}{z}\right)-\frac{5z^2}{2} f\left(\frac{1}{z}\right)=0
\eeq
This SPE is even more complicated than the rectangle case because now we can have more complicated reflected cuts from $f(1/z)$ like before, but also from $f(-z)$ and $f(-1/z)$. So once again, we cannot uniformize this Riemann surface with elliptic functions.
\section{Inversions and Double Inversions for Continuum Flows} \label{doubleInversionAp}
\subsection{Inversion for a Single Chamber} \la{directAp1}

Consider first the solution for the initial value problem (IVP). In this case, we are given $f(x,0)=F(x)$. Then $f(x,t)$ at any future time is simply given parametrically by 
\beq
\{x, f(x,t) \}_\text{fixed $t$} = \{X,Y\} \,, \qquad \left\{ \begin{array}{l}
X=u+t F(u)\\
\\
Y=F(u)
\end{array}
\right. 
\label{parametric}
\eeq
To check that this representation solves the flow (\ref{fluidEOM}), note that we have $u=X-t F(u)$ and so,
\beqa
    \frac{\partial f(X,t)}{\partial t} &=& F'(u) \frac{\partial u}{\partial t} = -F(u) F'(u)\\
    f(X,t)\frac{\partial f(X,t)}{\partial X}&=&F(u)F'(u)\frac{\partial u}{\partial X} = F(u)F'(u)
\eeqa
which clearly add up to zero. Another useful way to write the solution is in the following implicit form\footnote{Obtained by noting that $f(x,t) = F(u) = F(x-t F(u)) = F(x-t f(x,t))$}, 
\beq
f(x,t)=F(x-t f(x,t)) \,. \la{inversion}
\eeq

In practice, one is often interested in the flow at final time $t=1$. For this, let us define two function $G_\pm(x)$ as
\beq
    G_+(x)=x+f(x,t=0) \,, \qquad G_-(x)=x-f(x,t=1)
\eeq
Then, using (\ref{inversion}), one can easily check that these functions are inverses of each other, 
\beq
    G_+(G_-(x)) = G_-(G_+(x)) = x\la{GpGmInverses}
\eeq

Now, let us briefly address the boundary value problems (BVP). Here, we specify initial and final densities, without specifying the initial or final velocities. So, we know the  imaginary parts of $G_\pm(x)$ since we know $\text{Im}(G_+(x))=\pi \rho(x,0)$  and $\text{Im}(G_-(x))=-\pi \rho(x,1)$. Solving the flow equations reduces to a Riemann-Hilbert problem of finding the functions $G_\pm$ with prescribed imaginary parts knowing that they are inverses of each other.

Note that if we can find $G_+$ or $G_-$, we can then proceed to use (\ref{inversion}) to solve the full flow\footnote{With the caveat that the physical flow is sometimes obscured by the presence of velocity cuts, see discussion below (\ref{characterFinalFlow})}. Solving this Riemann-Hilbert problem is hard which is why the BVP is much harder than the IVP, whose solution is totally explicit. It would be great to develop new strategies to directly solve the BVP, perhaps using the integrable charges we found (\ref{newCharges}).

\subsection{The Double Inversion}

Our three point function problem in (\ref{action3}), (\ref{xVelGlue}) and (\ref{mVelGlue}) is clearly a BVP, which makes it hard. It also makes it interesting! 

Note that our fluid starts at the ``middle" with some density $\rho(x)$ and flows out to three legs ending at prescribed densities $\eta_n(h)$ with $n=1,2,3$ as represented in figure \ref{cChambers}. 

\begin{figure}[t]
\begin{center}
\includegraphics[scale=0.6]{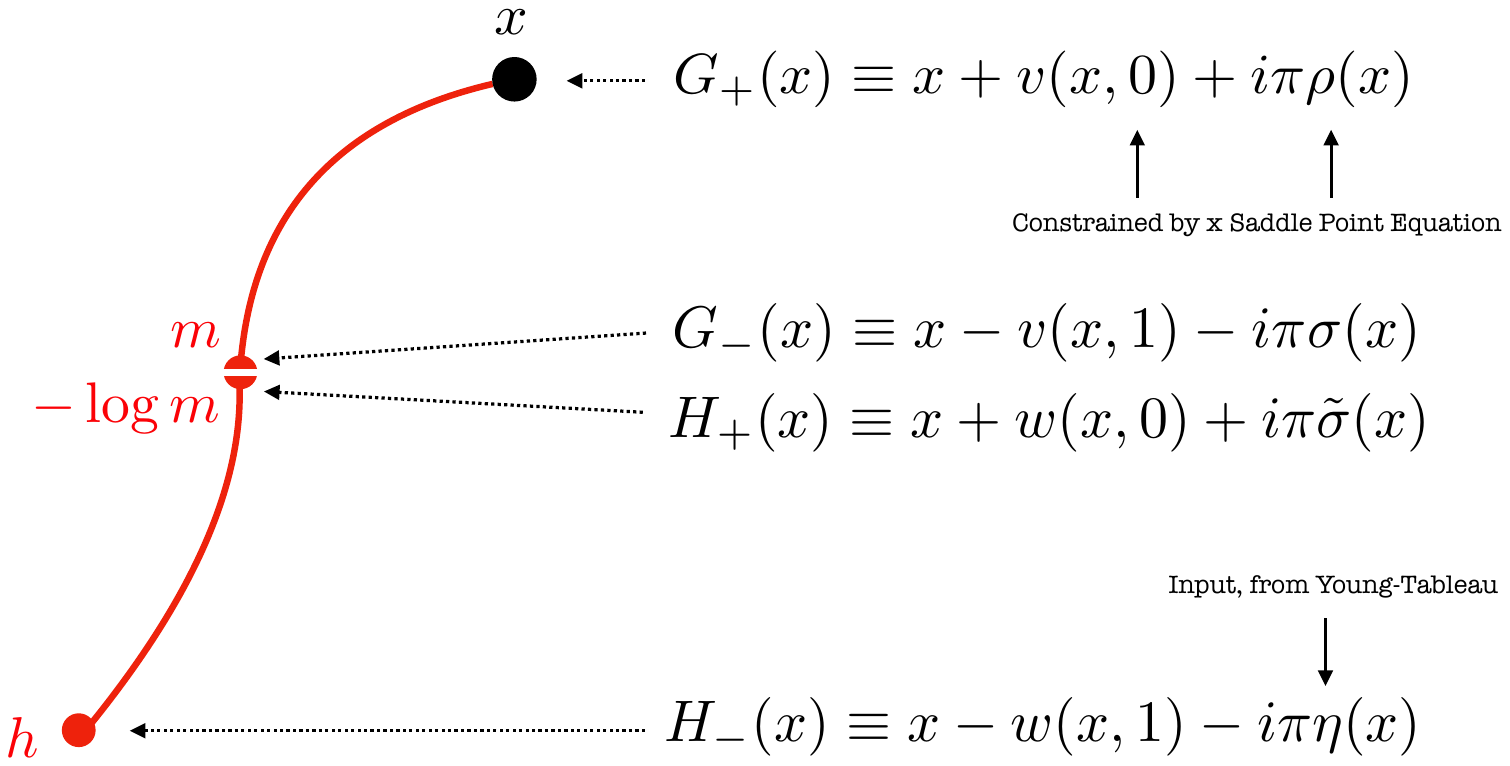}
\end{center}
\vspace{-5cm}
\caption{Each leg of figure \ref{cChambers} is broken into two: We flow from $x$ at the middle junction to $m$ -- which we take the lograrithm of to get $-\log m$ -- and then from there to the $h$ at the end. In flow language we have two flows, each from $t\in[0,1]$.  }  \la{cLeg}
\end{figure}

We use $G_\pm$ and $H_\pm$ to describe the flow at the various important points as indicated in the figure \ref{cLeg}. (Note that we dropped the index $n=1,2,3$ for the leg since we are focusing on each leg separately at the time.) These functions describe the flow at $t=0,1$ in each of the two segments making the full leg. 

We could want to \textit{integrate} out the intermediate functions $H_+$ and $G_-$ which describe the fluid at the intermediate junction to get a relation directly between $H_-$ (describing the $h$ variables) and $G_+$ (describing the $x$ variables). To do this, note that $H_+$ is related to $G_-$ as,
\begin{align}
    G_-(x)-x&=-v(x,1)-i\pi\sigma(x) &&\text{, definition of $G_-$}\nn\\
    &=\frac1x(-\log x+w(-\log x,0)+i\pi\tilde\sigma(-\log x))&&\text{, $\tilde \sigma(-\log x)= -x\sigma(x)$ \& eq.~(\ref{mVelGlue})}\nn\\
    &=\frac1x H_+(-\log x)&& \text{, definition of $H_+$}
\end{align}

and thus we arrive at the key junction equation
\beq
H_+(x)=e^{-x}(G_-(e^{-x})-e^{-x})
\eeq
Now that we have this we simply use that $G_+$ and $H_-$ are the inverses of $G_-$ and $H_+$ respectively to massage this equation. For example, we can apply $H_-$ to both sides of this equation and then replace $x$ by $x=-\log G_+(X)$. In this way we get what we call the \textit{double inversion formula} 
\beq
-\log G_+(X)=H_-\Bigl(G_+(X)(X-G_+(X))\Bigr) \,. \label{doubleInv}
\eeq
We do not know $H_-$ and we do not know $G_+$ but we do know the imaginary part of $H_-$ (it is the density of the Young-Tableau) and we similarly have a constraint relating real and imaginary parts of $G_+$ since the density at $x$ must obey the $x$ saddle point (\ref{xVelGlue}). So the Riemann Hilbert problem for this double flow is to find full functions $H_-$ and $G_+$ for which we have this partial information about each of them together with the inverse relation (\ref{doubleInv}). Would be terrific to develop a analytic, or numerical, method for solving this problem in full generality given arbitrary Young Tableaux densities. We hope to return to this problem in the future. 

Finally let $A$ and $B$ indicate the upper and lower branches of the leg in figure \ref{cLeg}. In other words, $A$ is the flow from $x$ to $m$ and $B$ is the flow from $-\log m$ to $h$. Then we can write the parametric solutions to $f(x,t)$ in both legs as,
\beqa
\{x, f_A(x,t)\} &=& \{u+t (G_+(u)-u), G_+(u)-u\} \,,  \nn\\
\{x, f_B(x,t)\} &=& \{u+t f_B(u,0), f_B(u,0)\}\label{fullF}\\
&=&\{(t-1)\log G_+(z) - G_+(z)(G_+(z)-z), \log G_+(z)-G_+(z)(G_+(z)-z)\}\nn
\eeqa
where in the last line, we used the gluing condition to write $f_B$ at $t=0$ in terms of $f_A$ at $t=0$ and then changed variables to $u=-\log G_+(z)$. Note that at $t=1$ in the second chamber, we recover the double inversion formula (\ref{doubleInv}).

\section{An Application of Flows -- Characters}\la{characterExample}
Here consider the evaluation of characters as an application of the technology developed in this paper. Albeit a simple toy model compared to the structure constant problem, the character problem allows us to formulate a general strategy for solving the BVP problem which, as we commented on earlier, is considerably more complicated than the IVP. The character of a Young tableau $\mathcal{R}$ is given by,
\beq
\chi_{\mathcal{R}}(\Phi)= \frac{\det\limits_{i,j\le N} \phi_j^{h_i}}{\Delta(\phi)} \,. \la{chiExpression}
\eeq
That is,
\beq
\chi_{\mathcal{R}}(\Phi)=\frac{N^{N(N-1)/2}}{G(N+1)}\times\frac{\Delta(h)\Delta(\log \phi)}{\Delta(\phi)}I(\log \phi,h) \,.\la{charAsHCIZ}
\eeq
We will now follow a simple strategy for evaluating these characters. We start from some simple examples for which the large $N$ limit of (\ref{chiExpression}) can be computed analytically without even using fluids. We then recast these examples in fluid language, as in \cite{Gross:1994ub}. Then, using these simple examples as seeds we can adiabatically deform them from the initial Young-Tableaux encoded by the seed $h_j$ to any desired YT. This deformation is done through a nice set of differential equations which exploit the underlying fluid integrability. 

This strategy will be followed pretty much verbatim for the three point correlators. The main difference is that there we have several flows in several chambers while here we deal with a single flow. This is thus a great laboratory to hone this simple strategy on a much simpler setup. 

\subsection{Two Simple Examples Solvable With or Without Fluids}
In this section we want to find some relevant flows for this problem for two special cases when~(\ref{chiExpression}) can be independently computed through  much more straightforward means, namely for 
\beq
h_j=K + n(j-1) \qquad \text{where} \qquad n=1,2 \text{ and } K\in \mathbb{Z}_+\,.\la{YTrectTrapz}
\eeq
For $n=1$, this corresponds to a rectangle of size $K\times N$ and $n=2$ is a trapezium. The characters for these cases are simply,
\beqa
\chi_{\mathcal{R}}(\Phi) = \prod_i \phi_i^K \times \left\{ 
\begin{array}{ll}
    1&, n=1\\ 
    \\
    \displaystyle\frac{\Delta(\phi^2)}{\Delta(\phi)} = \prod\limits_{i<j}(\phi_i+\phi_j) &,n=2
    \end{array}
\right.\la{charDirect}
\eeqa
Together with (\ref{charAsHCIZ}), this implies

\begin{multline}
    \lim_{N\to \infty}\frac{\log I(\log \phi,h)}{N^2} = -\frac34 -\frac{\log \Delta(h)}{N^2} + \frac KN \int d\phi \sigma(\phi) \log\phi\\
    +\frac12\int d\phi d\phi' \sigma(\phi)\sigma(\phi')\left(\log|\phi^n-\phi'^n|-\log|\log\phi - \log\phi'|\right)
\end{multline}
Equating this to Matytsin's large $N$ limit of HCIZ (\ref{matytsinAction}), we get
\begin{equation*}
S_{\texttt{fluid}}[\tilde\sigma,\eta]+\frac12 \int dx(\tilde\sigma(x)+\eta(x))x^2 = \frac KN \int d\phi \sigma(\phi) \log\phi + \frac12 \int d\phi d\phi' \sigma(\phi)\sigma(\phi')\log|\phi^n-\phi'^n|
\end{equation*}
where $\tilde\sigma(\log \phi)=\phi\sigma(\phi)$ is the density of $\log \phi$ and $\eta$ is the Young tableau density. Varying this equation with respect to $\tilde\sigma$ and using (\ref{velVariation}), we see that the initial velocity of the flow is determined in terms of $\tilde\sigma$,
\beqa
v(x,t=0) + x &=& \frac KN + n e^{nx} \dashint d z \frac{\tilde\sigma(z)}{e^{nz}-e^{nx}}\la{charVel}
\eeqa
Something nice happens when we add $i \pi \tilde\sigma(x)$ to this expression to construct the flow function $f(x,0)=v(x,0)+i\pi\tilde\sigma(x)$ -- the principal part integral comes a full complete integral, 
\beqa
f(x,0) 
= -x + \frac KN + n e^{nx}\int dz \ \frac{\sigma(z)}{z^n-e^{nx}}
\eeqa
To analytically solve the problem, let us pick a nice distribution for $\sigma$, 
\beqa
\sigma(\phi)d\phi = d\phi^n\ \frac{\sqrt{(\phi^n-a)(b-\phi^n)}}{\frac\pi2 \left(\frac{b-a}2\right)^2}\la{charRho}
\eeqa
so that the integral in $f(x,0)$ can be analytically evaluated leading to 
\beqa
f(x,0) &=&
    -x + \frac KN + \frac{8n e^{nx}}{(b-a)^2}\left(e^{nx}-\frac{a+b}2-\sqrt{(e^{nx}-a)(e^{nx}-b)}\right)
\eeqa
Note that the flow function has one ``physical" cut for $x\in [\log a,\log b]$ and in addition an infinite number of spurious branch cuts. These branch cuts can be thought of as coming from analytic continuation of the velocity on the physical cut. Now, we can use the inversion formula stating that $x+f(x,0)$ and $x-f(x,1)$ are inverses of each other -- see e.g. appendix~\ref{doubleInversionAp} -- to get
\beqa
f(x,1) &=& x-\frac1n \log\left(\frac{x-\frac KN}{x-\frac KN -n}\right) - \frac1n\log\left(\frac{a+b+\sqrt{\frac1n(x-\frac KN)(a-b)^2+4ab}}{4}\right)\la{characterFinalFlow}
\eeqa
The first term here has a branch cut with discontinuity $\frac{i\pi}n \Theta((h-\frac KN)(n+\frac KN-h))$, where $\Theta(x)=1$ for $x>0$ and zero otherwise. This is precisely the density of the Young Tableaux in (\ref{YTrectTrapz})! 

One could now go ahead and solve the flow for all $0<t<1$ and then compute the large $N$ limit of the character from the flow's action. However, let us note that there are a few shortcomings of the inversion method that are nicely illustrated in this simple character example
\begin{itemize}
    \item At intermediate times $0<t<1$, the inversion involves a transcendental equation and it cannot be done analytically
    \item The second term in the final flow function (\ref{characterFinalFlow}) has extra velocity cuts. In particular, when $4ab < 0$, we see that the square root cut from the second term overlaps with the physical cut obscuring the fluid's support. Resolving these overlapping cuts can be painful.
    \item We need the initial flow $f(x,0)$ on all its sheets. As seen in this example, the branch points can move around, so we need this information to compute the full flow.
\end{itemize}

\begin{figure}
    \includegraphics[width=0.5\textwidth]{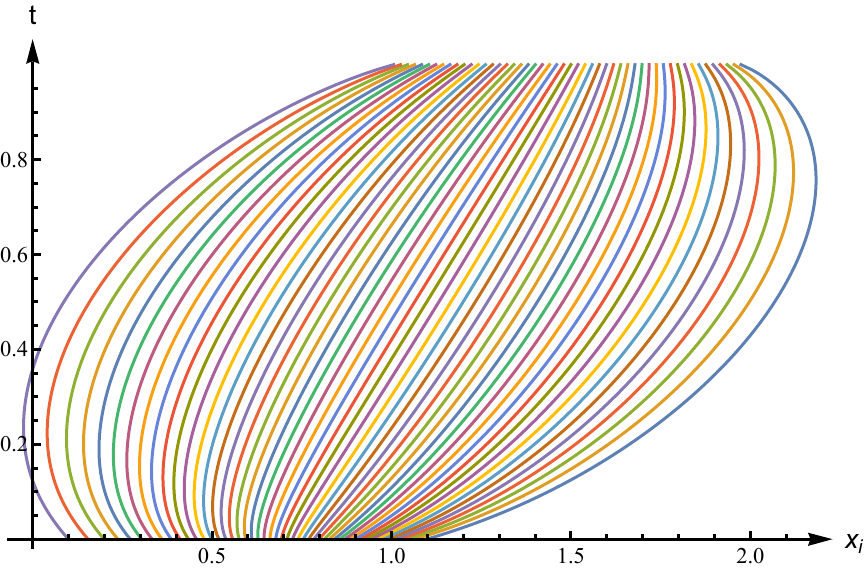}
    \includegraphics[width=0.5\textwidth]{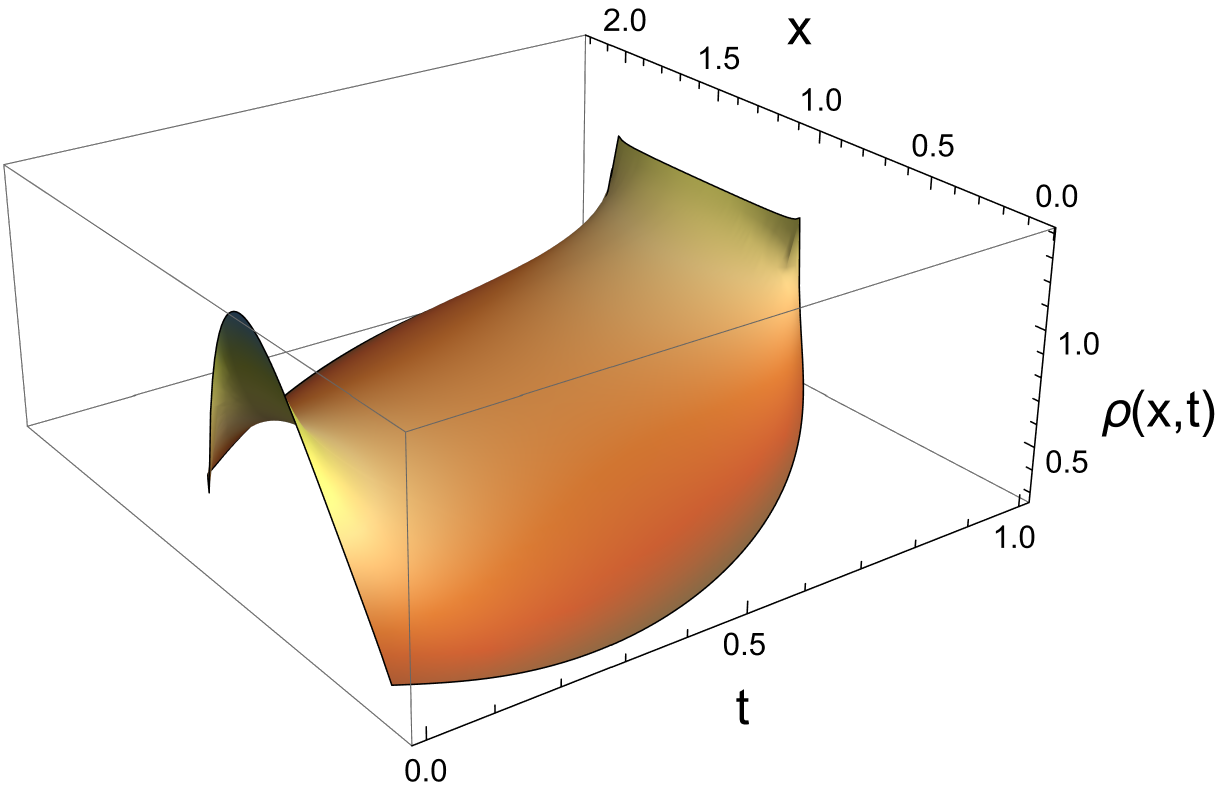}
    \caption{On the left, the discrete flow with $50$ particles for computing the character of a trapezium is shown. On the right, the eigenvalues density is plotted as function of time. It nicely interpolates between the initial density (\ref{charRho}) and the density of a logarithmic cut corresponding to the trapezium}
    \label{characterFlow}
\end{figure}

The discrete flow (\ref{discreteSol}) does not suffer from these shortcomings -- the support of the flow is manifestly the location of the eigenvalues, and the initial positions and velocities of the eigenvalues is all the information we need to solve the problem. 

Let us now solve the discrete problem with initial velocity and density following from (\ref{charVel}) and (\ref{charRho}) for the case of a rectangular YT (i.e. $n=1$). For $a=1$, $b=3$, $\frac{K}{N}=1$ and with $50$ eigenvalues $x_i$, the results are shown in figure \ref{characterFlow}. Note that in the discrete case, solving the flow at intermediate times is trivial since we can simply use (\ref{discreteSol}). The eigenvalue trajectories are shown on the left. The fluid starts out expanding but eventually the negative pressure takes over and we end up with equally spaced eigenvalues in the range $[1,2]$. Also in figure \ref{characterFlow}, we plot the density along the flow and it beautifully interpolates between (\ref{charRho}) and the constant YT density.

Using the discrete flow, we can compute the large $N$ limit of the character. The Calogero-Moser action is
\beq
    S_\texttt{CM}\,[x_i(t)] = \frac1{2N} \int_0^1 dt\left(\sum_i \dot{x}_i^2 + \sum_{j\neq k}\frac1{(x_i-x_j)^2}\right)
\eeq
The boundary terms for the flow are given by the obvious discretization of (\ref{bdyAction})\footnote{Explicitly, $S_\texttt{bdy} = \frac1{2N} \sum_i \left[x_i(0)^2 + x_i(1)^2\right] - \frac1{N^2}\log \Delta(x(0)) - \frac1{N^2}\log \Delta(x(1))$}. In figure~\ref{actionFit}, we show the convergence of $\frac1{N^2}\log \chi_R$ for the rectangle with as we change the number of particles $N$. The solid blue line is a fit obtained using only the first ten data points. The other data points lie perfectly on the curve, validating the fit. The dashed orange line is the exact result obtained by directly evaluating the character. The fit gives the correct large $N$ limit upto three decimal places.
\begin{figure}
    \includegraphics[width=0.75\textwidth]{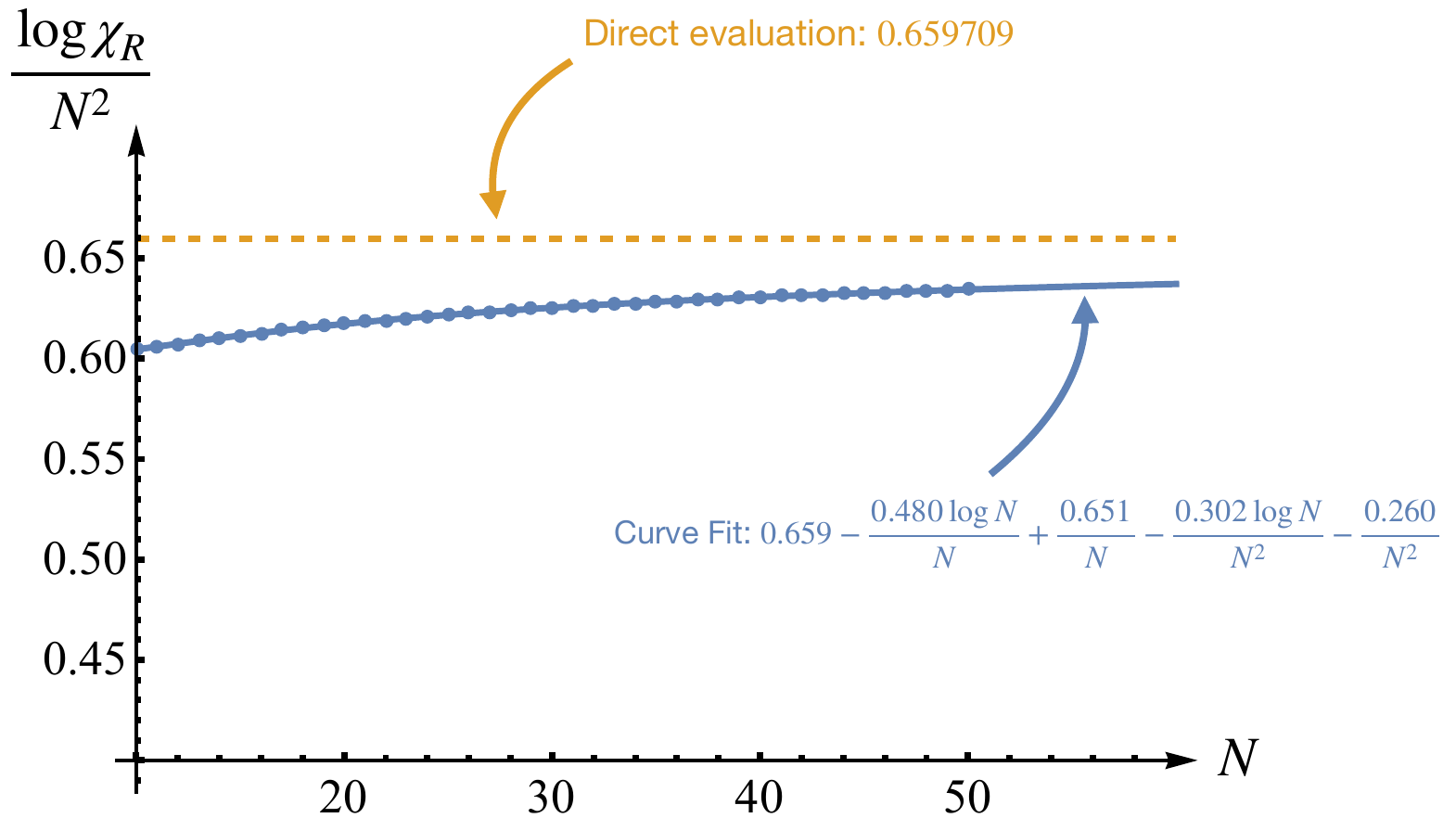}
    \centering
    \caption{Evaluating the character from flow action. The blue dots correspond to flows with various number of particles. The solid blue line is a fit obtained using only the first ten data points $N=10\ldots 20$. The dashed orange line is obtained directly from the definition (\ref{charDirect}). The fit agrees very well with the data points $N=20\ldots 50$. Furthermore, as $N\rightarrow\infty$, it agrees with the direct result upto three decimal places.}
    \label{actionFit}
\end{figure}

\subsection{Fluids And Their Deformations Then Yield The General Character.  
} \label{deformationChiSec}
Solving the discrete flow is straightforward once we have the initial positions and velocities. However, we are interested in solving a BVP and not an IVP. For instance, in the character example we managed to compute the initial velocity from the densities only for some special cases. In this section we will develop a numerical method using eigenvalue perturbation theory to deform a particular solution of the flow equations to obtain nearby flows\footnote{By ``nearby" flows, we mean flows that are continuously connected in the space of solutions. For instance, we cannot change filling fractions in a multi cut solution with this method.}. We will illustrate this method by deforming the rectangle YT in the character flow to a trapezium~YT.

Consider a flow that starts at positions $x_{0,i}$ and velocities $v_{0,i}$ at $t=0$. As usual, the position of the eigenvalues is given by (\ref{discreteSol}). Let $\ket{i, t}$ be the $i^{\text{th}}$ eigenvector at time $t$. A small deformation changes the flow as follows,
\begin{eqnarray}
    \delta x_i &=& \bra{i,t} \delta\biggl(\diag{x_0} + t P(x_0, v_0)\biggr)\ket{i,t}\\
    \delta \ket{i,t} &=& \sum_{j\neq i} \frac{\bra{j,t}\delta\biggl(\diag{x_0} + t P(x_0, v_0)\biggr)\ket{i,t}}{x_i - x_j} \ket{j,t}
\end{eqnarray}
Note that this is safe to do as long as there are no level crossings at the time $t$ under consideration. Using this we can immediately compute the velocity as,
\begin{eqnarray}
    v_i(t) &=& \bra{i,t} \frac{\partial}{\partial t} \left(\diag{x_0} + t P(x_0,v_0)\right)\ket{i,t}\\
    &=& \bra{i,t} P(x_0, v_0)\ket{i,t}
\end{eqnarray}
Later, we will also be interested in variations of the velocity. Defining $H\equiv \diag{x_0} + t P(x_0,v_0)$, this is given by,
\begin{eqnarray}
    \delta v_i &=& (\delta\bra{i,t})\ P \ket{i,t} + \bra{i,t}\delta P\ket{i,t} + \bra{i,t} P\ (\delta\ket{i,t})\\
    &=& \bra{i}\delta P\ket{i} + \sum_{j\neq i} \frac{\bra{i}P\ket{j}\bra{j}\delta H\ket{i}+\bra{j}P\ket{i}\bra{i}\delta H\ket{j}}{x_i - x_j}\label{velPert}
\end{eqnarray}
where the $t$ dependence is implicit in the last line. 

\begin{figure}[t]
    \centering
    \includegraphics[width=\textwidth]{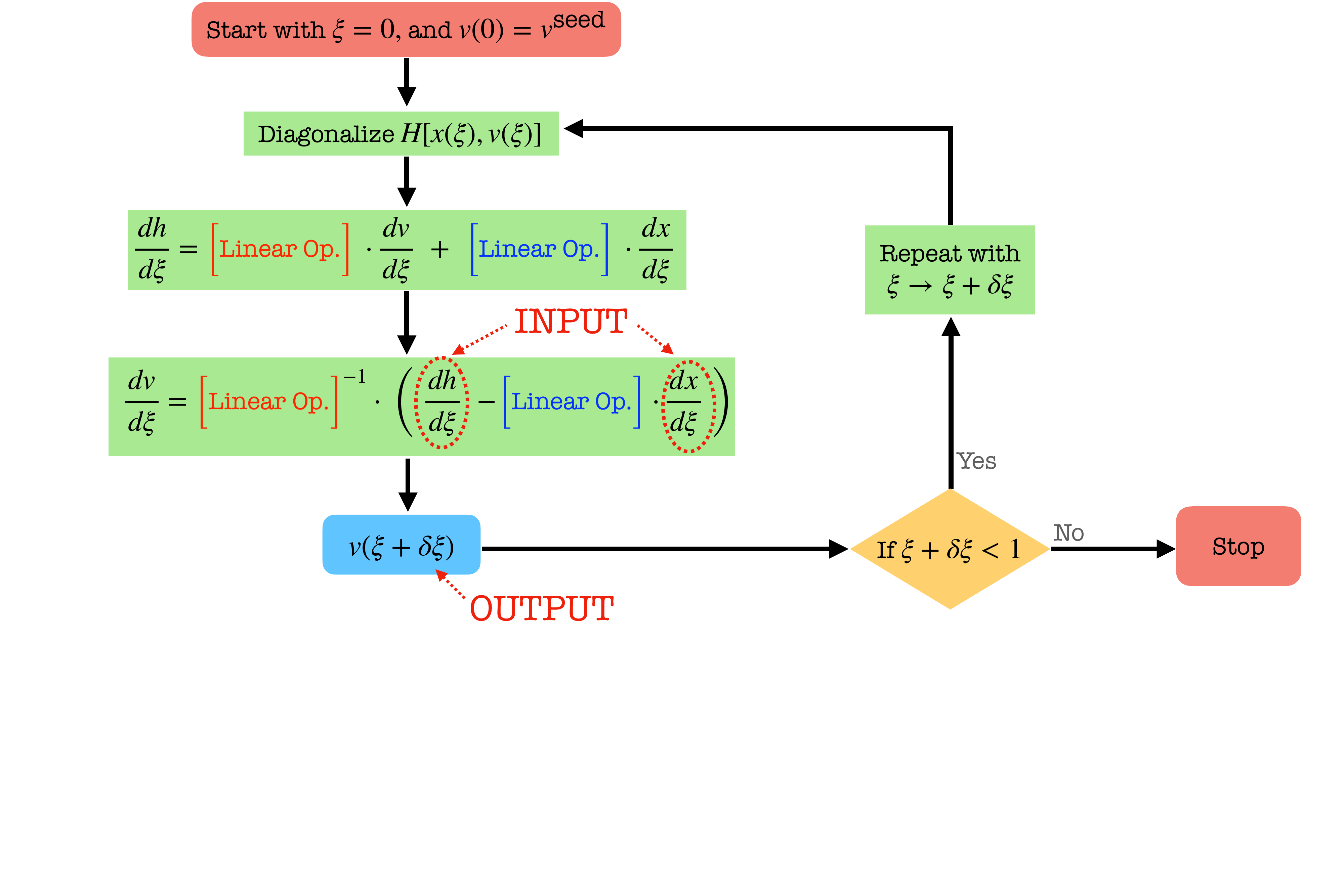}
    \vspace{-3.5cm}
    \caption{A flow chart depicting the algorithm we use to deform the flow for the character examples of this section. We start with a seed flow at $\xi=0$ and choose $h'(\xi)$ and $x'(\xi)$ as inputs according to (\ref{charDerivsChoice}). The algorithm then spits out the initial flow velocity at $\xi=1$.}
    \label{fig:flowChartChar}
\end{figure}

The main point here is that we can convert these first order variations into differential equations. One can then integrate to obtain finite deformations of the flow. 

Let us introduce a deformation parameter $\xi$ that interpolates between a known rectangle YT flow at $\xi=0$ and a new flow that we want to solve for at $\xi=1$. We pick the derivatives $\tfrac{dh}{d\xi}$ and $\tfrac{dx}{d\xi}$ such that 
\beq
\int_0^1 d\xi\ \frac{dh_i}{d\xi} = h^{\texttt{target}}_i - h^{(0)}_i \quad \text{ and }\quad\int_0^1 d\xi\ \frac{dx_i}{d\xi} = x^{\texttt{target}}_i - x^{(0)}_i\la{charDerivsChoice}
\eeq
Now, we have
\begin{eqnarray}
    \frac{dh_i}{d\xi} &=& \bra{i,\xi} \text{diag}\left(\frac{dx}{d\xi}\right) + \frac{dP(x(\xi),v(\xi))}{d\xi}\ket{i,\xi}\la{derivCharPert}
\end{eqnarray}
We can solve for $\tfrac{dv}{d\xi}$ and integrate to obtain the $v(\xi)$. The most expensive step here is to obtain the eigenvectors $\ket{i,\xi}$ at each $\xi$. The algorithm we use for deforming the flow is shown in figure \ref{fig:flowChartChar}.

\begin{figure}[t]
    \centering
    \includegraphics[width=\textwidth]{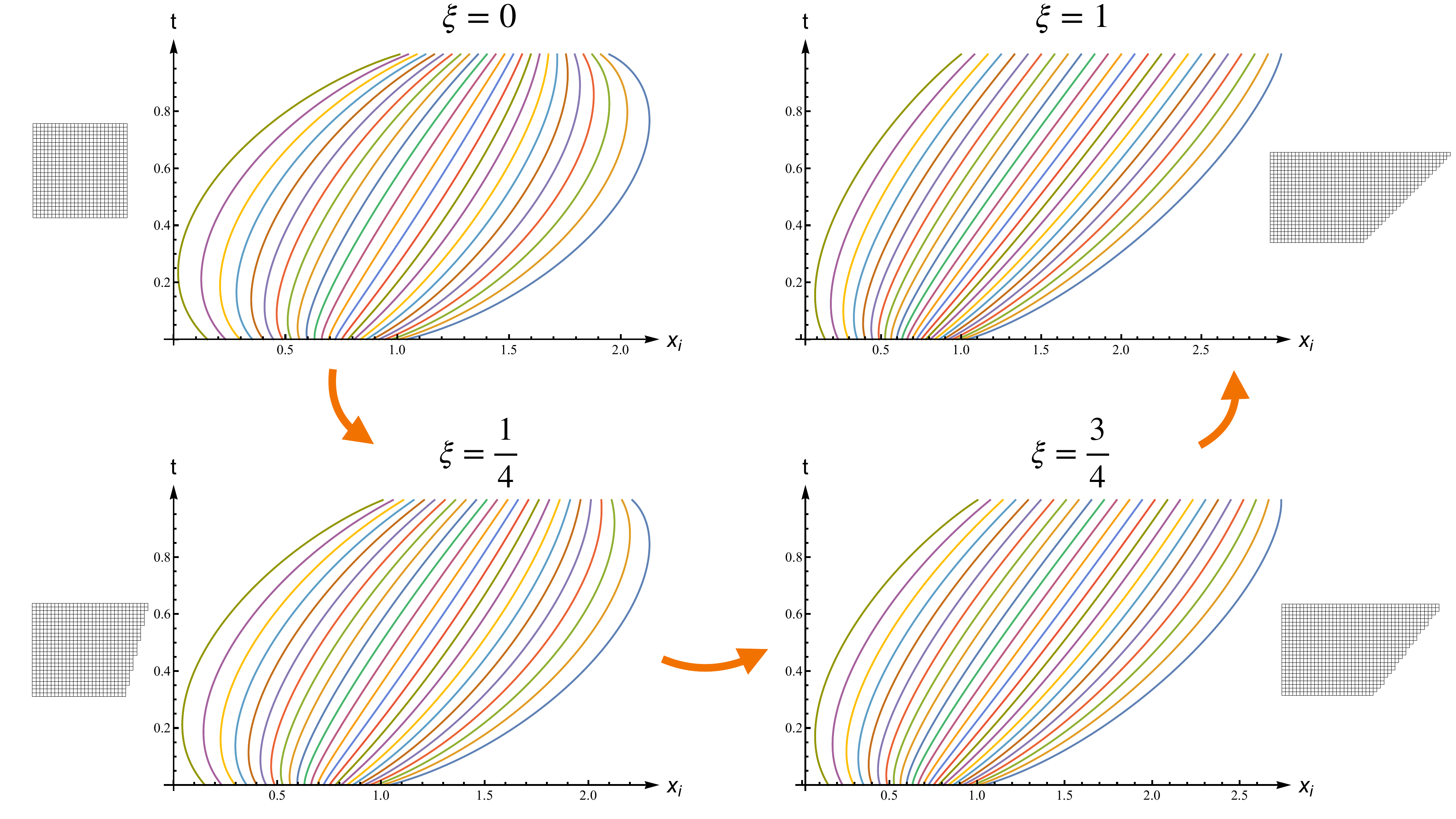}
    \caption{Deforming one flow into another. At $\xi=0$, we have the rectangle YT flow with $N=25$ eigenvalues. Using adiabatic deformation, we deform it to a trapezium YT at $\xi=1$ shown in the bottom right panel. This procedure gives us for free the flows for the intermediate YTs as illustrated by the $\xi=\frac14$ and $\frac 34$ cases here.}
    \label{rect2trapz}
\end{figure}
Let us now pick $h^{(0)}$ to be a rectangle YT with $\frac mN=1$ and the $h^{\texttt{target}}$ to be a trapezium also with $\frac{m}N=1$ in (\ref{YTrectTrapz}). For simplicity, let $x^{\texttt{target}} = x^{(0)}$. The deformation rates are chosen to be constants, $h_i'(\xi) = h_i^{\texttt{target}}-h_i^{(0)}$ and $x_i'(\xi)=0$. With this, we have 
\beq
h_i^{\texttt{target}}-h_i^{(0)} = \bra{i,\xi} \diag{v'(\xi)}\ket{i,\xi}
\eeq
Solving this ODE for $N=25$ particles, we obtain the results shown in figure \ref{rect2trapz}. In addition to the final trapezium YT, we also obtain the flows for all intermediate $\xi$'s which correspond to trapezia of various slopes as shown in the figure.

\section{Hermites: An Exact Solution}\label{Hap}
We claim that the roots $z_i(t)$ in (\ref{HermiteRoots}) provide an exact solution to the flow equations (\ref{discrete}) if the  radius $a(t)$ and center at $b(t)$ are chosen cleverly. 
Let $Z_i$ be a zero of the Hermite polynomials in (\ref{HermiteRoots}) so that
\beq
\sqrt{2N}\ \frac{z_i(t) - b(t)}{a(t)}=Z_i \,\la{Zi}
\eeq
is a constant. Taking up to two derivatives of this equation with respect to $t$ yields 
\beq
z_i''(t) = z_i(t) \times \frac{a''(t)}{a(t)}+\Big( b''(t)-b(t) \frac{a''(t)}{a(t)}\Big)= \frac{a''(t)}{a(t)} \Big(z_i(t)-b(t) \Big) \,.
\eeq
where in the second equality, we used $b''(t)=0$ because it is the acceleration of the center of mass. That means that we simply need to show that 
\beq
\sqrt{2N}\frac{z_i(t)-b(t)}{a(t)}=-\frac{\sqrt{2N}}{a''(t)} \frac{2}{N^2}\sum_{j\neq i} \frac{1}{(z_i(t)-z_j(t))^3} 
\eeq
or 
\beq
Z_i=-\frac{8}{a''(t)a(t)^3} \sum_{j\neq i} \frac{1}{(Z_i-Z_j)^3} \label{almost} 
\eeq
Here we use a known fact show in \cite{calogeroHermite,Agarwal:2019omb} which is that the roots of Hermite obey the Calogero equilibrium condition 
\beq
Z_i=2 \sum_{j\neq i} \frac{1}{(Z_i-Z_j)^3} 
\eeq
So all we need to impose is for the prefactor in (\ref{almost}) to be equal to $-2$. This leads to a differential equation for $a(t)$ which we can easilly solve. We thus conclude that if 
\beqa
    a(t)&=& \sqrt{a(0)^2(1 + \alpha t)^2-4t^2/a(0)^2} \,,\\
    b(t)&=&b(0)+t(\beta + \alpha b(0))
\eeqa
the roots (\ref{HermiteRoots}) provide an exact solution to the flow equations. This is true for any value of $\alpha$ and $\beta$. Taking the derivative of (\ref{Zi}) at $t=0$ yields,
\beq
    z_i'(0) = \alpha z_i(0) + \beta
\eeq
For our flows we have the extra regularity condition (\ref{xGlueSymm}) which fixes these constant once we use an even more well known identity for the Hermite polynomial roots, namely that
\beq
Z_i= \sum_{j\neq i} \frac{1}{Z_i-Z_j} \, .\label{hermiteSP}
\eeq
which plugged into (\ref{xGlueSymm}) forces 
\beq
\alpha = -\frac23 \left(1-\frac1{a(0)^2}\right), \quad \beta=-\frac{2b(0)}{3a(0)^2}
\eeq

\section{Flows for Very Long Young Tableaux}\la{apLongYT}
In this appendix, we will derive the flow described in figure \ref{fig:longYTFlow}. As explained in the main text, when the first $K$ eigenvalues are such that
\beq
    |x_i+t v_i| \gg \frac1{|x_j-x_k|}\ , \qquad \text{for } i<K \text{ and } j\neq k,
\eeq
they decouple from the rest and evolve freely as $x_{n,i}(t) = x_i+t v_{n,i}$. Recall that the gluing condition at $X$ is given by,
\beqa
    2x_i + \sum_n v_{n,i} = \frac1N \sum_{j\neq i} \frac1{x_i-x_j} \approx 0
\eeqa
Together with the free evolution, $m_{n,i}=x_i+v_{n,i}$, this implies
\beq
x_i = \sum_{n=1}^3 m_{n,i}\ .
\eeq

Similarly, in the second chamber the tube eigenvalues also evolve freely if we have $\left|-\log m_{n,i}+t w_{n,i}\right| \gg |\log m_{n,i}-\log m_{n,j}|^{-1}$. Again, as explained in the Section \ref{longYTsec}, the final positions are a permutation of the YT weights, $y_{n,i} = h_{n,\pi_n(i)}$. The gluing condition between first and second chamber reads,
\beqa
    w_{n,i} &=&  \log m_{n,i} - m_{n,i} v_{n,i}\\
    &=& \log m_{n,i} + y_{n,i}
\eeqa
where the last line follows from the free evolution. Plugging in the expressions for $x_i$ and $v_{n,i} = m_{n,i}-x_{n,i}$, we see that
\beqa
    y_{n,i} &=& m_{n,i}\left(-m_{n,i} + \sum_p m_{p,i}\right)\ ,\qquad n=1,2 \text{ and }3
\eeqa
Solving the above equations,
\beqa
    m_{n,i} &=& \frac1{\sqrt2} \sqrt{\frac{(y_{n,i}+y_{n+1,i}-y_{n-1,i})(y_{n-1,i}+y_{n,i}-y_{n+1,i})}{y_{n-1,i}+y_{n+1,i}-y_{n,i}}}\\
    x_i &=& -\frac1{\sqrt 2}\frac{\sum_{n=1}^3 (y_{n,i}^2-2y_{n-1,i}\, y_{n+1,i})}{\prod_{n=1}^3 \sqrt{y_{n-1,i} + y_{n+1,i}-y_{n,i}}}
\eeqa
where $n$ is defined modulo three. Notice that this is precisely the solution to the SPEs of a three Hermites (\ref{threeHermiteSPE}), hinting at the result that the full three point function is simply a product of $N$ copies of the symmetric YT three point function.

Now, let us revisit the assumptions and check for self consistency. Let the YT weights be $h_{n,i} \sim L\gg 1$ for $i=1\ldots K$. Then, $m_{n,i} \sim x_i \sim \sqrt{L}\gg 1$ and we have,
\beqa
x_i + t v_{n,i} &=& (1-t)x_i + t m_{n,i} \sim L^{\frac12} \\
&\gg& \frac1{x_j-x_k} \sim L^{-\frac12}
\eeqa
So, in the first chamber, decoupling holds. Similarly, in the second chamber, it is clear that $|-\log m_{n,i}+t w_{n,i}| \gg 1 \gg |\log m_{n,i}-\log m_{n,j}|^{-1}$.

The asymptotics of the three point function in this limit can be computed by plugging in this solution to the flow action. Plugging in the simple free flow and dropping subleading terms, we see that the action is given by
\begin{multline}
    \frac1{N^2L}\log Z \approx \frac1{2NL}\left[- \sum_i x_i^2 - \sum_{n ,i}m_{n,i}^2 \underbrace{-\sum_{n,i}(m_{n,i}-x_i)^2}_{S_{\texttt{fluid-1}}} + \underbrace{\sum_{n,i}\left(x_i^2 + m_{n,i}^2\right)}_{S_{\texttt{bdy-1}}}\right.\\
    \left. \underbrace{-\sum_{n,i} (\log m_{n,i} + y_{n,i})^2}_{S_{\texttt{fluid-2}}}+\underbrace{\sum_{n,i}\left((\log m_{n,i})^2+y_{n,i}^2\right)}_{S_{\texttt{bdy-2}}}\right]
\end{multline}
Plugging in the expression for the $m$'s and $x$'s, and normalizing the three point function, we obtain 
\beqa
\frac1{N^2L} \log Z \simeq \max_{\pi,\pi'\in S_K}\left[\frac1{NL}\sum_i \texttt{tube}(h_{1,i}, h_{2,\pi(i)}, h_{3,\pi'(i)})\right]
\eeqa
Note here that we dropped one of the permutations to avoid overcounting the flows.

\bibliographystyle{JHEP}
\bibliography{Plane}


%
%
%
%
%
%

\end{document}